\documentclass[fleqn,usenatbib]{mnras}

\usepackage{newtxtext,newtxmath}

\usepackage[T1]{fontenc}
\usepackage{ae,aecompl}

\usepackage{graphicx}	
\usepackage{amsmath}	
\usepackage{amssymb}	

\newcommand{\pdv}[2]{\frac{\partial #1}{\partial #2}}

\newcommand{\rot}{\mathbf{\nabla}\times}

\newcommand{\dotM}{\dot{M}}

\newcommand{\Tfrac}[2]{\left(\frac{#1}{#2}\right)}
\newcommand{\divi}{\mathbf{\nabla}\cdot}
\newcommand{\vgrad}[1]{\mathbf{\nabla}{#1}}
\newcommand{\ephi}{\mathbf{e_{\phi}}}
\newcommand*\diff{\mathop{}\!\mathrm{d}}
\newcommand{\sub}[1]{_{\mathrm{#1}}}

\DeclareMathAlphabet\mathbfcal{OMS}{cmsy}{b}{n}

\newcommand{\aderiv}[1]{\frac{\diff #1}{\diff{a}}}
\newcommand{\upo}{\mathbf{u_{p}}}
\newcommand{\Bpo}{\mathbf{B_{p}}}
\newcommand{\Int}{\int\limits}
\newcommand{\spacio}{\hspace{4px}}

\title[Jets and Winds from accretion disks]{Magnetically-driven jets and winds from weakly magnetized accretion disks}

\author[J. Jacquemin-Ide et al.]{
J. Jacquemin-Ide,$^{1}$\thanks{E-mail: jonatan.jacquemin@univ-grenoble-alpes.fr}
J. Ferreira,$^{1}$
G. Lesur$^{1}$
\\
$^{1}$Univ. Grenoble Alpes, CNRS, IPAG, 38000 Grenoble, France
}

\date{Accepted 23/09/2019. Received 16/09/2019; in original form 22/07/2019}

\pubyear{2019}

\begin{document}
\label{firstpage}
\pagerange{\pageref{firstpage}--\pageref{lastpage}}
\maketitle

\begin{abstract}
Semi-analytical models of disk outflows have successfully described magnetically-driven, self-confined super-Alfv\'enic jets from near Keplerian accretion disks. These Jet Emitting Disks are possible for high levels of disk magnetization $\mu$ defined as $\mu=2/\beta$ where beta is the usual plasma parameter. In near-equipartition JEDs, accretion is supersonic and jets carry away most of the disk angular momentum. However, these solutions prove difficult to compare with cutting edge numerical simulations, for the reason that numerical simulations show wind-like outflows but in the domain of small magnetization. In this work, we present for the first time self-similar isothermal solutions for accretion-ejection structures at small magnetization levels. We elucidate the role of MRI-like structures in the acceleration processes that drive this new class of solutions. The disk magnetization $\mu$ is the main control parameter: massive outflows driven by the pressure of the toroidal magnetic field are obtained up to $\mu \sim 10^{-2}$, while more tenuous centrifugally-driven outflows are obtained at larger $\mu$ values. The generalized parameter space and the astrophysical consequences are discussed. We believe that these new solutions could be a stepping stone in understanding the way astrophysical disks drive either winds or jets. Defining jets as self-confined outflows and winds as uncollimated outflows, we propose a simple analytical criterion based on the initial energy content of the outflow, to discriminate jets from winds. We show that jet solution are achieved at all magnetization level, while winds could be obtained only in weakly magnetized disks that feature heating. 
\end{abstract}



\section{Introduction}

Jets are observed from a wide variety of astrophysical objects. They are emitted from the central regions of young stellar objects where a protostar is being born \citep{Burrows1996,hirth_spatial_1997,ray_hst_1996,hirth_spatial_1997,dougados_t_2000,bally_observations_2007}, from the central core of active galactic nuclei and quasars (\cite{merloni_fundamental_2003} and references therein ), and also from the compact object of a binary system, be it a black hole, a neutron star or even a white dwarf \citep{mirabel_sources_1999,corbel2000,gallo_universal_2003,gallo_black_2005,coppejans_novalike_2015}. These jets are detected by different means in each environment, mostly in radio around compact objects (interpreted as self-absorbed synchroton emission) and in emission lines (from radio to optical) in young stellar objects. While jets from compact objets are relativistic, those from young forming stars have speeds ranging from 100 to 600 km/s. They do however share some properties: they are both supersonic, have small opening angles already close to the source and they exhibit a tight correlation with the underlying accretion disk \citep{cabrit_forbidden-line_1990,hartigan_disk_1995,serjeant_radio-optical_1998,markoff_exploring_2003,ferreira_which_2006}. These are evidences for two important aspects: (i) the acceleration process must also be related to the jet confinement issue, calling therefore for an initial self-collimation instead of an external confinement; (ii) the mass and/or also power that are feeding the jets must be related to the accretion activity.  

It is now accepted that large scale magnetic fields anchored on a rotating object are a necessary ingredient for launching self-confined outflows \citep{konigl_disk_2000,Ferreira_theory_2002, Pudritz07,Hawley2015}. This rotating object could be either the central object or the surrounding accretion disk. However, given the universality of the process, it sounds reasonable to rely on their common denominator, namely the accretion disk as proposed by \citet{blan82}. The physical ingredients that need to be included in a self-consistent steady-state accretion-ejection model are then the following: (1) a near-keplerian disk of plasma surrounding a central mass; (2) a large scale vertical magnetic field threading the disk; (3) the possibility of mass diffusion through the field, so that the disk material can accrete onto the central object while leaving the magnetic field behind. Accretion is then driven by two possible mechanisms: the turbulent torque as proposed by \citet{shak73} and the laminar torque due to the jets themselves \citep{blan82}. While the latter is a natural consequence of the presence of magnetic jets, the former requires the existence of some self-sustained turbulence within the disk. 

Since the seminal work of \citet{balb91}, it is now well known that magnetized disks are unstable to the magnetorotational instability (hereafter MRI). The saturated state of the MRI is a 3D magnetohydrodynamic (MHD) turbulence giving rise to an anomalous radial transport of angular momentum, that can indeed be described by a turbulent viscosity (see \citep{balb03} and references therein). 
While it is now quite commonly argued that magnetic fields are present in accretion disks, the question of their topology remains open. 
Indeed, 3D global numerical simulations show that as long as there is no large scale vertical magnetic field, MRI is present (thus accretion proceeds) but no jet is launched (\citep{beck08}, see however \citet{lisk18}). 
Thus, the process of launching jets from accretion disks, that will carry away mass, energy and angular momentum, requires the presence of a large scale vertical field. 
In that case then, whatever the relative importance of the jet torque to the turbulent torque, steady accretion is achieved only if the disk material is allowed to diffuse through the magnetic field. 
The origin of this diffusion remains one of the less studied aspects of accretion-ejection theory. One possibility is the influence of non-ideal MHD processes, such as Ohmic resistivity, ambipolar diffusion and Hall effect \citep{koni89, ward93,Fle2000, salm07, salm11, gres15, beth16}. However, while relevant in outer regions of protostellar accretion disks, innermost disk  regions and disks around active galactic nuclei and X-ray Binary are ionized enough and these effects vanish. One needs therefore to rely on another source for diffusion, the MHD turbulence itself. 

Building upon this idea, \cite{ferr93a} analyzed the general conditions for designing self-similar steady-state models of accretion-ejection structures. Within this model, most of the disk plasma accretes in a resistive (turbulent) MHD region around the disk midplane, while a fraction is deviated vertically and is smoothly connected to an ideal MHD zone, where it crosses the usual MHD critical points. \citet{ferr95} obtained the first outflow solutions becoming super-Slow Magnetosonic and showed, for the first time, that the required magnetic field needs to be smaller than but close to equipartition with the total (gas plus radiation) pressure, see also \cite{Li1995}. This result was then generalized to super-Alfv\'enic \citep{ferr97} and super-Fast Magnetosonic \citep{ferr04} jets. In these highly magnetized solutions, termed Jet Emitting Disks (hereafter JED), the inclusion of a turbulent viscous torque appears to play no significant role \citep{cass00a}, most of the disk angular momentum being extracted by the jets. On the other hand, the mass loaded in the jets happens to be highly dependent on the thermodynamic conditions at the disk upper layers: allowing for some heat deposition (coronal heating) is shown to lead to a significant enhancement of the ejected mass \citep{cass00b}. 

Most results shown in these early works have been confirmed by other groups, in particular using 2D numerical simulations of "alpha"  disks, where viscosity and magnetic diffusivity are prescribed using an alpha law (see eg. \cite{cass02, zann07, tzef09, tzef13}). The main caveat of these alpha-disk simulations is their possible inconsistency with MHD turbulence, since all anomalous transport coefficients are parametrized. In order to probe the analytical results on accretion-ejection structures, 3D global MHD simulations of turbulent accretion disks with large scale magnetic fields must be done. 
But achieving reliable simulations of this kind is a fantastic task. Indeed, MHD turbulence in a vertically stratified disk needs to be properly followed to make sure that the simulations have converged. Moreover, when a vertical magnetic field is included, mass loss is systematically observed from the disk surface, loaded field lines become bent and ejection is obtained. As a consequence, the size of the computational domain needs to be large enough so that boundary conditions do not affect (or not too much) the outcome of the simulation. These are the main reasons why shearing box simulations could hardly address MRI with a non zero net magnetic flux (see for instance \cite{lesu13,from13,bai13} and references therein). Therefore, for quite a long time, the main focus of MRI studies was the measurement of the Shakura-Sunayev alpha parameter describing the turbulent viscosity and the influence of non ideal MHD effects. It became clear only recently that these MRI-driven outflows would carry away some angular momentum as well, possibly affecting the structure of weakly ionized accretion disks \citep{bai11b, bai16, Scepi2018}. 

Converged global simulations of accretion disks threaded by a weak vertical magnetic field have been obtained for a plasma beta around $10^4$ \citep{suzu14,beth17,Zhu_Stone}. Super-Alfv\'enic flows are systematically obtained and, in the case of \citep{Zhu_Stone}, there are even indications of some collimation occurring within the domain. Despite the presence of the vertical laminar torque due to these outflows, most of the disk angular momentum is transported outwardly in the radial direction. As a consequence, the power carried away by these "winds"  remains a small fraction of the released accretion power. But the existence of super-Alfv\'enic outflows from weakly magnetized accretion disks is in contradiction with the analytical (JED) model. Although self-similarity introduces some biases in the flow solutions, it does allow to take into account all dynamical terms. Thus, the results of those global simulations motivated us to revisit the analytical theory of accretion-ejection structures and to seek for new outflow solutions at low disk magnetization levels.  
 
The paper is organized as follows. Section~2 provides the governing equations and assumptions allowing to describe steady-state accretion disks driving jets. The JED parameters are introduced and those describing the MHD turbulence are discussed in the framework of MRI simulations. It will be shown that a condition, used to obtain the previously published solutions, must be disregarded in order to be consistent with global MRI simulations. New solutions, obtained at low magnetization levels, are then indeed naturally obtained. Section~3 describes the new parameter space of super-Slow Magnetosonic (SM) flows. Although there is no MRI in our steady-state calculations, it is shown that these winds are a natural outcome of MRI-like modes or MRI channel flows in stratified unbounded flows. Super-Alfv\'enic flows are then a subset of these super-SM solutions and their properties are shown in Section~4. Section~5 analyses the effects of the turbulence parameters on the solutions and in particular on the disk mass loss, with a possible clear distinction between jets and winds. Some caveats of our study and comparison with other works are then presented in Section~6. We conclude in Section~7 by discussing some astrophysical implications.

\section{Describing accretion-ejection structures}

\subsection{Governing equations}
\label{sec:Eq_para}

Accretion-Ejection structures are described in the framework of axisymmetric MHD. The plasma velocity and magnetic field can be decomposed into poloidal and toroidal components, $\mathbf{u}=\upo+\Omega r\ephi$ and $\mathbf{B}=\Bpo+B_\phi\ephi$ respectively. The poloidal magnetic field can then be written 
\begin{equation}
    \Bpo = \frac{1}{r}\vgrad a \times \ephi
\end{equation}
where $a(r,z)$ is related to the vector potential $\mathbf{A}$ by $a=rA_\phi$ and is the magnetic flux function. A poloidal magnetic surface is defined by a constant vertical flux, namely $a(r,z)=a(r_o,0)$ where $r_o$ is the cylindrical anchoring radius of the magnetic surface. The magnetic field topology is then assumed bipolar with an even symmetry with respect to the disk equatorial plane. This translates into an even function $a(r,z)$ in $z$ and an odd function $B_\phi$.  The set of MHD equations are then the following \citep{ferr97,cass00a}
\begin{align}
	\label{eq:Scon}
& \divi(\rho \mathbf{u})=0 \\
	\label{eq:Smonp}
& \rho(\mathbf{u_p}\cdot\vgrad{})\mathbf{u_p}=-\vgrad{P}+\rho\vgrad{\Phi_{G}}+ J_\phi \frac{\vgrad a}{r} - \frac{\vgrad (r B_\phi)^2}{2 \mu_o r^2} \\
	\label{eq:Smonto}
&	\divi\left[\rho \Omega r^2 \upo - \frac{ rB_\phi}{\mu_0}\Bpo -r\mathbfcal{T}\right]=0\\
	\label{eq:Sohm}
&	\nu_{m}J_{\phi}\ephi=\frac{1}{\mu_0}\upo\times\Bpo\spacio \\
	\label{eq:Sinduc}
&	\divi\left(\frac{\nu'_{m}}{r^{2}}\vgrad{rB_{\phi}}\right)=\divi \frac{1}{r}  \left(B_{\phi}\upo-\Bpo\Omega r\right)
\end{align}
with $\rho$ the total mass density, $P$ the thermal pressure of the plasma, $\mathbf{J}=\frac{1}{\mu_0}\rot \mathbf{B}$ the plasma electric current density, $\nu_{m}$ and $\nu'_{m}$ are the anomalous resistivities (poloidal and toroidal respectively), $\Phi_{G}= - GM/\sqrt{r^2+z^2}$ the gravitational potential of the central object of mass $M$ (ignoring the self gravitation of the disk) and $\mathbfcal{T}= \mathcal {T}_{r\phi} \mathbf{e}_r$ where $\mathcal{T}_{r\phi}$ is the radial stress of turbulent origin, associated to an anomalous viscosity $\nu_v$ \citep{shak73}.

This set of equations is closed using the equation of state for a perfect gas. In this paper, the temperature will be assumed to be remain constant along each magnetic surface (isothermal solutions). The specific form of this equation of state is detailed in Appendix \ref{A:self_eq}.

\subsection{MHD turbulence and transport coefficients}

As discussed earlier, the disk is assumed to be fully turbulent and that such a turbulence can be described using a mean field approach with anomalous transport coefficients. This is the alpha-disk description introduced by \citet{shak73}. The disk is then defined as the densest region around the equatorial plane, of scale-height $h(r)$, inside which accretion and turbulence are taking place. 
The real disk scale height is provided by the vertical balance between gravity, magnetic forces and plasma pressure support. It is thus only known once a full solution (including the energy equation) is computed. It is however practical to define the hydrostatic scale height such that $C_s = \Omega_K h$, where $C_s$ is the midplane isothermal sound speed and  $\Omega_K = \sqrt{GM/r^3}$ the Keplerian angular velocity. While accurately providing the scale height of standard accretion disks, it is a slight overestimate in the case of strongly magnetized disks, usually by a factor 2 or so \citep{ferr95}. In this work, the disk aspect ratio $\epsilon= h/r = C_s/\Omega_Kr$ will be used as a free parameter for prescribing the temperature at the disk equatorial plane.   

Whatever the instability that triggers and sustains the MHD turbulence, we assume that it translates into a viscosity as well as a magnetic diffusivity, namely that the turbulent electromotive force is proportional to the mean electric current density. All transport coefficients are then assumed to vanish outside the disk, the jet region being described within the ideal MHD regime. Our description must then allow for a smooth transition from a resistive viscous MHD regime (the disk) to an ideal MHD regime (the jet) on a few disk height scales. For simplicity, will use the same gaussian profile for the vertical behavior of all transport coefficients (see Appendix \ref{A:self_eq}).   

The amplitude of the turbulent transport coefficients is then prescribed as follows: 
\begin{itemize} 
\item Viscosity $\nu_v$: following  \citet{shak73} it is chosen as $\nu_v = \alpha_v C_s h$, where $C_s$ is the midplane sound speed and $\alpha_v$ is the usual turbulence parameter. 
   
\item Poloidal diffusivity  $\nu_m$: it is the magnetic diffusivity acting on the poloidal magnetic field (or $J_\phi$), responsible for the bending of the magnetic field (as measured for instance by the ratio $B_r^+/B_z$ at the disk surface) and allowing steady accretion through the magnetic field. Following the initial prescription made in \citet{ferr93a}, we use $\nu_m= \alpha_m V_A h$, where $V_A$ is the Alfv\'en velocity at the disk midplane and $\alpha_m$ a constant.  

\item Toroidal diffusivity  $\nu'_m$: it is the magnetic diffusivity acting on the toroidal magnetic field (or $J_p$). It is responsible for the magnetic shear (as measured by the ratio $-B_\phi^+/B_z$ at the disk surface) and allowing for a steady rotation. Given our lack of knowledge on these aspects in turbulent MHD disks, we follow  \citet{ferr95} and use $\nu'_m= \nu_m/ \chi_m$, where $\chi_m$ is a measure of a possible anisotropy. 

\end{itemize}

In the first studies of JEDs \citep{ferr95,ferr97}, only the jet torque was taken into account and the relevant turbulent parameters were $\alpha_m$ and $\chi_m$. When the viscous torque was included in the equations \citep{cass00a}, the Shakura-Sunyaev $\alpha_v$ parameter was computed using the effective magnetic Prandtl number  $\mathcal{P}_m = \nu_v/\nu_m$, namely 
\begin{equation} 
\alpha_v= \alpha_m \mathcal{P}_m \mu^{1/2}
\label{eq:av1}
\end{equation}
where $\mu= V_A^2/C_s^2= B^2/\mu_o P$ is the disk magnetization measured at the disk midplane (note that $\mu=2/\beta$ where $\beta$ is the usual plasma beta only in accretion disks dominated by the gas pressure). Hence, assuming a constant $\mathcal{P}_m$ leads to a Shakura-Sunyaev $\alpha_v$ parameter scaling with the disk magnetization, whereas the magnetic diffusivity $\alpha_m$ remains a constant. All results published so far on JEDs verify this property. It is striking to realize that MRI studies actually provide 
\begin{equation} 
\alpha_v = \alpha_o \mu^{1/2}
\label{eq:av2}
\end{equation}
where $\alpha_o \simeq 7$ according to \cite{Salvesen_16} and the scaling law $\alpha_v \propto \mu^{1/2}$ previously identified \citep{hawl95}. Identifying Eq.(\ref{eq:av1}) to Eq.(\ref{eq:av2}) leads to $\alpha_m = \alpha_o/\mathcal{P}_m$ which is indeed a constant. Now, measuring the magnetic diffusivity in turbulent accretion disks is a tricky task, especially in global simulations. To our knowledge, only few works attempted it and found, using different approaches, $\mathcal{P}_m$ and $\chi_m$ both slightly larger than, but of the order, unity \citep{lesu09,guan09,from09}, consistent with estimates done in global simulations \citep{Zhu_Stone}. Note however that the anisotropy parameter $\chi_m$ has been measured only in one configuration (radial diffusion of a vertical field compared to the vertical diffusion of a toroidal field). Because of the absence of data in the more general case (vertical diffusion of a radial field for instance) the value of the anisotropy parameter $\chi_m$ will be used here as a free parameter.

To conclude, using $\mathcal{P}_m$, $\chi_m$ and $\alpha_m \geq 1$ as free constants and $\alpha_v= \alpha_m \mathcal{P}_m \mu^{1/2}$ is actually consistent with our current knowledge of MRI-driven turbulence.

\subsection{Parameters and method of resolution}

The full set of MHD equations (\ref{eq:Scon})-(\ref{eq:Sohm}) is solved using a self-similar Ansatz. Since gravity is expected to be the leading energy source, all other quantities will have to follow the same mathematical dependency. Such an approximation brings of course a lot of caveats but it allows to take into account all dynamical terms in the equations. For a newtonian potential, this translates into seeking solutions of the form
 \begin{equation}
    A(r,z) = A_o \left(\frac{r}{r_o}\right)^{\zeta_A}f_A(x) 
\end{equation}
where $f_A(x)$ is the profile of any quantity $A$, expressed with the self-similar variable $x= z/h(r)= z/\epsilon r$. In this radial self-similarity, a constant $x$ corresponds to a cone and the values of the exponents $\zeta_A$ are obtained by solving algebraic equations. Note that, as with many other disk models, all quantities are then a power-law of the radius. In a JED, disk mass loss must be allowed and quantified. This is done by allowing a radial dependence of the disk accretion rate 
 \begin{equation}
\dot M_a (r)= - 2\upi r \int_{-h}^{+h} dz \rho u_r \propto r^\xi
\end{equation}
where $ 1\geq \xi >0$ is the disk ejection efficiency. The bigger $\xi$ the larger the amount of ejected matter, while $\xi=0$ describes a standard accretion disk. The other exponents $\zeta_A$ can then be expressed as function of $\xi$, including the magnetic field distribution \citep{ferr93a}. See Appendix~\ref{A:self_eq} for more details. 

The global energy budget of geometrically thin accretion-ejection structures, established between an inner disk $r_i$ and an outer disk $r_e$ 
writes $P_{acc}= 2 P_{jet} + P_{diss}$. In this expression, $P_{jet}$ is the power leaving the disk and carried away by each jet while $P_{diss}$ is the power that is released within the disk through turbulent dissipation and giving rise to the disk luminosity. The accretion power is
\begin{equation}
P_{acc} = \left [ \frac{G M \dot M_a(r)}{2r} \right ]^{r_i}_{r_e} = \frac{G M \dot M_a(r_i)}{2r_i} -  \frac{G M \dot M_a(r_e)}{2r_e}
\end{equation}
and its amplitude depends thereby on how much mass is leaving the disk.  In this global budget, advection of energy into the central object scales as $P_{adv} \propto (h/r)^2 P_{acc}$ and has thus been neglected, as well as any external source of energy (such as irradiation from a central source). As a consequence, requiring that jet launching and disk luminosity are both powered by the release of mechanical energy leads to the constraint $\xi <1$. We will come back to this constraint later. 

The self-similar antsatz allows for a full description in the poloidal plane and is therefore required when dealing with the deviation of the flow from accretion to ejection. One consequence of self-similarity is that all local dimensionless quantities must be real constants, defining thereby the parameters of the solution. The list of the 7 JED parameters, evaluated at the disk midplane, is then
 \begin{equation}
 \begin{array}{lclcl}
    \epsilon = \frac{h}{r}                     &    &    \alpha_m = \frac{\nu_m}{V_Ah}   &  &    \mu = \frac{B_0^2}{\mu_0P_0}  \\ 
 \xi =\frac{d\ln{\dotM_a}}{d\ln r}        &     &    \mathcal{P}_m  = \frac{\nu_v}{\nu_m} & &p = \frac{J_{\phi 0}}{\frac{B_0}{\mu_0 h}}  \\   
                                                         &     &      \chi_m  = \frac{\nu_m}{\nu'_m}          & &  
 \end{array}
\end{equation}

Here, $p$ controls the toroidal electric current density at the disk mid plane. It is a measure of the bending of the magnetic surface at the disk mid plane, resulting from the interplay between advection and turbulent diffusion. It provides also a rough estimate of the bending of the field lines at the disk surface, namely $B^+_r/B_z \sim h \mu_o J_{\phi 0}/B_z \sim p$.  

In the above list, three parameters ($\alpha_m, \mathcal{P}_m, \chi_m$) are unavoidable as they describe the MHD turbulence. We will use $ \mathcal{P}_m=1$ in this work and explore the other two, for the values $\chi_m=[0.01,0.1,1,2]$ and $\alpha_m=[0.8,1,2,8]$. Our reference set of parameters will be $(\alpha_m=1, \chi_m=1)$. The disk aspect ratio $\epsilon$ should be computed using the energy equation. But, as said before, this is not done here and we will instead fix it to the common value $\epsilon=0.1$.

Since two parameters will be constrained by the crossing of two critical points (see below), this leaves one free parameter and we choose the disk ejection efficiency $\xi$. Thus, for a given set ($\epsilon, \alpha_m, \chi_m, \mathcal{P}_m$), we compute the values of the toroidal current $p$ and disk magnetization $\mu$ that are necessary to allow for a JED with the desired value $\xi$. The ejection index $\xi$ will thus be varied from the smallest value allowing for a solution to $\xi=1$. For illustrative purposes, we will display the resulting parameter space showing $\xi$ as function of the disk magnetization $\mu$. Table \ref{tab:parameter} contains a list of the disk parameters evaluated at the disk mid-plane as well as their type (constrained or free). We have also included other useful quantities that might be needed for the comprehension of the dynamical properties. 

\begin{table}
\centering
\caption{List of all the dimensionless parameters used in this work. Even though the parameters like $\xi$, $\alpha_m$ and $\chi_m$ are free their possible values are going to be constrained by the underlying physics, see section \ref{sec:relation_SA_SM}}
\begin{tabular}{lll} 
\hline
Name                                                                                            & Symbol           & Type                         \\ 
\hline
\begin{tabular}[c]{@{}l@{}}Disk geometrical\\ thickness \end{tabular}                           & $\epsilon$       & Fixed to $0.1$               \\ 
\hline
\begin{tabular}[c]{@{}l@{}}Magnetic Prandtl\\ number \end{tabular}                              & $\mathcal{P}_m$  & Fixed to $1$                 \\ 
\hline
\begin{tabular}[c]{@{}l@{}}Level of\\ turbulence \end{tabular}                                  & $\alpha_m$       & Free                         \\ 
\hline
\begin{tabular}[c]{@{}l@{}}Anisotropy of \\ tubulence \end{tabular}                             & $\chi_m$         & Free                         \\ 
\hline
\begin{tabular}[c]{@{}l@{}}Disk \\ ejection index \end{tabular}                                 & $\xi$            & Free                         \\ 
\hline
\begin{tabular}[c]{@{}l@{}}Disk \\ magnetization \end{tabular}                                  & $\mu$            & SM regularity condition      \\ 
\hline
\begin{tabular}[c]{@{}l@{}}Toroidal current \\ at the disk\\ mid-plane \end{tabular}            & $p$              & Alfven regularity condition  \\ 
\hline
\begin{tabular}[c]{@{}l@{}}Ratio between \\ the vertical and \\ the radial torque \end{tabular} & $\bar{\Lambda}$  & Calculated                   \\ 
\hline
\begin{tabular}[c]{@{}l@{}}Rotation of the\\ magnetic surfaces \end{tabular}                    & $\omega$         & Calculated                   \\ 
\hline
\begin{tabular}[c]{@{}l@{}}Magnetic\\ lever-arm \end{tabular}                                   & $\lambda$        & Calculated                   \\ 
\hline
\begin{tabular}[c]{@{}l@{}}Jet mass \\ load \end{tabular}                                           & $\kappa$         & Calculated                   \\ 
\hline
\begin{tabular}[c]{@{}l@{}}Bernoulli\\ invariant \end{tabular}                     & $e$    & Calculated                   \\
\hline
\begin{tabular}[c]{@{}l@{}}Initial\\ jet magnetization \end{tabular}                     & $\sigma$    & Calculated                   \\
\hline
\end{tabular}
\label{tab:parameter}
\end{table}
Thanks to the method of variable separation, the set of PDEs is transformed into a set of ODEs on the functions $f_A$ (see Appendix \ref{A:self_eq} for their expressions). These equations can then be numerically solved from the disk mid plane ($x=0$) to infinity using a Burlish-Stoer method for stiff equations. The resolution is done in the same way as in \cite{ferr97}. The integration starts at $x=0$ with a guess for the parameters $(\mu, p)$. This guess of parameters allows us to define the vertical boundary conditions at the disk mid-plane, for example the toroidal and radial current as well as the accretion speed ($u_r(x=0)<0$). To be consistent with the assumption of bipolar magnetic topology we need to choose $B_r(x=0)=B_\phi(x=0)=0$. The initial conditions for all fields are explicitly defined in Appendix \ref{A:self_eq}. 

After properly setting up the boundary conditions the integration is propagated upwards using the resistive viscous MHD equations. As we move upward, the accretion flow is deviated and becomes parallel to the poloidal magnetic field. When this is achieved with enough accuracy, we switch to ideal MHD equations. In this regime, magnetic forces are more effective and try to accelerate the flow up to a super-slow magnetosonic (SM) speed. 

Fulfilling the regularity condition is not necessarily achieved for our initial choice of parameters. This condition is going to constrain the magnetization $\mu$ for a given value $p$. If $\mu$ is too large the flow will be accelerated too efficiently, which results in a shock. If $\mu$ is too small the acceleration will not be efficient enough and the flow falls back to the disk. By fine-tuning the value of $\mu$, one can approach the critical point enough to safely make a leapfrog. This is done by conserving the various MHD invariants (see their definition in section \ref{sec:Ideal_MHD_jet}). Once super-SM, the flow is still accelerated by the magnetic force and needs to become super-Alfv\'enic (A). 

This condition is going to constrain the parameter $p$, in the same way as before. If $p$ is too small, the magnetic tension overcomes the centrifugal push and the magnetic surface closes back to the axis ($B_r \rightarrow 0$). If $p$ is too large, centrifugal acceleration is now too efficient leading to a vanishing toroidal field. By fine-tuning the parameter $p$, one can approach the Alfv\'en critical point close enough to jump beyond it and propagate the solution farther out (again by conserving the MHD invariants). It is important to note that a full super-SM solution needs to be computed from the origin (hence a new critical $\mu$ found) each time $p$ is changed. This can be computationally demanding when the size of the explored parameter space is considerable.

 \begin{figure*}
    \centering
    \includegraphics[width=\textwidth,height=0.35\paperheight]{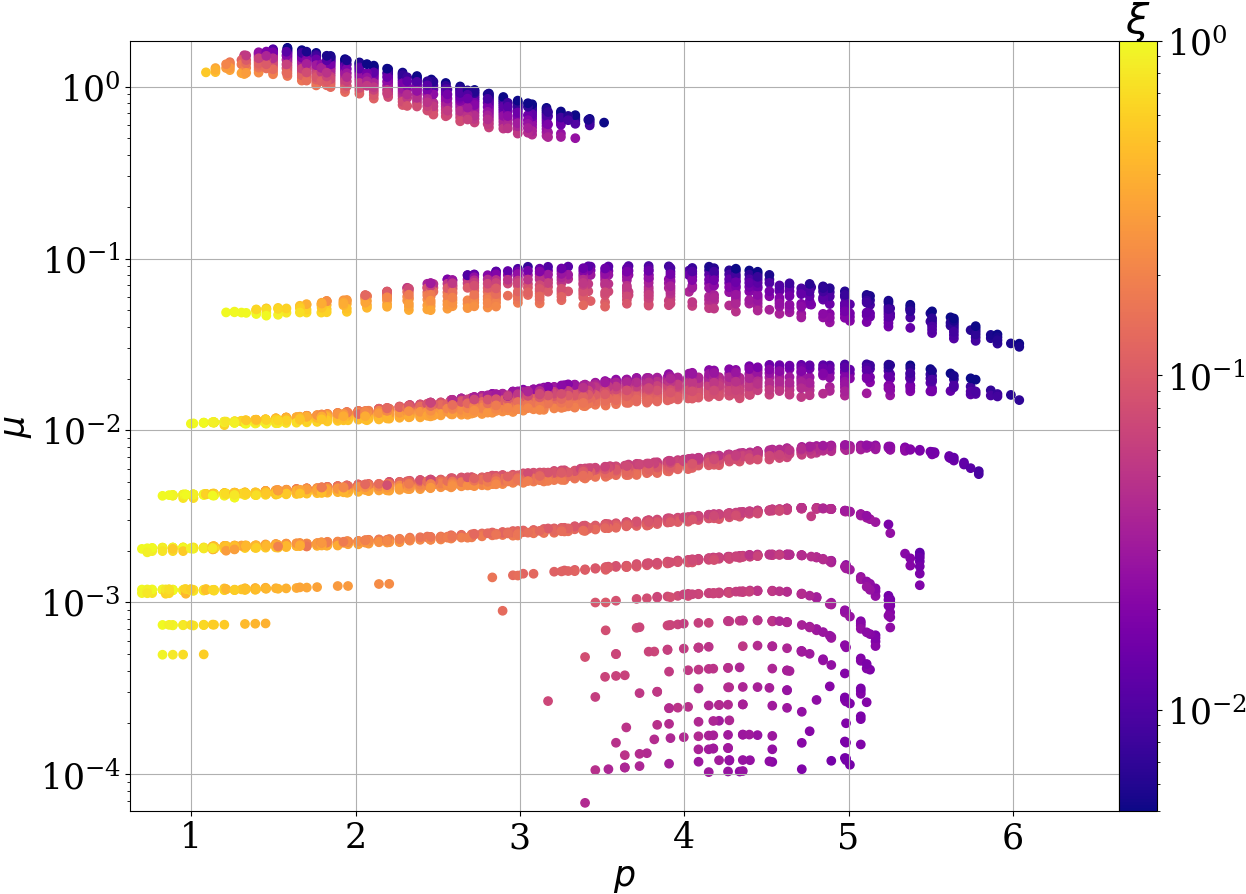}
    \caption{Parameter space $\mu(p)$ for super-SM isothermal solutions in our fiducial case $\alpha_m=1$, $\chi_m=1$, $\mathcal{P}_m=1$ and $\epsilon=0.1$. Each point in this plane corresponds to a solution characterized by an ejection index $\xi$ whose value is shown in color. The old near-equipartition solutions found by \citet{ferr95} correspond to the top island (see for instance their Fig.3, with $\mathcal{R}_m=p/\epsilon$).}  
    \label{fig:Sm_Space}
\end{figure*}

While previously published JED solutions were found for a magnetization $\mu \in[0.1;0.8]$, we now wish to reproduce the results of global simulations and achieve super-A jets with magnetization values as low as $10^{-4}$. In this regime, MRI is active and should be the source of the required MHD turbulence. However, around $\mu \sim 10^{-4}$ and bellow, MRI dynamo becomes significant  \citep{Scepi2018}. Since such an effect is not included in our calculations, we restrict ourselves to solutions with a magnetization no smaller than $\mu \sim 5\times10^{-4}$ (see however \cite{Stepa2014} and \cite{Dyda2018} for the study of the influence of a mean field dynamo).

When seeking for the critical value of $\mu$ for obtaining super-SM flows, we realized that our previous numerical procedure was explicitly forbidding spatial oscillations in the magnetic field within the disk. The physical justification for this choice was that oscillating magnetic fields would give rise to an oscillating velocity field as well, leading most probably to an unstable situation (through e.g. Kelvin-Helmholtz instability). Since we were interested only in steady-state configurations, such solutions have been simply disregarded. But a careful look at global simulations (ie Fig.12 in \citet{beth17} or Fig.6 in \citet{Zhu_Stone}) shows that this situation is actually realized, with $B_r$ first becoming negative in the disk upper layers before becoming positive at higher altitude. To be consistent with these simulations, we thus relaxed our previous constraint and allowed now for negative radial fields within the disk. The fact that all previous JED solutions have been obtained only for $\mu>0.1$ is a direct consequence of the explicit requirement (within our code) that the poloidal magnetic field has a monotonous vertical behavior within the resistive MHD disk zone. As will be shown in the next sections, relaxing this constraint (ie, removing any condition on $B_r$), allows for new solutions at much smaller magnetization levels.   Although we still recover the previous ones at near equipartition fields, we will mainly focus our attention on the new ones. It turns out that the asymptotic behavior of the associated jets is not different than that described in \citet{ferr97} and \citet{ferr04}, namely with a recollimation towards the axis. Our main interest will therefore be on the disk physics unveiled by these new solutions.

\section{Super-SM flows} 

\subsection{The super-SM parameter space}
\label{sec:SM_space}

Figure~\ref{fig:Sm_Space} shows the parameter space for our fiducial case, obtained without any restriction imposed neither on $p$ nor on $\mu$. Each point corresponds to a flow that goes smoothly from the resistive MHD disk to the ideal MHD flow regime and becomes super-SM.  We have been able to extend the parameter space in $\mu$ by 4 orders of magnitude. The new enlarged parameter space entails several features:
\begin{enumerate}
\item The existence of distinct and well separated islands, namely zones in the plane $p-\mu$ where solutions can be found. The old parameter space found in \citet{ferr95} corresponds to the top island (with their $\mathcal{R}_m=p/\epsilon$).

\item A monotonous behavior $p(\xi)$, as discussed in \citet{ferr95}. This is mostly due to the vertical disk balance which leads to smaller ejected mass ($\xi$) when $p$ is increased. Although a deeper examination shows that the rate at which $p(\xi)$ varies with $\xi$ depends also on $\mu$ and the turbulent parameters, we will not discuss further this already known trend.  

\item For a given $\xi$, the function $\mu(p)$ is bi-valued in some islands for $\mu<0.1$. This is a signature of two distinct vertical equilibria. One branch is associated with a dominant toroidal field at small $\mu$, whereas the other has a dominant radial field at larger $\mu$. This will be further discussed in section \ref{sec:SA_param}.

\item The range $[\xi_{min},\xi_{max}]$ of possible super-SM solutions varies with the magnetization: both $\xi_{max}$ and $\xi_{min}$ increase at small magnetization. This will also be further detailed in section \ref{sec:SA_param}. Note that we restrained ourselves to values $\xi \leq 1$ because isothermal flows with $\xi >1$ would have a negative energy and could not describe unbounded flows \citep{ferr97}. Putting aside this issue, super-SM flows could nevertheless be achieved with larger $\xi$ (up to 2 or more, \cite{ferr95}).   
\end{enumerate}

Figure~\ref{fig:Sm_Space} illustrates also the reason why solutions at small magnetization were difficult to find. The existence of forbidden zones in $\mu$ between islands (in particular at high magnetization levels) introduces a discontinuity that requires to jump to much smaller values in $\mu$ for a given $\xi$. In this forbidden zone, magnetic fields exhibit spatial oscillations without allowing for super-SM flows. This discontinuity made it hard to believe that the parameter space continued beyond what was already explored.


\begin{figure*}
\centerline{\includegraphics[width = 0.31\paperwidth]{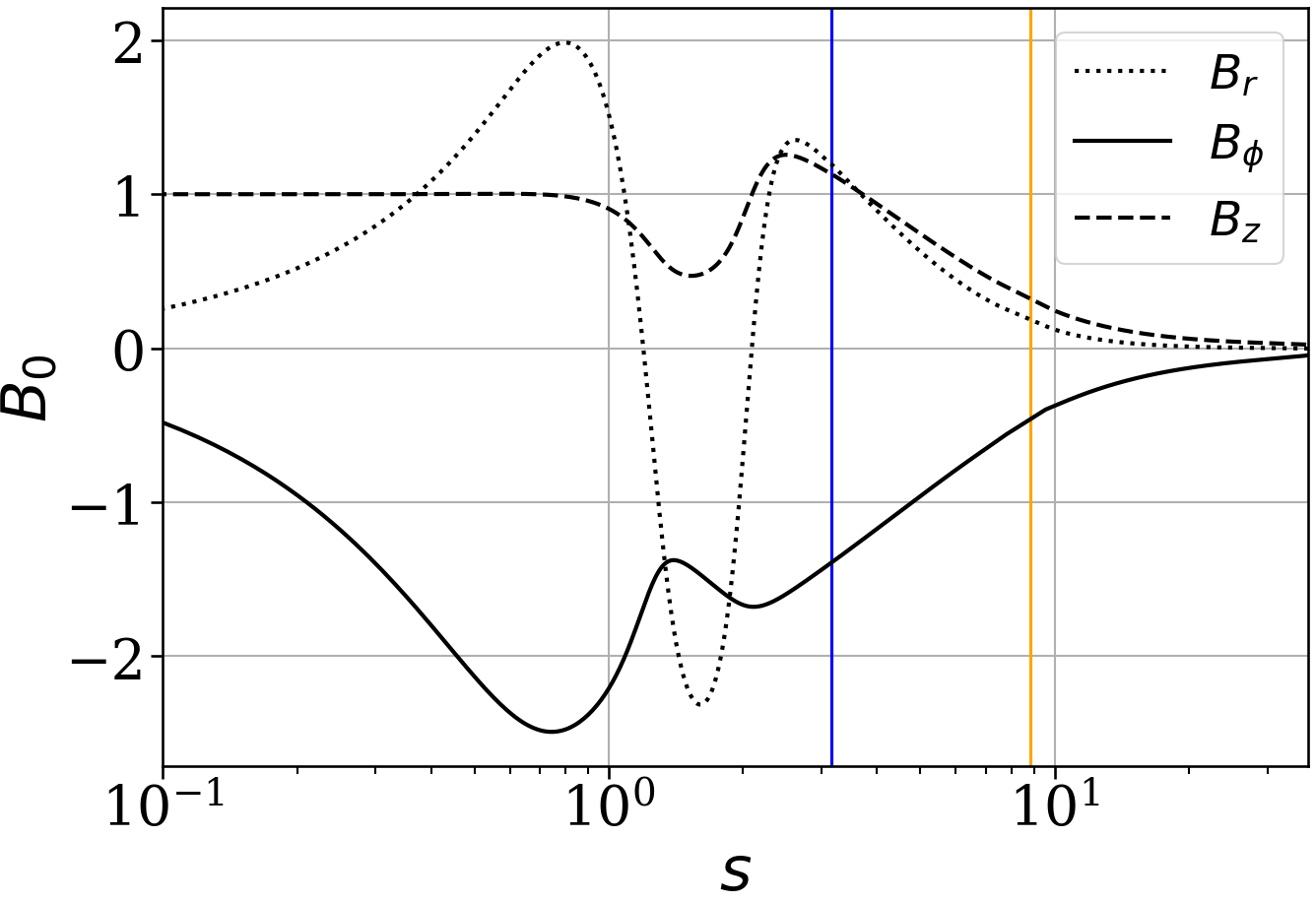}
\includegraphics[width = 0.31\paperwidth]{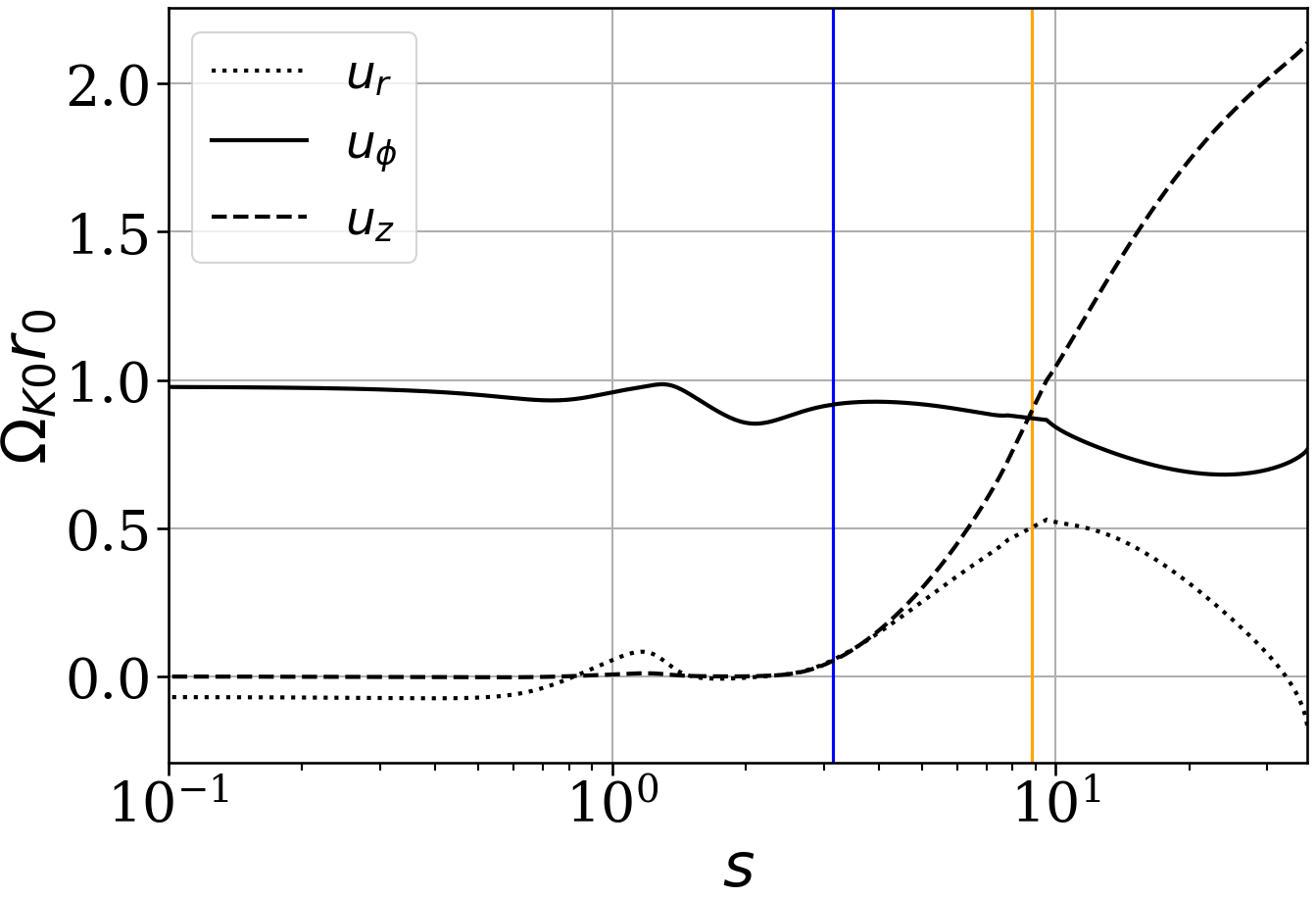}
\includegraphics[width = 0.31\paperwidth]{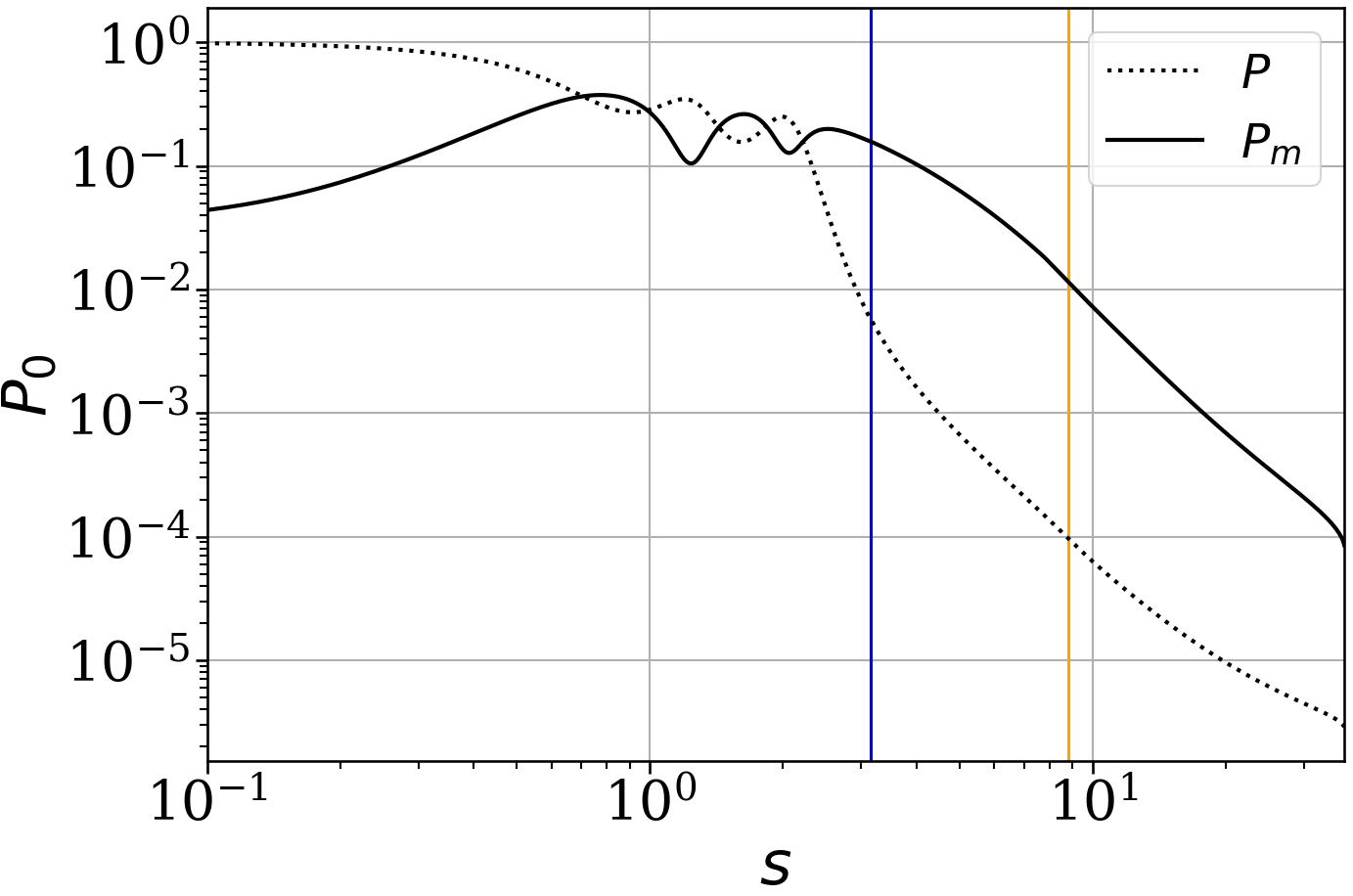}}

\centerline{\includegraphics[width = 0.31\paperwidth]{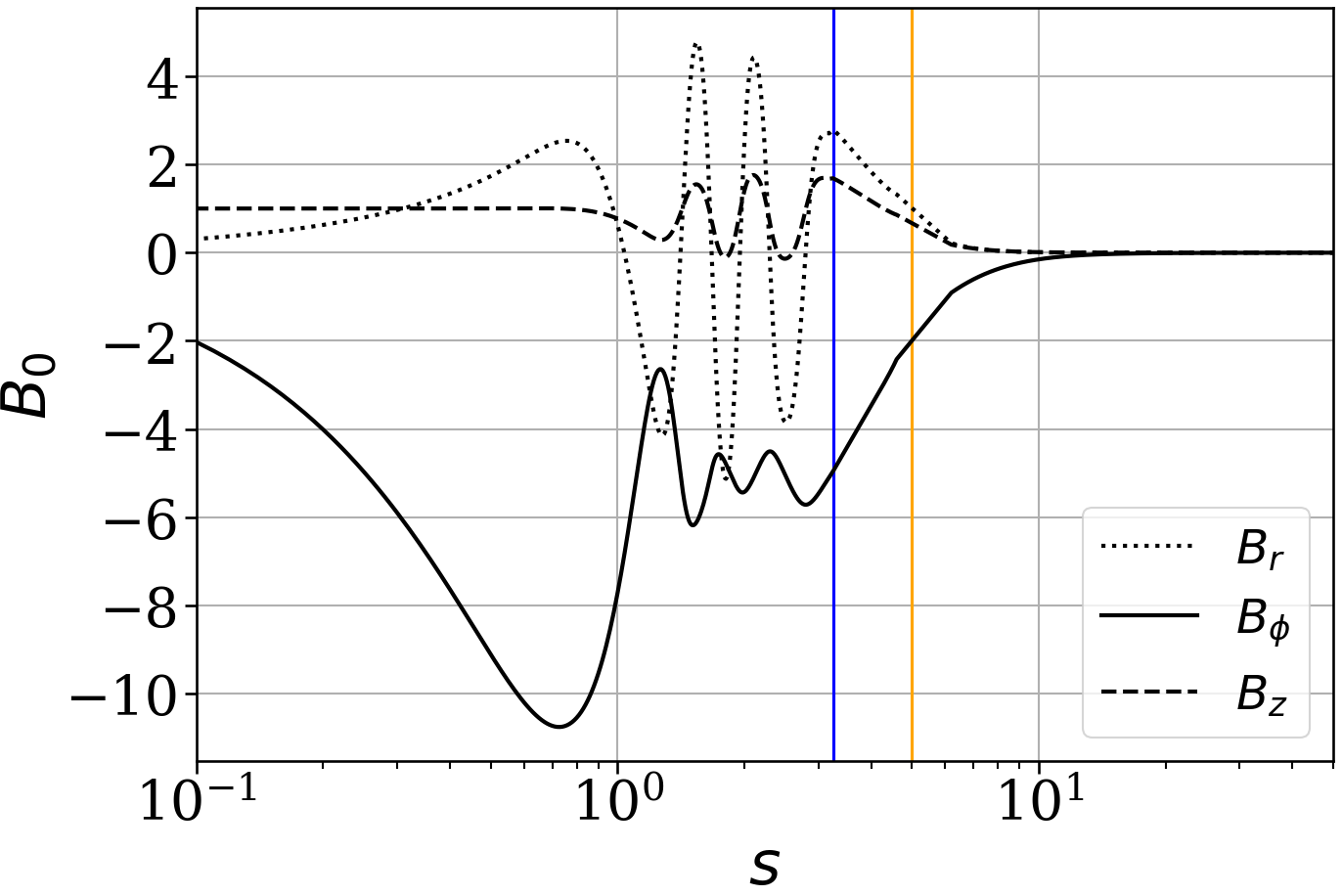}
\includegraphics[width = 0.31\paperwidth]{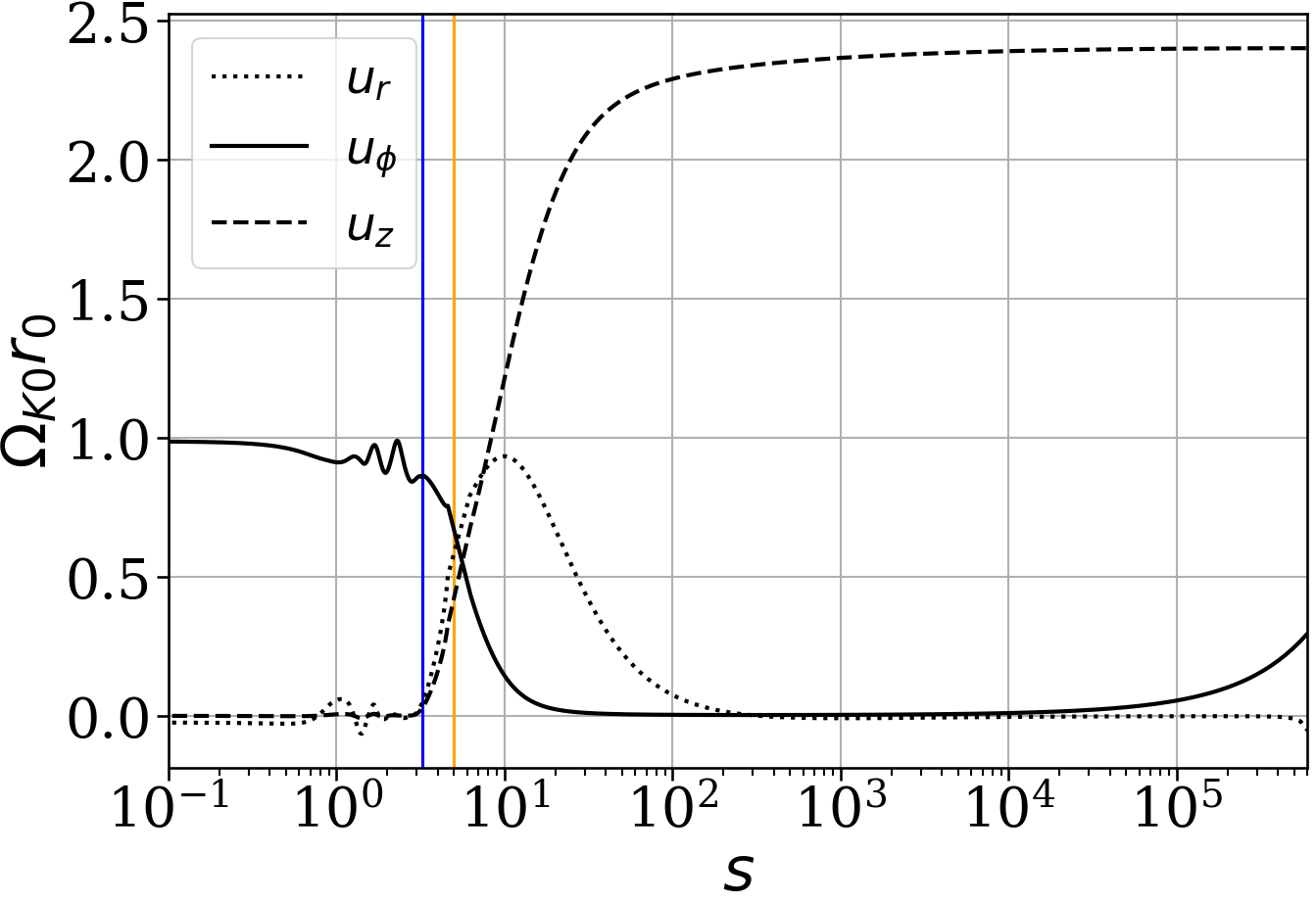}
\includegraphics[width = 0.31\paperwidth]{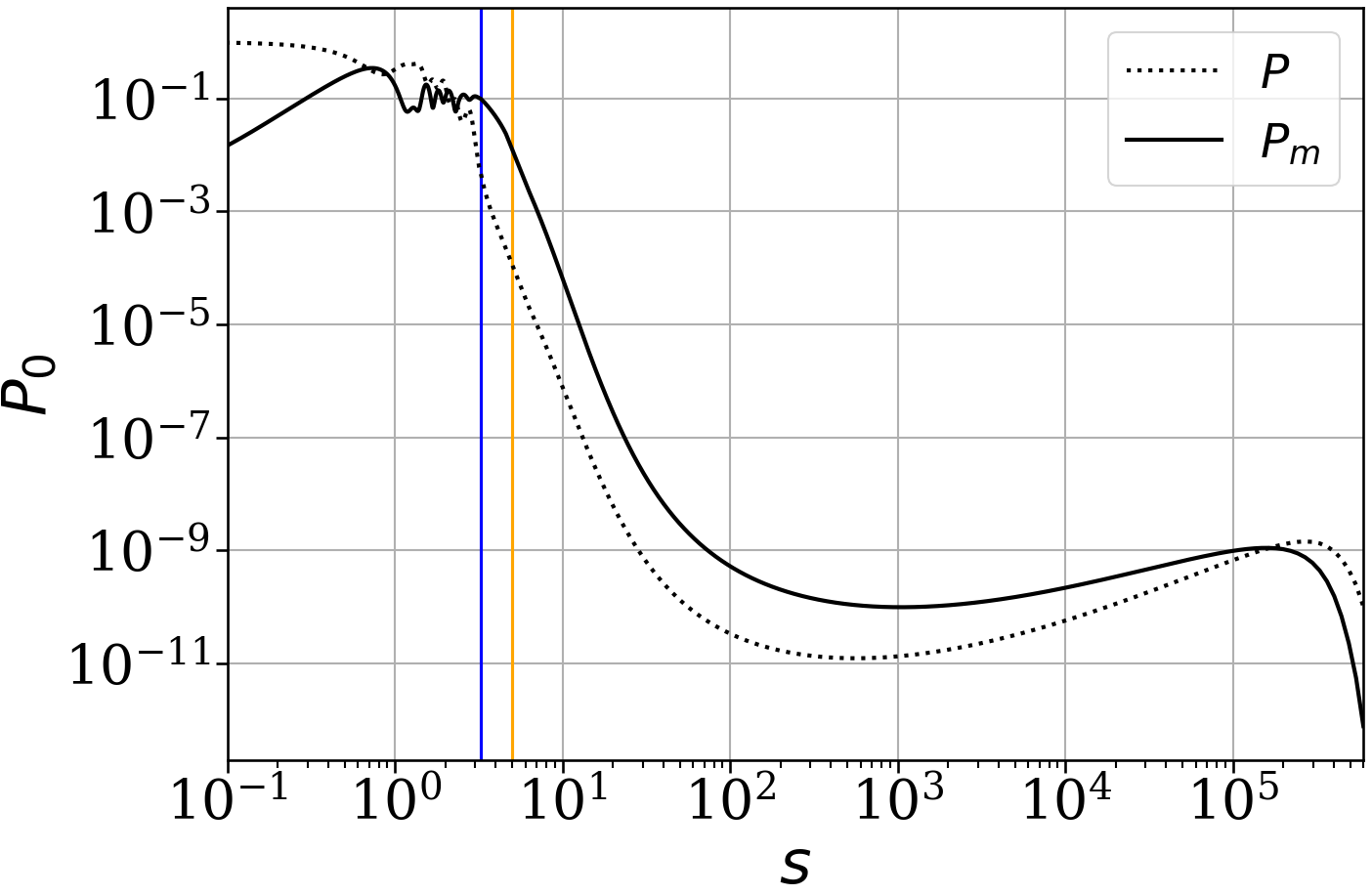}}
\caption{Vertical profiles for several quantities as function of the variable $s=z/h$ along a magnetic surface for two solutions obtained with 
$\alpha_m=1, \chi_m=1, \mathcal{P}_m=1, \epsilon=0.1$, the same ejection index $\xi=0.1$ and $\mu=6.7\times10^{-2}$ (Top), $\mu=5.7\times10^{-3}$ (Bottom). The magnetic field components (left) are normalized to the vertical field at the disk midplane $B_o$, the velocity components (middle) to the keplerian velocity $\Omega_{Ko}r_o$ and the kinetic $P$ and magnetic $P_m$ pressures (right) to the kinetic pressure at the disk midplane. The blue and orange vertical lines represent respectively the SM and Alfv\'en critical points. 
Clearly, the smaller the magnetization $\mu$ the larger the magnetic shear $-B_\phi^+/B_z$ at the disk surface.}
\label{fig:profiles}
\end{figure*}

In order to better understand these new super-SM solutions, we plot the vertical profiles of several quantities as function of the variable $s=z/h$ along a magnetic surface for two solutions (Fig~\ref{fig:profiles}). They have the same parameter set ($\alpha_m=1, \chi_m=1, \mathcal{P}_m=1, \epsilon=0.1$), the same ejection index $\xi=0.1$ but were obtained with $\mu=6.7\times10^{-2}$ (top row) and $\mu=5.7\times10^{-3}$ (bottom). While the former exhibits one spatial oscillation in $B_r$, the second has 3 spatial oscillations. The different islands correspond therefore to different spatial oscillation modes in the radial magnetic field, starting with $n=0$ (no oscillation) for $\mu>0.1$. For example the island located at $3\times10^{-2}<\mu<0.1$ corresponds to $n=1$ spatial oscillation while $3\times10^{-3}<\mu<8\times10^{-3}$ corresponds to $n=3$ spatial oscillations. Furthermore, we can see that for decreasing $\mu$ both the extent of the islands, $\delta \mu$, and the distance between the islands, $\Delta\mu$, get smaller and smaller.  

This oscillatory behavior is also visible in all the other disk quantities namely, the other components of the magnetic field, the density and the velocities. These spatial oscillations start above the disk mid-plane, they seem to exhibit a constant wavelength and decay very rapidly before the SM critical point (shown as a vertical blue line). These spatial oscillations are therefore localized at the disk surface, overriding the resistive and ideal MHD regimes.  Putting aside the spatial oscillations, these solutions behave like the previous ones. A fraction of the disk mass (controlled by $\xi$) is deflected vertically by the combined effect of the thermal and magnetic (toroidal) pressure gradients and is ejected vertically. Within our isothermal situation with $\epsilon=0.1$, the flow is energetically "cold" and the \cite{blan82} criterion for cold ejection applies. As a consequence, the poloidal field lines are indeed bent by more than 30 degrees with respect to the vertical axis. This bending is actually more pronounced for these oscillating solutions, especially since magnetic compression is less of a danger for the disk vertical equilibrium at low magnetization. As analyzed in Appendix~\ref{A:Gra_Shaf}, this larger initial bending allows solutions to meet the Alfv\'en point at smaller altitudes. Once they become super-A, nothing seems to distinguish solutions at large $n$ from the previously published $n=0$ solutions.

\subsection{MRI-driven magnetic winds}
\label{sec:MRI_driven}

The oscillatory solutions obtained at low magnetization are actually a generalization of the "exotic solutions" of \cite{ogilvie1997}. 
The existence of spatial oscillations in all quantities, occurring at low magnetization levels, is a manifestation of saturated MRI-like modes or channel flows, as described for instance in \cite{Latter_2010}. However, a word of caution is appropriate. 
The solutions found in our work are not linear unstable MRI modes as the equations we solve are stationary. They are exact non-linear solutions of the MHD equations, which exhibit physics similar to that of MRI modes in the disc. This is not surprising since MRI modes are known to spontaneously saturate into wind-like solutions \citep{lesu13}. 
Nevertheless, as will be shown below, a linear approach allows to grasp the complex non-linear physics.
Since MRI is an ideal MHD instability, these modes will tend to develop only when the Alfv\'en time scale $L/V_{Az}$ becomes smaller than the diffusion time scale $L^2/\nu_m$  over a length $L$. A second necessary condition is that there is enough room to allow for spatial oscillations on that scale. This requires that the fastest growing MRI mode has a wavelength $\lambda\sub{MRI} \sim L/n$, where $n$ is the number of spatial oscillations. The fact that the spatial oscillations need to stop when the radial magnetic field is positive imposes an integer number of spatial oscillations. Taking $L \sim h$ as an estimate of the relevant vertical dynamical scale provides the following crude conditions for the appearance of spatial oscillations \begin{align}
    R\sub{mag}\equiv \frac{hV_{Az}}{\nu_m} > 1\\
    \frac{\lambda\sub{MRI}}{h} \equiv 2\upi\frac{B_z}{B_0}\sqrt{\mu\frac{\rho_0}{\rho}}\sim \frac{1}{n}
\end{align}
We checked that these two conditions are indeed always verified in our solutions. The disk mid plane is always too diffusive for $\alpha_m > 1$ and no spatial oscillations are present regardless of the magnetization $\mu$. But since both the density and diffusivity decrease vertically, $R\sub{mag}$ becomes large enough and spatial oscillations can then develop on a length scale of order $h$. Indeed, spatial oscillations are possible at the disk surface ($z\simeq h$) since the flow is already in ideal MHD. Hence, at the disk surface the channel mode kicks in and produces the oscillatory behavior, as can be observed in figure \ref{fig:profiles}.
The expression of $\lambda\sub{MRI}$ harbors several features: 
\begin{itemize} 

\item The number of possible spatial oscillations depends mostly on $\mu$, the smaller $\mu$ the larger $n$. However, this is only an order of magnitude estimate and there is some interval $\delta \mu$ around an average value $\mu_n$ allowing for super-SM solutions. This can be done by playing with the vertical profile of the density, namely the toroidal current parameter $p$ and the ejection index $\xi$.     

\item As the density decreases, the wavelength increases and becomes eventually larger than the local dynamical scale. At the SM point, $V_A >> C_s$ and the plasma beta writes $\beta_{SM} \simeq V_{Az}^2/2 V_{SM}^2 \ll 1$, showing that the magnetic tension becomes too large and quenches the instability. Spatial oscillations are therefore limited between the disk upper layers and the SM point, located a few disk scale heights ($x_{SM}\sim 2 - 3$). 
\end{itemize}

\begin{figure}
    \centering
    \includegraphics[width=0.45\textwidth]{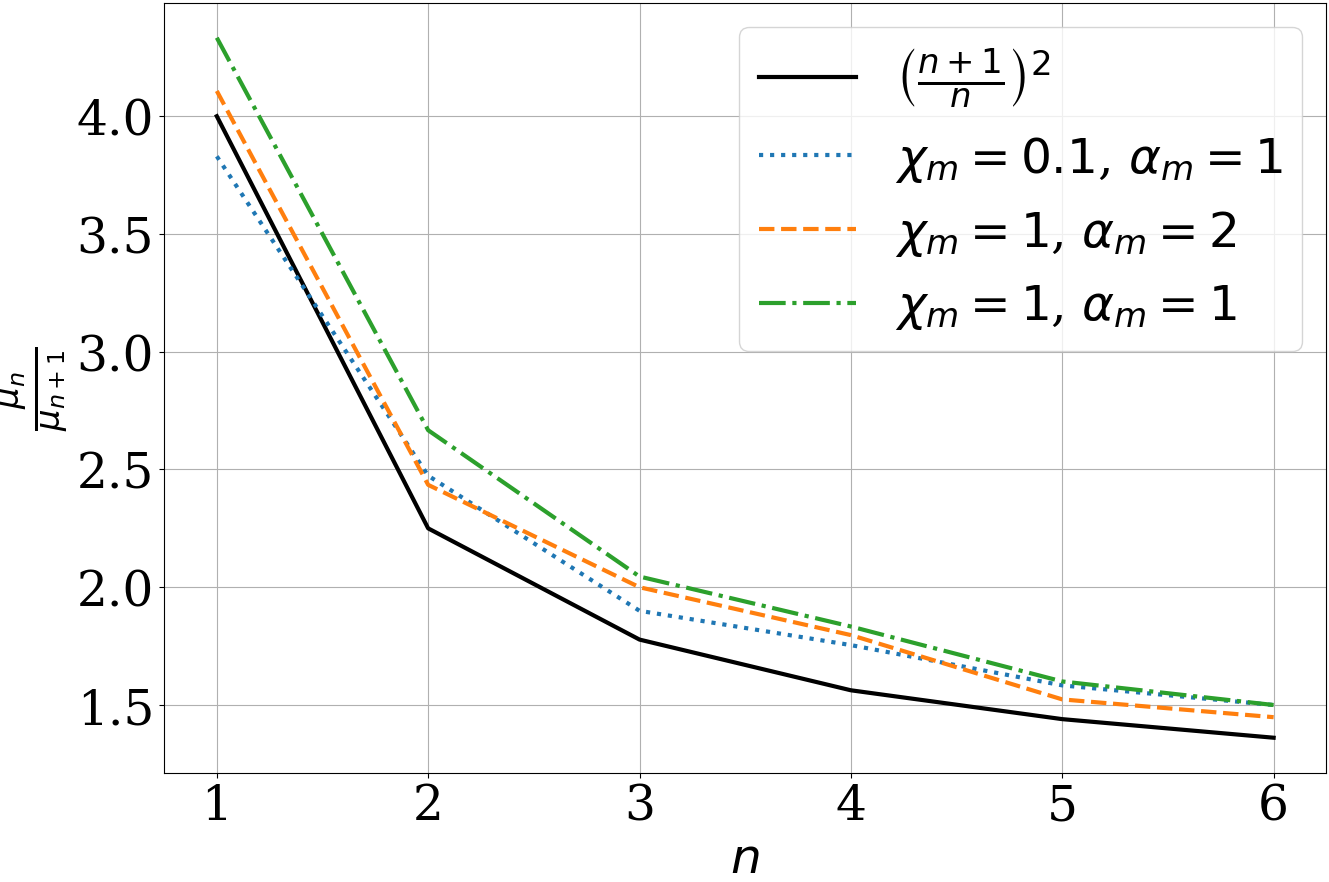}
    \caption{Ratio $\mu_n/\mu_{n+1}$ as function of the number $n$ of spatial oscillations for three different sets of super-SM solutions obtained with 
  $\mathcal{P}_m=1, \epsilon=0.1$ and a constant $p$. The value $\mu_n$ is obtained as the regularity condition for a super-SM flow exhibiting $n$ spatial oscillations (see text). The black solid line is our simple analytical estimate Eq.(\ref{eq:mu_n}) and the colored curves are for the following parameter sets: $\alpha_m=1, \chi_m=1$ (green),  $\alpha_m=2, \chi_m=1$  (red) and $\alpha_m=1, \chi_m=0.1$ (blue).}
    \label{fig:mu_n}
\end{figure}

The above properties explain the existence of the islands seen in Fig.~\ref{fig:Sm_Space}, as well as the fact that their spacing in $\mu$ decreases with $\mu$. Indeed, the MRI wavelength and the SM point can be approximately related by $n \lambda\sub{MRI} \sim h x_{SM}$ which leads to $\mu_{n}=\frac{f(x_{SM})}{n^2\alpha_m^2}$, where $f(x_{SM})$ is a complicated function of the altitude $x_{SM}$  and $\mu_n$ is the value of $\mu$ for a given $n$. Since $x_{SM}$ is weakly dependent of the magnetization $\mu$ (it is mostly related to the diffusivity scale height), this condition writes
\begin{equation}
\label{eq:mu_n}
    \frac{\mu_n}{\mu_{n+1}} \sim \Tfrac{n+1}{n}^2
\end{equation}
In order to test the generality of this expression, we took three different MHD solutions obtained with $\mathcal{P}_m=1, \epsilon=0.1$ but with different values for ($\alpha_m, \chi_m$). For a fixed value of the parameter $p$, we varied the disk ejection efficiency $\xi$ and computed super-SM solutions, spanning thereby the various islands in $\mu$. We could therefore associate the number of spatial oscillations $n$ to a precise value $\mu_n$, which is the critical value required to get a super-SM outflow. We could thus obtain the ratio $\mu_n/\mu_{n+1}$ as function of $n$, as shown in Figure~\ref{fig:mu_n}. This plot demonstrates that the above simple analytical estimate is actually accurate and provides further evidence that the origin of the spatial oscillations it is indeed a saturated vertically stratified MRI-like mode, or channel mode, in an unbounded flow.  It explains also why the islands become closer and thinner as $\mu$ decreases (i.e. $\Delta \mu= \mu_{n} - \mu_{n+1}$ decreases). The size of the islands becomes very small when $\mu$ approaches $10^{-4}$. 

Since \citet{blan82} it is well known that cold (negligible jet enthalpy) ejection requires at the disk surface a radial magnetic field component comparable to or larger than the vertical field. For near equipartition solutions with $n=0$, the generation of the magnetic field geometry at the disk surface is a natural outcome of the transfer of the disk angular momentum to the jet base, as described in \citet{ferr95}. This process can also be seen within the framework of MRI: as $\mu$ increases so does the MRI wavelength, resulting in the transport of the disk angular momentum along the field lines and fueling the jets \citep{lesu13}. There is therefore a possible continuous transition when the disk is thread by a large scale vertical field: from a turbulent (mostly radial) angular momentum transport at low $\mu$ to a laminar (mostly vertical) transport at large $\mu$. Near equipartition field solutions with $\mu \geq 0.1$ give rise to mostly centrifugally-driven super-A outflows \citep{ferr97,cass00a}. This does not hold anymore at low magnetization levels and the existence of these channel modes in the disk upper layers is of utmost importance for cold ejection. Indeed, it is the existence of these spatial oscillations that actually leads to the generation of the radial and toroidal magnetic field components required for ejection. In other words, MRI-like spatial oscillations provide the bending necessary for ejection. This will be further discussed in Section~\ref{sec:relation_SA_SM}.

The spatial oscillations exhibited by our solutions may seem in contradiction with the resistive profile used. Indeed, parasitic instabilities (such as Kelvin-Helmholtz) may be triggered and lead to a disruption of these channel modes (see for instance \cite{goodman_parasitic_1994}). It is therefore possible that our solutions will ultimately lead to some internal rearrangement, namely a modification of the turbulent profiles. However, only numerical simulations could elucidate this issue. This will be further discussed in sec \ref{sec:Bias}.

To summarize, channel modes triggered above the disk mid plane build up the magnetic field components allowing the ejection of cold disk material in the form of a super-SM ideal MHD flow. These oscillating modes are confined on a few disk scale heights, between $R\sub{mag}>1$ and $\lambda\sub{MRI}/h\leq 1$. But these modes are also constrained by the imposed boundary conditions, both at the SM point and at the disk mid plane. While the conditions at the SM point are quite general  ($B_r$ and $B_\phi$ must be respectively positive and negative), those imposed at the disk mid plane are questionable. Our self-similar solutions have been computed assuming a symmetric magnetic structure such that at $z=0$ (i) $B_r= B_\phi=0$ and (ii) $u_r <0$ (inward accretion motion). This clearly forbids other modes like for instance those leading to an outward decretion motion at the disk midplane. Allowing for such a boundary condition would lead to a supplementary half wavelength for instance ($n+ 1/2)\lambda\sub{MRI} \sim h x_{SM}$). More importantly, breaking the z-symmetry could also allow other modes with $B_\phi=0$ located above or below $z=0$, while not significantly modifying the physics involved.  

Computing such solutions is beyond the scope of the present paper. We nevertheless argue that the parameter space shown in Fig.~\ref{fig:Sm_Space} is actually a subset of the real parameter space of super-SM accretion-ejection structures. Indeed, since these new solutions would make use of the same type of modes, we believe that they would simply fill-in the forbidden zones between the islands. This property will be used in Section~\ref{sec:turb}, where the effects of the turbulence parameters on the parameter space will be explored.

\section{From super-SM to super-A flows} 

\subsection{MHD jet invariants}
\label{sec:Ideal_MHD_jet}

Not all of the super-SM solutions shown in Fig.~\ref{fig:Sm_Space} lead to steady-state outflows. In order to achieve that, they need to become super-Alfv\'enic (super-A) as well. The theory of steady-state MHD jets makes use of the existence of MHD invariants defined in ideal MHD along each magnetic surface of flux $a$. An axisymmetric, isothermal magnetic surface requires 6 boundary conditions at the base and features 4 integrals of motion (not counting the temperature): the two remaining quantities are thus determined by the SM and A regularity conditions.   

In ideal MHD mass conservation Eq.(\ref{eq:Sohm}) becomes
\begin{equation}
    \label{eq:mag_flux}
    \upo = \frac{\eta(a)}{\mu_0\rho}\Bpo
\end{equation}
where $\eta(a)$ is the first invariant and describes the mass to magnetic flux ratio ($\eta= \mu_o \partial \dot M_j/\partial \Phi$, where $\dot M_j$ is the mass flux in one jet and $\Phi$ its magnetic flux). 
The induction equation (\ref{eq:Sinduc}) writes
\begin{equation}
    \label{eq:OM_star}
    \Omega_*(a) = \Omega -\eta\frac{B_\phi}{\mu_0\rho r}
\end{equation}
where $\Omega_*$ is the rotation of the magnetic surface. Since the field lines are anchored on the accretion disk, they rotate 
at roughly the same rate as the disk material at the jet base, namely $\Omega_*\simeq \Omega_{SM}$. 
The disk angular momentum conservation Eq.(\ref{eq:Smonto}) becomes
\begin{equation}
    \label{eq:Angu_Con}
    L=\Omega_*r_A^2 = \Omega r^2-\frac{r B_\phi}{\eta}
\end{equation}
where $L$ is the total specific angular momentum carried away by both matter and the magnetic field and $r_A$ is the cylindrical radius where the flow becomes super-A.  
Finally, the projection along the magnetic surface of Eq.(\ref{eq:Smonp}) leads to 
\begin{equation}
    \label{eq:Bernoulli}
    E(a) = \frac{u^2}{2}+H+\Phi_G-\Omega_*\frac{rB_\phi}{\eta}
\end{equation}
where $E(a)$ is the Bernoulli integral and describes the total specific energy carried away along the magnetic surface. The enthalpy, defined as $\vgrad H =\vgrad p/\rho$ namely $H= C_s^2 \ln \rho$ for isothermal flows, can be safely neglected in analytical estimates (since $-H/\Phi_G \propto \epsilon^2$). Note however that our numerical resolution solves the full set of MHD equations including all terms (see Appendix \ref{A:self_eq}).

Since our super-SM solutions are in ideal MHD regime, the MHD invariants are already determined. It is therefore convenient to express them as function of the underlying disk parameters. To do so, we normalize these 4 invariants by quantities defined at the anchoring radius $r_o$ of the magnetic surface, at the disk equatorial plane. This leads to the following dimensionless parameters for cold jets launched from thin accretion disks\footnote{To derive the expression of $\kappa$, mass conservation Eq.(\ref{eq:Scon}) is written as $\frac{d \dot M_a}{dr} = 2 \frac{d \dot M_j}{dr}$, leading to the useful relation $\rho^+ u_z^+ \simeq \xi \epsilon \rho_o u_o$ valid at the disk surface, while Eq.(\ref{eq:Sohm}) gives $u_o= \alpha_m p V_A$.}     
\begin{align}
    \omega &\equiv \frac{\Omega_*}{\Omega_{Ko}} \\
   \label{eq:Def_kap}
    \kappa &\equiv \eta\frac{\Omega_{Ko}r_o}{B_o} \simeq \xi \alpha_m \frac{p}{\mu^{1/2}} \\
    \label{eq:Def_lam}
    \lambda &\equiv \frac{\Omega_* r_A^2}{\Omega_{Ko} r_o^2} \simeq 1 + \frac{1}{2\xi} \frac{ \bar{\Lambda}}{ 1+ \bar{\Lambda} }\\
        \label{eq:Def_en}
    e& \equiv \frac{E}{\Omega_{Ko}^2 r_o^2/2} \simeq 2\lambda - 3 + \Theta
\end{align}
where $\Omega_{Ko}= \sqrt{GM/r_o^3}$ is the Keplerian angular velocity and $\Theta=2\frac{H_{SM}}{\Omega_{Ko}^2 r_o^2}$ is the normalized enthalpy at the SM point. In geometrically thin disks, $\omega$ is always close to unity, but a significant deviation may occur for thicker disks \citep{cass00a}. For simplicity, we assumed $\omega=1$ in the expression of the dimensionless specific energy $e$ (see its exact expression in Appendix \ref{A:Gra_Shaf}).

The term $ \bar{\Lambda}= M_{z\phi}/M_{r\phi}$ is the ratio of the vertical (jet) torque 
\begin{equation}
    M_{z\phi} =  2 r \frac{B_z B_\phi}{\mu_0}\Bigg|_{SM}
    \label{eq:vert_torque}
\end{equation}
exerted at the disk surfaces to the total radial torque 
\begin{equation}
    M_{r\phi} = 2\Int_{0}^{z_{SM}}\frac{1}{r}\pdv{}{r}\left[r^2\left(T_{r\phi}+\frac{B_rB_\phi}{\mu_0}\right)\right]\,\diff{z}
\end{equation}
acting within the disk. The latter torque includes thereby both the jet (laminar) and turbulent (viscous) contributions to the radial transport of angular momentum. The disk angular momentum conservation equation (\ref{eq:Smonto}) writes $M_{z\phi} + M_{r\phi} = 2\Int_{0}^{z_{SM}} \rho \mathbf{u_p} \cdot \vgrad \Omega r^2 \diff{z} \simeq - \dot M_a \frac{\Omega_K}{4\upi}$. Using the definitions of the magnetic lever arm and $\xi$ leads then to Eq.(\ref{eq:Def_lam}). Note that it is a generalization of the relation found in \citet{cass00a}, where the radial transport of angular momentum by the laminar torque was negligible (parameter $\Lambda$ in their Eq.~33).

 \begin{figure*}
    \centering
    \includegraphics[width=0.8\textwidth]{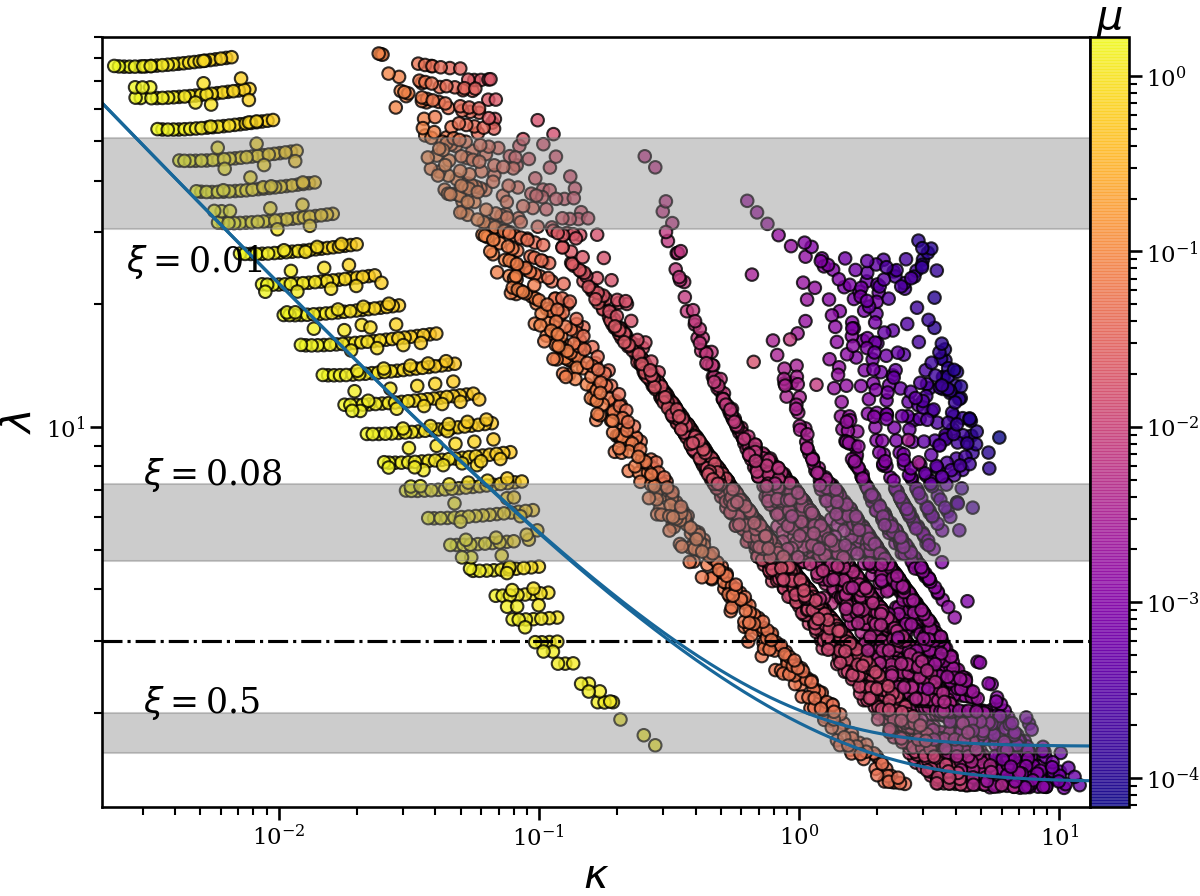}
    \caption{Projection on the usual $\kappa-\lambda$ plane of all the super-SM solutions shown in Fig.~\ref{fig:Sm_Space}. The color scale is  the disk magnetization $\mu$, while the grey areas correspond to zones with approximately a constant ejection index $\xi$ whose value is indicated. Note that the stripes (best seen in the yellow high magnetization zone with $n=0$) are an effect of our numerical procedure for seeking solutions. The blue solid lines correspond to Eq.(\ref{eq:Grad_BP}) while the dashed line corresponds to $\lambda=3$ (see text).} 
    \label{fig:lam_kap}
\end{figure*}

The mass load $\kappa$ and magnetic lever arm parameter $\lambda$ have been first introduced by  \cite{blan82} and extensively used since then. They are related by $\lambda \simeq 1 + | B_\phi/B_o |_{SM}/\kappa$ \citep{ferr97}. The value of the toroidal magnetic field depends on the underlying disk physics and is discussed next section.  In any case, the outflow must have a positive energy $e >0$, which requires $\lambda > 3/2$ when $\omega \simeq 1$ and  for cold flows with $\Theta\ll1$, namely when thermal effects (such as a warm corona) can be neglected.

\subsection{The magnetic shear as a function of the disk magnetization $\mu$}
\label{sec:relation_SA_SM}

Using Eq.(\ref{eq:Def_kap}) and (\ref{eq:Def_lam}), we can easily compute the jet invariants for all our super-SM solutions and put them in the classical $\kappa-\lambda$ plane of super-A solutions. This is shown in Fig.~\ref{fig:lam_kap}. The islands seen in Fig.~\ref{fig:Sm_Space} appear as well here, with a clear trend in $\kappa$: the smaller $\mu$ the larger $\kappa$. The super-SM constraint, that determines $\mu(p)$ for a given $\xi$, allows to obtain almost all possible values of $\xi$ up to 1, with $\kappa \propto \xi$. But reaching large values of $\kappa$ can only be done by switching to another island. The jet mass load is thus a function $\kappa (\xi,\mu)$ with an approximate linear dependence on $\xi$. For a given mass loss $\xi$, increasing $\kappa$ can be done by decreasing the magnetic field strength $\mu$. Matter dominated super-SM flows with $\kappa >1$ become thus achievable.   

\begin{figure}
    \centering
    \includegraphics[width=0.45\textwidth]{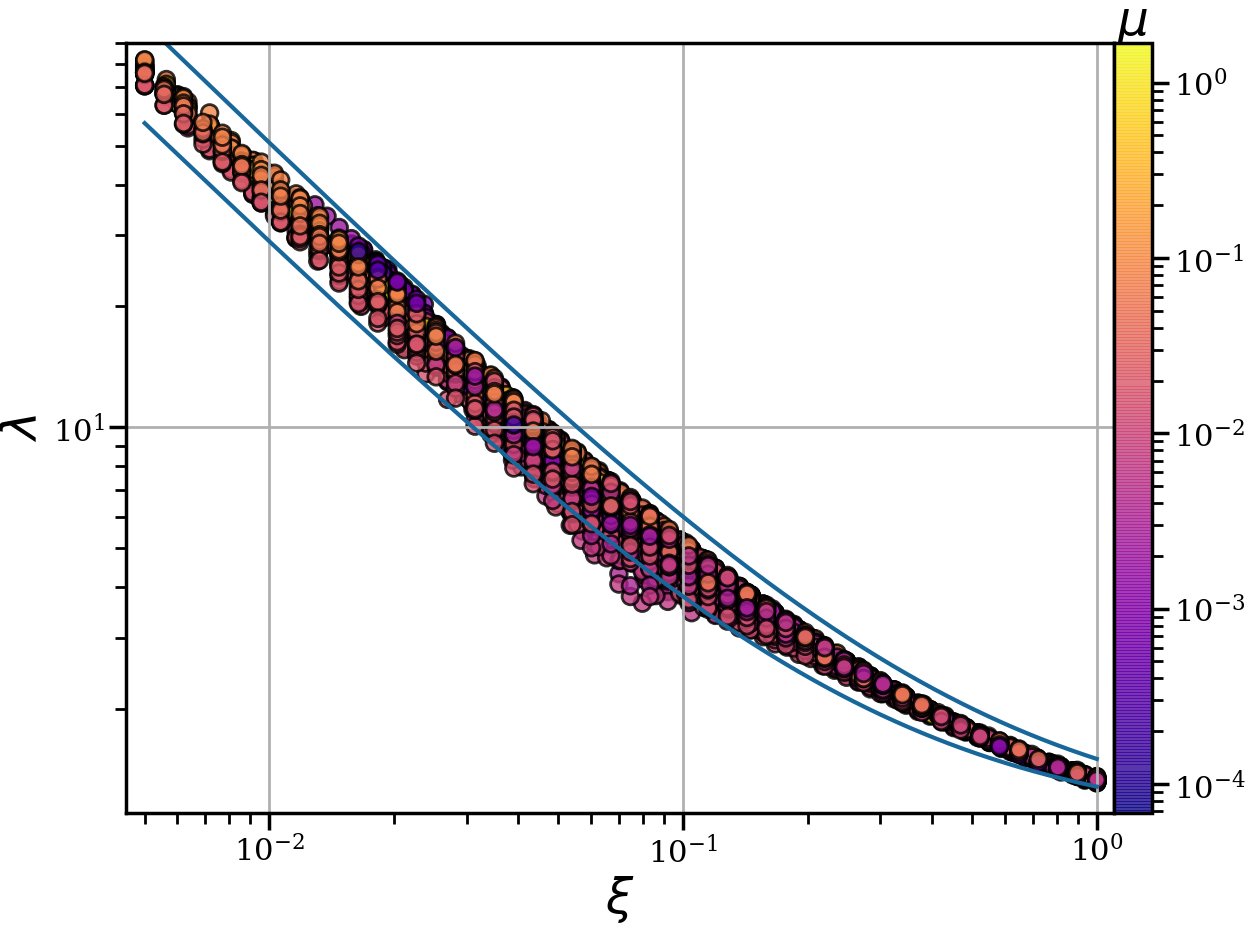}
    \caption{Magnetic lever arm parameter $\lambda$ calculated using Eq.(\ref{eq:Def_lam}) as a function of $\xi$ for all super-SM solutions appearing in Fig.~\ref{fig:Sm_Space}. The solid lines correspond to curves $\lambda=1+\frac{R}{2\xi}$, computed using either $R=1$ (top) or $R=0.6$ (bottom).}
    \label{fig:lam_xi}
\end{figure}

On the other hand, the magnetic lever arm $\lambda$ does not appear to be strongly dependent on $\mu$ but mostly on $\xi$. This can be clearly seen in Fig.~(\ref{fig:lam_xi}), where $\lambda$ is plotted as function of $\xi$ for all our super-SM solutions. Indeed, $\lambda = 1 + R/2\xi$, where $R= \bar{\Lambda}/(1+ \bar{\Lambda})$ is a rather weak function of $\mu$ and $\xi$ (the small dispersion in $\lambda$ doesn't seem to depend on $\mu$). This is remarkable as $\xi$ and $\mu$ span respectively 2.5 and 4 decades. Such a behavior must therefore be the outcome of some intrinsic physics. Using this result and Eq.(\ref{eq:Def_kap}) leads to the necessary constraint on the  magnetic shear at the disk surface 
\begin{equation}
    \label{eq:Bphi_SM}
\left |\frac{B_\phi}{B_0}\right|_{SM} = \frac{p \alpha_m}{2 \mu^{1/2}} R  \propto \mu^{-1/2}
\end{equation}
since $R$ is a weakly varying function and $p$ has a small range. 

This scaling of the toroidal magnetic field can be understood the following way. For a rather wide range in disk conditions, the flow must become super-SM near the surface, namely $u_z^+ \sim V_{SM} \sim C_s V_{Az}/V_{A}$ where $V_A$ is the total Alfv\'en speed. The vertical velocity $u_z^+$ is provided by the unbalance in the vertical forces around the dis surface, which is quite difficult to estimate. Another way to grasp it is to look at the Ohm's law (Eq.~\ref{eq:Sohm}) at the turning point where the radial velocity vanishes, right before the SM point. At this particular point $u_z^+ B_r^+= \nu_m^+ \partial B_r/\partial z$, which provides the scaling $u_z^+ \sim \nu_m^+/h$. This simple relation tells us that mass loading in jets is a diffusion process and that the initial jet velocity is directly related to the strength of the poloidal magnetic diffusion. Using now $u_z^+ \sim V_{SM} $ leads to 
\begin{equation}
\alpha_m^2\mu \simeq F^2_{SM} \frac{1}{1+\Tfrac{B_{r,SM}}{B_0}^2+\Tfrac{B_{\phi,SM}}{B_0}^2}
    \label{eq:SM_constraint_simp}
\end{equation}
where $F^2_{SM}$ is a function depending on the vertical profiles of the temperature and the magnetic diffusivity. It stems from this expression that, in order for this condition to remain valid at all $\mu$, the magnetic shear $|B_{\phi,SM}|/B_0$ must indeed scale as $\mu^{-1/2}$. It is therefore the SM constraint itself that imposes such a scaling: it guarantees that, whatever $\mu$, cold super-SM solutions can be found.

\begin{figure}
    \centering
    \includegraphics[width=0.45\textwidth]{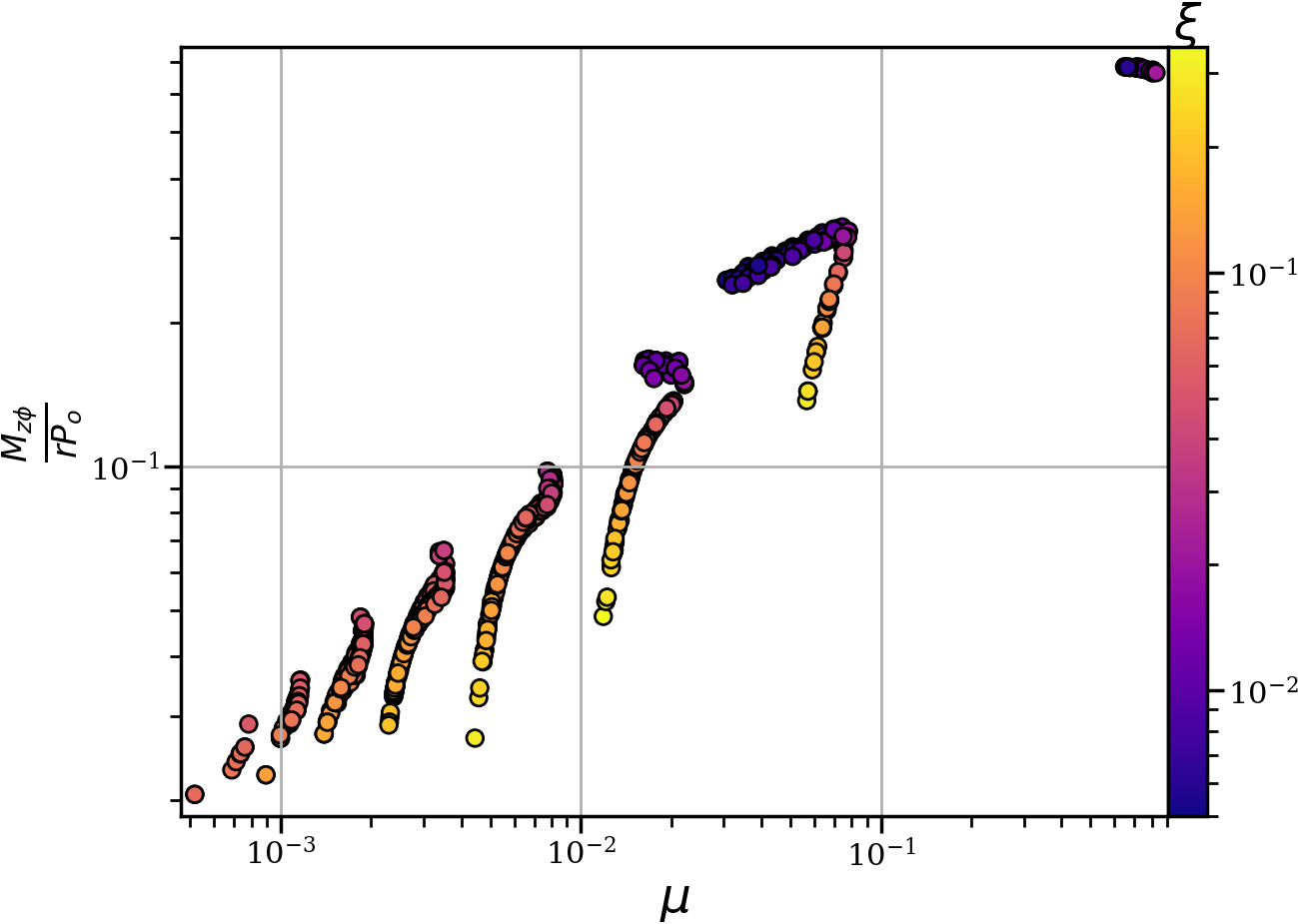}
    \caption{Vertical torque defined by Eq.(\ref{eq:vert_torque}) as a function of the magnetization $\mu$ and the mass ejection index $\xi$.
    Every point corresponds to a super-Alfv\'enic solution, section \ref{sec:SA_param} }
    \label{fig:vert_torque}
\end{figure}

We can use the SM constraint on the magnetic shear, Eq.(\ref{eq:Bphi_SM}), to derive an approximate expression for the vertical torque, Eq.(\ref{eq:vert_torque}):
\begin{equation}
    \frac{M_{z\phi}}{r P_o} \simeq \mu \left|\frac{B_\phi}{B_0}\right|_{SM}\simeq \kappa (\lambda-1)\mu\propto\mu^{1/2}
\end{equation}
this scaling is consistent with Figure \ref{fig:vert_torque}. Therefore, the wind stress can be easily modeled as a function of the magnetization and the plasma pressure at the disk mid-plane. This prescription could be useful for including the effects of wind driven accretion in hydrodynamic models.

\subsection{The super-A parameter space}
\label{sec:SA_param}


As discussed previously, steady-state solutions are only those that have the capability to produce super-Alfv\'enic flows. Magnetic acceleration can be seen as some centrifugal effect, the frozen-in jet plasma being accelerated because magnetic field lines are rotating faster than the jet material. This can be illustrated using Eq.(\ref{eq:OM_star},\ref{eq:Angu_Con}), leading to $\Omega = \Omega_*(1-g)$ where
\begin{equation}
    g=\frac{m^2}{m^2-1}\left(1-\frac{r_A^2}{r^2}\right)
\end{equation}
with $r_A$ the Alfv\'en radius and $m= u_p/V_{Ap}$ the poloidal Alfv\'en Mach number. The function $g$ measures the discrepancy between the two angular velocities and is related to the poloidal current flowing in the jet \citep{ferr97}. Starting from a tiny value at the disk surface, this function increases as the flow gets accelerated. It can then be seen that when the flow becomes super-A, namely $m=1$, a regularity condition $r=r_A$ must be fulfilled. 

\begin{figure*}
    \centering
    \includegraphics[width=0.8\textwidth]{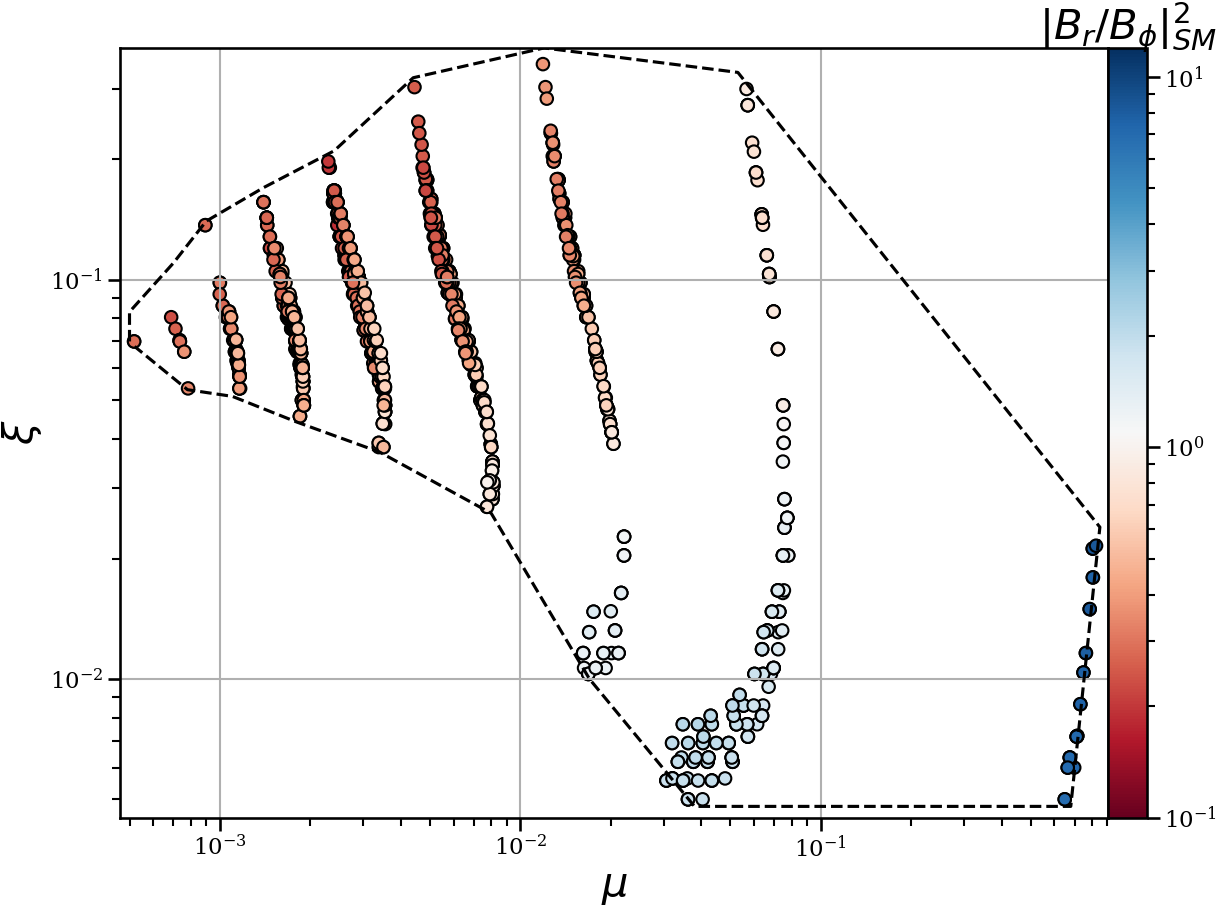}
    \caption{Parameter space $\xi(\mu)$ for isothermal super-A solutions with $\alpha_m=1$, $\chi_m=1$, $\mathcal{P}_m=1$ and $\epsilon=0.1$. In colors are shown the ratio of the radial to the toroidal magnetic field at the SM point. A clear trend emerges, with small magnetizations leading to highly wound magnetic fields, whereas larger magnetizations correspond to more bent structures. This can be seen as an evolution from a vertical pressure lift at small $\mu$ to a magneto-centrifugal push at large $\mu$. The black dashed line is a sketch of the border of the parameter space.} 

    \label{fig:Sa_space}
\end{figure*}

The Bernoulli equation can be interpreted as providing the velocity that matter reaches for a given "magnetic funnel". At infinity, if acceleration is so efficient that the magnetic energy becomes negligible, Eq.~(\ref{eq:Bernoulli}) gives the maximum jet velocity $u_{p\infty} \simeq \sqrt{2 E}\simeq \Omega_{Ko}r_o \sqrt{2 \lambda-3}$ for cold flows. But the shape $r(z)$ of this magnetic funnel, or more precisely the jet transverse equilibrium, is provided by the Grad-Shafranov equation
\begin{multline}
    \label{eq:Grad_Shaf}
    \vgrad \cdot \left[\left( m ^ { 2 } - 1 \right) \frac { \vgrad a } { \mu _ { o } r ^ { 2 } }\right] = \rho \left\{ \aderiv { E } - \Omega \aderiv {\Omega _ { * } r _ { A } ^ { 2 } } + \left( \Omega r ^ { 2 } - \Omega _ { * } r _ { A } ^ { 2 } \right) \aderiv {\Omega _ { * } } \right\}\\+ \frac { B _ { \phi } ^ { 2 } + m ^ { 2 } B _ { p } ^ { 2 } } { \mu _ { o } } \aderiv { \ln \eta }
\end{multline}
where $d/da = \vgrad{a}\cdot\vgrad{}/\vgrad{a}\cdot\vgrad{a}$ \citep{ferr97}. This equation is obtained by projecting Eq.~(\ref{eq:Smonp}) across the magnetic surfaces and, in practice, is not used to solve the jet physics. It does however provide a useful tool to derive the analytical constraint to get trans-Alfv\'enic flows (see Appendix \ref{A:Gra_Shaf}). 
 
A lengthy calculation allows to express $g_A$ explicitly as function of the MHD invariants and the Alfv\'en position angle $\Psi_A$. There are two branches that can be both connected to the accretion disk, one of them being chosen according to the interplay between $\lambda$ and  $\kappa$.  \citet{blan82} pointed out that, for a given mass load $\kappa$, there must be a minimum $\lambda\sub{min}$ that they estimated such that $\kappa \lambda\sub{min} (2\lambda\sub{min}- 3)^{1/2} = 1$. The generalization of this estimate is that, for a given magnetic lever arm $\lambda$, there must be a minimum mass load $\kappa\sub{min}$ for a cold flow such that  
\begin{equation}
        \kappa\sub{min}^2\lambda^{3}g_{B}^2= 1 \quad \mathrm{with}\quad g_B^2 = 1 - \frac{3}{\lambda}+  \frac{ 2 \sin \Psi_A}{\lambda^{3/2}}
        \label{eq:Grad_BP}
\end{equation}
(Eq.~(\ref{eq:def_g0}) in Appendix \ref{A:Gra_Shaf}, with $\omega=1$). This expression provides the two solid blue curves shown in Fig.~\ref{fig:lam_kap}, computed in two extreme cases for the location of the Alfv\'en surface: near the SM surface with $x_{SM}=2$, namely  $\Psi_A=\upi/2-\arctan\left(2h/r\right)$ (lower curve) and much further out with $\Psi_A=\upi/3$ for the upper one (as in typical near-equipartition cold flows). For a given magnetic lever arm $\lambda$ (mostly determined by $\xi$), $\kappa$ must be large enough. This rules out all super-SM solutions located at the left-hand side of these curves.   

It can be seen right away that the parameter space of super-A solutions for near-equipartition fields ($n=0$) will be very small, with mass loads $\kappa$ smaller than 0.1, disk ejection efficiencies $\xi$ smaller than 0.08 and magnetic lever arms larger than $\sim 7$, in agreement with \citet{ferr97}. On the other hand, MRI-like driven flows at small $\mu$ seem to allow mass loads larger than unity with small $\lambda$. These isothermal flows could be of great interest for reproducing dense outflows ($\xi >0.1$) with low asymptotic speeds ($u_{p\infty} \sim \Omega_{Ko} r_o$). In other words a weakly magnetized disk is better suited for a a massive ejection than a near-equipartition disk, not only by providing the necessary bending at the disk surface through the MRI-like mechanism, but also by bringing the Alfv\'en surface closer to the disk (see discussion on the role of $\Psi_A$ in Appendix \ref{A:Gra_Shaf}).  
    
Nevertheless, not all super-SM solutions located at the right hand side of the solid curves in Fig.~\ref{fig:lam_kap} can become super-A. For a given $\xi$, $\lambda$ is roughly determined and so is the cylindrical radius $r_A=r_o (\lambda/\omega)^{1/2}$ of the Alfv\'en point. But Eq.~(\ref{eq:Def_kap}) shows that $\kappa$ depends also on $p$ and $\mu$. For a given toroidal current density $p$, the SM constraint provides $\mu$ so that $\kappa$ is fixed. On the other hand, $p$ determines also  the radial magnetic field component at the disk surface and thereby the initial jet bending (see fig.~\ref{fig:Br_SM}). As a consequence, playing with $p$ not only affects the disk vertical equilibrium (SM point) but also this initial jet angle. Not all couples $(\kappa, \lambda)$ fulfill the Grad-Shafranov equation (\ref{eq:Grad_Shaf}). If a solution is not possible, this means that there is no altitude $z_A$ of the Alfv\'en point that can be found starting from the conditions provided at the base of the jet (SM point). Changing the value of $p$ leads to a slight modification of $\kappa, \lambda$ as well as the jet angle at the SM point, allowing thereby to (possibly) meet the A condition. This translates into an adaptation of the altitude $z_A$ of the Alfv\'en point (thus the angle $\Psi_A$) according to the disk conditions. 

The necessary condition $g_B^2>0$ highlights this aspect. When $\lambda$ is large, $r_A/r_o$ is large and gravity plays a negligible role so that $g_B^2>0$ is satisfied whenever $\lambda > 3$ (dashed line in Fig.~\ref{fig:lam_kap}). But gravity cannot be neglected anymore at lower values of $\lambda$, as can be seen in Eq.~(\ref{eq:Grad_BP}). In that case, providing $g_B^2>0$ requires to increase $\sin \Psi_A$, namely to bring the Alfv\'en surface closer to the disk surface. The closer it is to the disk and the less energy is been consumed to reach it. We have been able to find super-A solutions with $\lambda$ as small as 1.6 from low magnetized accretion disks.  

Figure~\ref{fig:Sa_space} shows the parameter space $\xi(\mu)$ for  isothermal super-A solutions in our fiducial case. It is a subset of 
the SM parameter space shown Fig.~\ref{fig:Sm_Space}. The islands are now seen as almost vertical stripes in $\mu$ with a range in ejection index $\xi$. We recover the same results as \cite{ferr97} for near equipartition fields ($n=0$, right) but with a significant enlargement in $\mu$ by almost 4 orders of magnitude ($n=8$, left). 

The color scale indicates the ratio of the radial to the toroidal magnetic field components at the SM point. While $n=0$ solutions are clearly dominated by the radial component, the toroidal field becomes gradually dominant as $n$ increases ($\mu$ decreases). This is of course consistent with the scaling $|B_\phi/B_o|\propto \mu^{-1/2}$ imposed by the SM regularity condition. However, it highlights a possible dichotomy between "magnetic tower" jets \citep{lynd94,Sheik2012}, where ejection is due to a dominant $B_\phi$ field, and "centrifugally-driven" jets \citep{blan82}, where a dominant radial field is of utmost importance. As already pointed out in \cite{ferr97}, these are two expressions of the same magnetic process. However, the dependence $\xi(\mu)$ is quite different for the two cases within each island. It can be seen for instance that for $n=0$ the ejection index $\xi$ increases when $\mu$ increases (although in a very limited range and for $\xi < 0.08$). On the other hand, above $n=3$ ($\mu < 10^{-2}$ for our fiducial case), it is the other way around: $\xi$ decreases for increasing $\mu$ (although for $ 0.4 > \xi > 0.08$). The functional dependence $\xi(\mu)$ can thus be seen as a fingerprint of the dominant ejection mode. 

As discussed earlier, we expect to find other solutions by changing the boundary conditions at the disk mid plane. These solutions should be  located between the islands appearing in Fig.~\ref{fig:Sa_space}, possibly filling-in the actual forbidden area. However, and for the same reasons, the general contour of the parameter space (shown as the solid line) should not be modified. The upper and lower contours describe the curves $\xi\sub{max}(\mu)$ and $\xi\sub{min}(\mu)$ respectively.      

The minimum value of the ejection index $\xi\sub{min}(\mu)$ increases when the magnetization decreases. This a fossil feature of the SM constraint (Fig.~\ref{fig:Sm_Space}). Indeed, when $\mu$ decreases the increasing toroidal magnetic field (Eq.~\ref{eq:Bphi_SM}) leads to a stronger vertical push and to a larger quantity of ejected plasma. Furthermore, as $\mu$ decreases, the number $n$ of spatial oscillations before the SM point increases as well, enforcing thereby the disk to be subjected to them at deeper and more massive layers.
The terminal velocity of the outflow is going to be linked to the value of $\lambda$, which is a function of mostly $\xi$. Thus, the maximal terminal velocity will be determined by $\xi\sub{min}$, which depends mostly on the disk magnetization $\mu$. 

As shown in \citet{ferr97}, the maximum ejection index $\xi\sub{max}$ is determined by the Alfv\'enic constraint. It is interesting to see that it has a non-monotonous behavior, first increasing with $\mu$ until the maximum value $\xi\sub{max}=0.35$  for $n=2$, and then decreasing down to $\xi\sub{max}=0.08$ for $n=0$. As illustrated in  Fig.~\ref{fig:lam_xi}, increasing $\xi$ leads to a decrease in $\lambda$. Low values of $\lambda$ are possible only for large values of $\kappa$, which are accessible only by decreasing $\mu$. As a consequence, the Alfv\'en surface comes closer to the disk ($z_A/r_A$ decreases as $\xi$ increases, see Fig.\ref{fig:A_surface}). However, if we keep decreasing $\mu$, the magnetic energy available in the jet becomes also smaller and jet acceleration less efficient. The Alfv\'en surface moves away from the disk ($z_A/r_A$ increases as $\xi$ increases), requiring thereby a larger magnetic lever arm $\lambda$ to get super-A flows so that $\xi\sub{max}$ decreases.

\section{General accretion-ejection properties}
\label{sec:turb}

We showed that, for our fiducial parameter set, there is a maximum value $\xi_{max}= 0.35$, obtained with $n=2$ for $\mu \sim 10^{-2}$, while the minimum value $\xi_{min}=5\times 10^{-3}$ is obtained with $n=0$ for $\mu= 0.5$. In this section we explore the effect of the turbulence parameter $\alpha_m$ on the existence of cold super-A flows. More specifically, we investigate how turbulence affects the contours of the parameter space, namely the curves $\xi\sub{min}(\mu)$ and $\xi\sub{max}(\mu)$. To avoid confusion, we will not plot the points corresponding to each solutions found, but mark only the contours of the parameter spaces.   

The exploration of the anisotropy parameter $\chi_m$ is done in Appendix \ref{A:chim}. While the turbulence level parameter $\alpha_m$ affects all magnetic field components, $\chi_m$ affects only the toroidal field. Besides, as will be shown, the condition for jet launching from a thin disk introduces the extra link  $\chi_m \sim \alpha_m^2$, that can also be written as $\nu'_m \sim \alpha_m^{-1} V_A h$. We thus focus here only on the effect of $\alpha_m$.

\subsection{Effect of the turbulence level $\alpha_m$}
\label{sec:am_effect}

\begin{figure}
    \centering
    \includegraphics[width=0.45\textwidth]{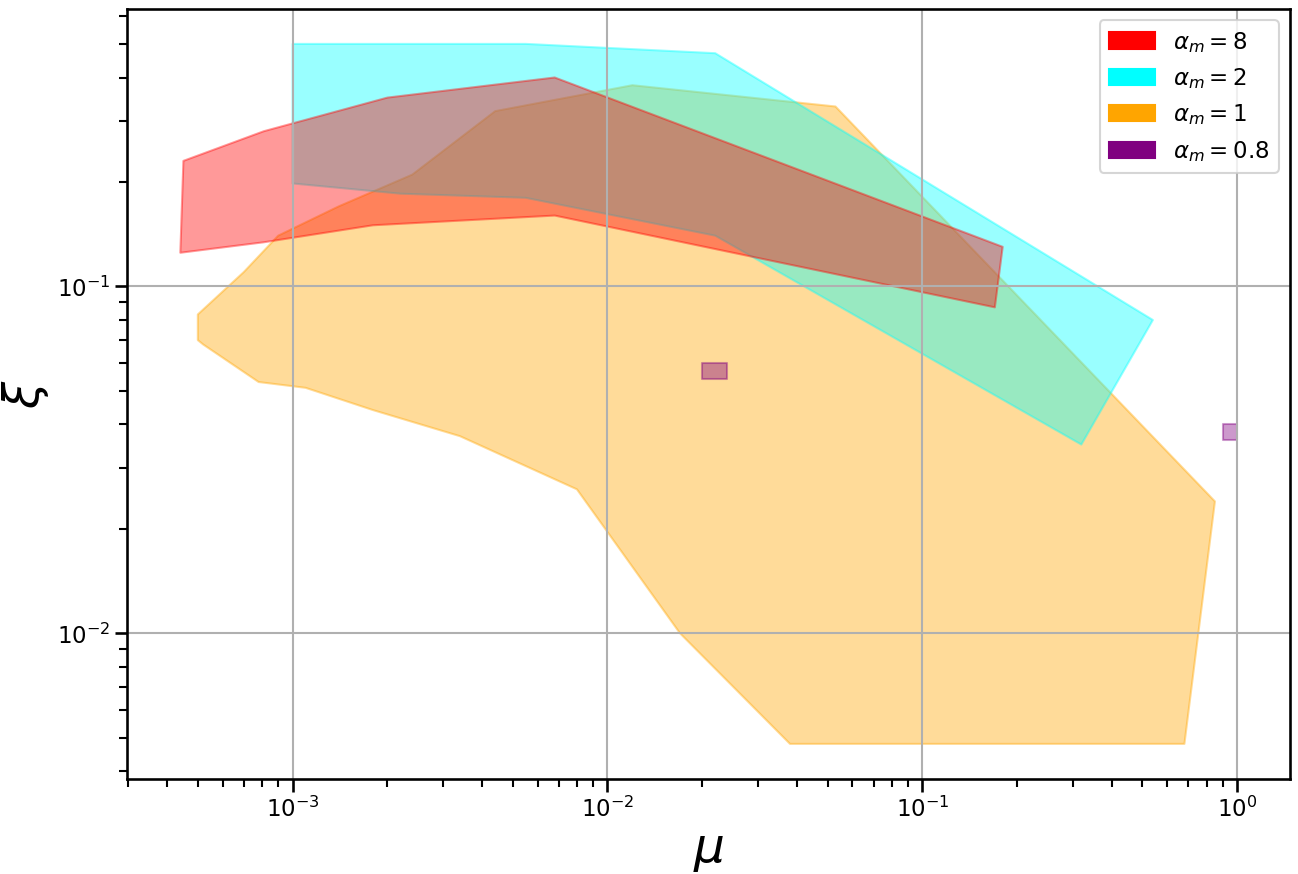}
    \caption{Effect of the MHD turbulence level $\alpha_m$ on the parameter space of super-A flows for $\epsilon=0.1, \chi_m=1, {\cal P}_m=1$. Only the contours of the parameter spaces are shown. Note that $\alpha_m$ affects all three coefficients ($\nu_v,\nu_m, \nu'_m$).}
    \label{fig:Sa_eta}
\end{figure}

Figure~\ref{fig:Sa_eta} shows that $\alpha_m$ has a huge impact on the parameter space of super-A outflows, obtained here with $\epsilon=0.1, \chi_m=1, {\cal P}_m=1$. Two important trends arise with $\alpha_m$:

\begin{itemize}
    \item When $\alpha_m$ increases above unity, $\xi\sub{max}$ is barely modified while the curve $\xi\sub{min}$ increases, leading to a shrinking of the parameter space. Furthermore, as $\alpha_m$ increases, solutions of same $\xi$ are displaced to smaller $\mu$. Note that we did not explore values larger than $\alpha_m=8$, as it corresponds to the scaling deduced from shearing box simulations \citep{Salvesen_16}.
    \item When $\alpha_m$ decreases below unity most of our super-A solutions disappear. For $\alpha_m=0.8$ only two solutions are found,  one with $n=0$ and the other with $n=3$. The fact that no solution can be found for $\alpha_m$ smaller than unity has been already reported in \citet{ferr95,ferr97}. 
\end{itemize}

The displacement to smaller magnetizations as $\alpha_m$ increases arises naturally from the SM constraint (Eq.~\ref{eq:SM_constraint_simp}). Since the velocity at the disk surface is related to the poloidal diffusion, increasing $\alpha_m$ requires to decrease $\mu$ (so that $\mu \alpha_m^2$ remains approximately constant). The drastic diminution of the parameter space as $\alpha_m$ varies is related to the jet launching condition and requires a deeper examination.  

Magnetic ejection occurs only if the jet torque $F_\phi= J_z B_r - J_r B_z \sim - J_rB_z$ switches sign and becomes positive at the disk surface \citep{ferr95}. This requires therefore that $J_r$ decreases on a disk scale height, which can be guaranteed only if
\begin{equation}
    \Gamma = \frac{3}{2}\frac{\chi_{m}}{\alpha_m^{2}}\frac{p}{p-\mathcal{P}_m\epsilon }\lesssim1
\end{equation}
where $\Gamma$ controls the vertical scale of the emf in the induction equation (see Eq.~(\ref{eq:Jr_Ga}) and Appendix~\ref{A:ind_eq} for more details). This condition implies that for solutions with $p$ of order unity, the toroidal current density must adapt to the turbulent properties of the disk with $p \sim \chi_m /\alpha_m^2$ (in turn, this also implies $\chi_m \sim \alpha_m^2$). Thus, when $\alpha_m$ increases $p$ needs to decrease. Since $p$ controls the radial magnetic field at the disk surface (see fig.~\ref{fig:Br_SM}), a decrease in $p$ leads to a decrease of the magnetic vertical compression, thus to a larger mass loss rate from the disk. This feedback on the disk vertical balance explains why the curve $\xi\sub{min}$ increases when $\alpha_m$ increases.  

In order to keep $\Gamma$ near unity, as $\alpha_m$ increases one gets $p \rightarrow p\sub{min}={\cal P}_m \epsilon$ which could be very small. This might be an indication that, for larger values of $\alpha_m$, the MHD solution would eventually try to reverse the sign of the accretion speed, with an outward motion at the disk mid plane ($u_r >0$ and $J_\phi <0$). Such a situation, seen in numerical simulations, is actually forbidden by our assumed boundary condition. This is a general symptom that was discussed in section \ref{sec:MRI_driven} and will also be touched upon in section \ref{sec:Bias}. On the contrary when $\alpha_m$ decreases $\Gamma$ quickly becomes larger than unity since $\frac{p}{p-\mathcal{P}_m\epsilon }$ is bounded by one. Hence, toroidal field induction becomes highly inefficient and the torque $F_\phi$ remains negative, providing no magnetic acceleration. As a consequence, solutions are mostly wiped out when $\alpha_m$ becomes smaller than unity.

\subsection{From jets to winds}
\label{sec:disk_properties}

\begin{figure*}
    \centering
    \includegraphics[width=0.45\textwidth]{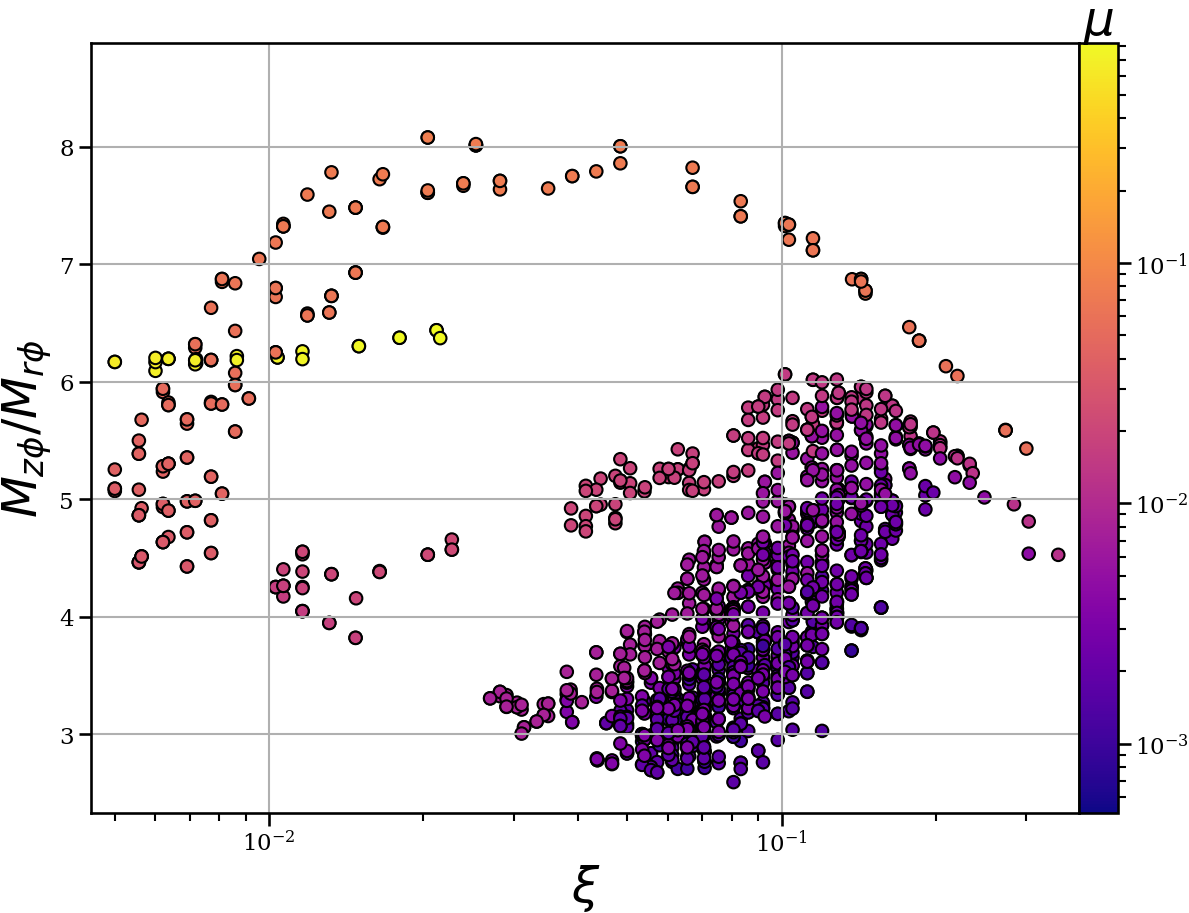}
    \includegraphics[width=0.45\textwidth]{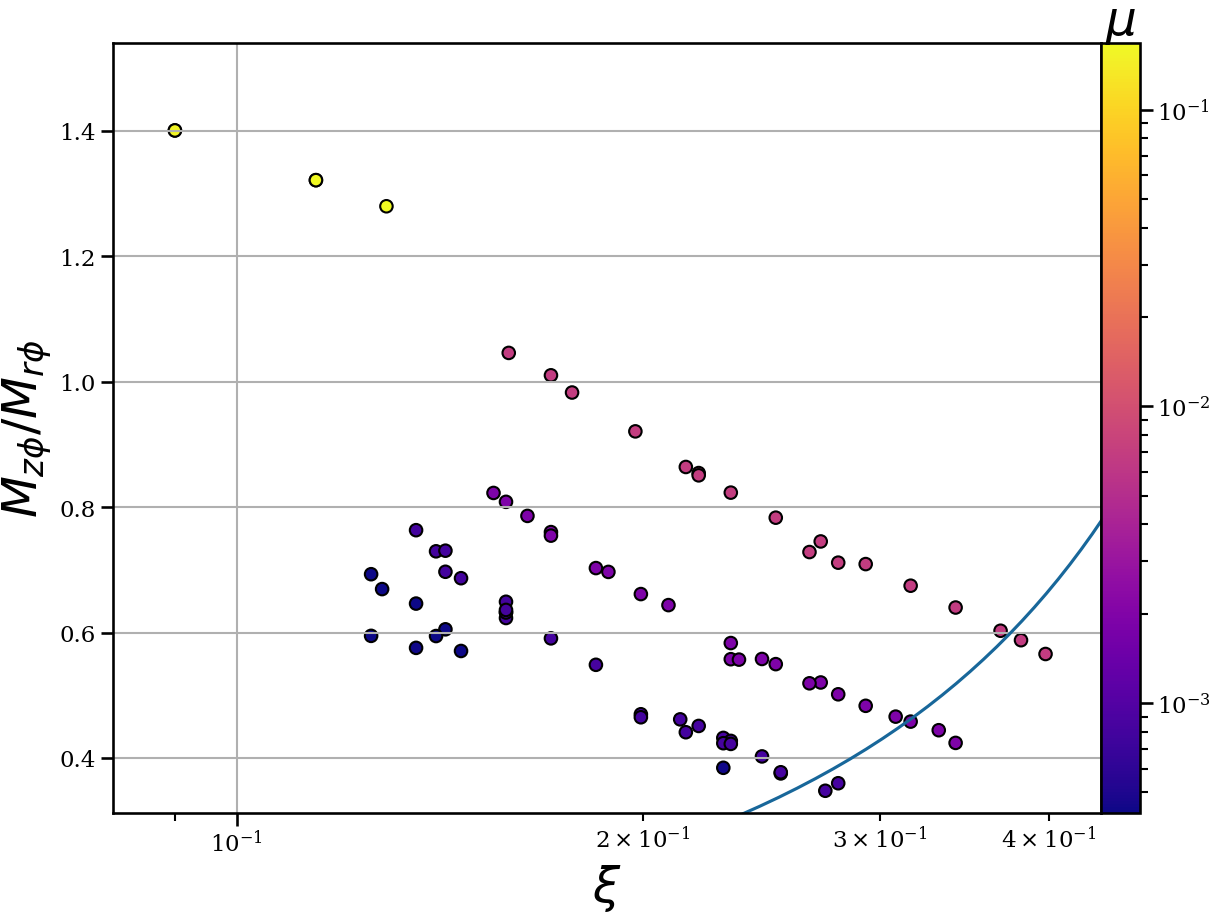}
     \caption{Ratio $ \bar{\Lambda}= M_{z\phi}/M_{r\phi}$ of the vertical stress $M_{z\phi}$ to the radial stresses $M_{r\phi}$ acting on the disk for solutions obtained with $\epsilon=0.1, {\cal P}_m=\chi_m=1$ and $\alpha_m=1$ (left), $\alpha_m=8$ (right). Every point represents a super Alfv\'enic solution and the color is the disk magnetization $\mu$. The non monotonous behavior with $\xi$ and $\alpha_m$ is a consequence of the complex feedback of $\alpha_m$ on $B_r$ and $B_\phi$ and the associated laminar torques (see text). The blue solid line on the right panel is the $\sigma=1$ contour. All solutions located above it display $\sigma >1$ (this is the case for all solutions on the left panel).}
    \label{fig:And_transport}
\end{figure*}

Despite the existence of biases introduced by our prescription of MHD turbulence, we would like to summarize here how the disk magnetization $\mu$ affects some properties such as (1) the disk ejection efficiency, (2) the accretion Mach number, (3) the angular momentum transport and global energy budget and (4) some jet/wind properties.   

The disk ejection efficiency $\xi$ is one of the most important quantities in Jet Emitting Disks (JED) as it provides the link between disk and jet properties. It is defined as $\dot M_a \propto r^\xi$, so it can be measured for instance in numerical simulations that converge to power-law accretion disks (assuming they reach a steady state). Since the local disk magnetization $\mu$ is also easily measurable, simulations can be directly compared to our parameter space $\xi(\mu)$. However, as shown above, one needs to make sure that the turbulence parameters $\alpha_m, \chi_m, {\cal P}_m$ are comparable, as they do influence considerably the final outcome. Note also that our parameter space has a maximal size for $\alpha_m \sim \chi_m \sim {\cal P}_m \sim 1$. For these values, $\nu_v \sim \nu_m \sim \nu'_m \sim V_A h$ and both modes of ejection are at play, allowing thereby to maximally explore the parameter space.

Also, our calculations have been done for cold outflows (isothermal magnetic surfaces) only and it is known that thermal effects may drastically enhance $\xi$ \citep{cass00b}. While the largest value of $\xi$ is imposed by the jet physics (Alfv\'en constraint), the smallest value depends on the disk physics (vertical equilibrium and toroidal field induction). For cold jets, we found a minimum $\xi\sub{min}\simeq 5\, 10^{-3}$ and a maximum $\xi\sub{max}\simeq 0.47$, with a clear tendency of reaching larger $\xi$ with low-$\mu$ solutions. Contrary to previous near-equipartition solutions, cold massive outflows with $\xi\sim 0.1$ are possible as long as spatial oscillations are allowed. But cold super-A solutions with $\xi > 0.5$ remain out of reach.   

The accretion Mach number is defined at the disk equatorial plane as $m_s= - u_r/C_s$, where $u_r$ is the radial accretion speed due to the turbulent and jet torques. However, $m_s$ is also related to turbulence by
\begin{equation}
    m_s = p \alpha_m \mu^{1/2}
    \label{eq:ms}
\end{equation}
through Ohm's law (Eq.~\ref{eq:Sohm}). Given the small range in $p$, it becomes obvious that $m_s \propto \mu^{1/2}$ whatever the dominant torque. No wonder then that supersonic accretion becomes possible only for near-equipartition $\mu>0.1$ solutions. Such a high accretion speed has profound consequences: not only accretion time scales are much shorter than in usual standard accretion disks, but it may lead to optically thin accretion disks with observable features in young stellar objects \citep{comb08,comb10} or X-ray Binaries \citep{marc18b}. One might however question if defining $m_s$ at the disk equatorial plane remains relevant in the case of MRI-like active disks. Indeed, not only the radial velocity is prone to spatial oscillations, but we expect a larger accretion speed at higher altitude. We checked this by computing the density weighted Mach number within the disk,  
\begin{equation}
    \tilde{m}_s =  \frac{\dot M_a}{2\upi r C_s \Sigma}   =  \frac{-1}{C_s}\frac{\int_0^{x_{SM}}u_r\rho\,\diff{x}}{\int_0^{x_{SM}}\rho\,\diff{x}},
\end{equation}
from the disk mid plane up to the SM point. For all solutions found, $\tilde{m}_s$ is never larger than $m_s$ by more than a factor 3. Looking at fig.~(\ref{fig:profiles}) this may seem surprising, but the spatial oscillations tend to compensate each other leading to very little difference with $m_s$. This important result confirms that only $n=0$ solutions provide supersonic accretion. A second interesting aspect is the influence of the turbulence strength $\alpha_m$ on $m_s$. While Eq.~(\ref{eq:ms}) seems to imply that $m_s$ increases with $\alpha_m$, it behaves in the opposite way. This is because, as the diffusion of the poloidal field ($\alpha_m$) increases, the toroidal current density $J_\phi$ decreases with $p \propto 1/\alpha_m^2$ and so $m_s$ decreases.  

Accretion is due to the vertical and radial torques acting in conjunction. To better understand the disk angular momentum transport  it is interesting to look at the ratio $ \bar{\Lambda}= M_{z\phi}/M_{r\phi}$ of the vertical (jet) torque to the radial (laminar and turbulent) torques acting on the disk (see Sect.~\ref{sec:Ideal_MHD_jet} for their expression). It can be analytically estimated by neglecting the radial laminar  contribution, leading to 
\begin{equation}
\bar{\Lambda} \sim \left . \frac{- 2 B_\phi B_z}{\mu_o \alpha_v \epsilon P_o} \right |_{SM} \sim \frac{2 \mu^{1/2}}{\alpha_m {\cal P}_m \epsilon} \left |  \frac{B_\phi}{B_z}\right  |_{SM}   \propto \frac{2}{\epsilon {\cal P}_m \alpha_m}
\end{equation}
where the last expression comes from the SM constraint $| B_\phi/B_z| \propto \mu^{-1/2}$. It can be seen that $\bar{\Lambda}$ is only weakly dependent on $\mu$ and decreases when $\alpha_m$ increases. This is illustrated in Fig.~\ref{fig:And_transport}. Previous cold solutions at near-equipartition fields with $\alpha_m=\chi_m= {\cal P}_m=1$ had $\bar{\Lambda} \sim \epsilon^{-1} = 10$.  Here, as $\mu$ decreases $\bar{\Lambda}$ decreases also because of the increasing effect of the radial torque due to the laminar magnetic field. Large spatial oscillations of the magnetic field within the disk lead to radial transport of the disk angular momentum within the resistive layers that will not be carried away by the jets (channel modes). However, because of the MRI scaling of $\alpha_v$ with $\mu$ (Eq.~\ref{eq:av2}), $\bar{\Lambda}$ remains larger than unity even at $\mu \sim 10^{-3}$ for $\alpha_m=1$. This is no more the case for $\alpha_m=8$, where  solutions with $\bar{\Lambda}$ as low as 0.3 can be obtained (Fig.~\ref{fig:And_transport}, right). Further increasing $\alpha_m$ would thus allow to produce accretion-ejection structures with massive winds ($\xi >0.1$) that carry away a negligible fraction of the released accretion power. We should however remain cautious as our results depend on our assumed vertical profiles for the turbulent coefficients (see next section).  

Finally, there is a need to address the asymptotic properties of the new low-$\mu$ solutions, namely jet speed and collimation. In any case, the fraction of the initial energy that remains stored within the magnetic structure depends on the jet transverse equilibrium (Eq.~\ref{eq:Grad_Shaf}). For cold jets, the maximum asymptotic speed is reached only if the magnetic structure converts almost all its energy into kinetic energy, so that $u_{p\infty}= \sqrt{2E}$. Even in that case, the value of the Bernoulli invariant $E(a)$ is given by Eq.~(\ref{eq:Def_en}) and depends on the magnetic lever arm parameter $\lambda$. For $\lambda >3$, the position of the Alfv\'en point has no impact and the asymptotic jet speed can be estimated with the usual expression $\sqrt{2\lambda-3}$ (in units of the Keplerian speed at the anchoring radius $r_o$). But for smaller $\lambda$, the Alfv\'en surface is much closer to the disk which introduces a deviation from this expression, as shown in Fig.~\ref{fig:up_final}. The difficulty with  low-$\mu$ solutions is to relate $\lambda$ to the disk ejection efficiency $\xi$, since it requires the knowledge of the vertical to radial torques ratio $\bar{\Lambda}$ (see Eq.~\ref{eq:Def_lam}). If we take for instance our fiducial parameter set, Fig.~(\ref{fig:lam_xi}) shows that $\lambda(\xi)$ deviates the most from the usual expression $\lambda = 1 + \frac{1}{2\xi}$ for $\xi \sim 0.1$, corresponding neatly to where $\bar{\Lambda}$ is the smallest. It is thus more problematic to derive $\xi$ directly from the jet asymptotic speed in the case of low-$\mu$ JEDs. 

A clear distinction between MHD winds and jets, which are both super-A MHD flows, can nevertheless be made. Indeed, $\lambda$ determines the importance of the initial magnetic reservoir feeding the jets. Defining the initial jet magnetization $\sigma$ as the ratio of the MHD poloidal Poynting flux  to the kinetic plus thermal (enthapy) energy flux (measured at the jet base taken as the SM point) leads to the general useful relation for cold flows
\begin{equation}
\sigma = \left . \frac{- \Omega_* r B_\phi B_p}{\left(\frac{u^2}{2}+H\right) \rho u_p \mu_o} \right |_{SM} \simeq 2\omega (\lambda -1) \simeq \frac{\omega}{\xi}  \frac{\bar{\Lambda}}{1+\bar{\Lambda}}
\label{eq:sigma}
\end{equation}
Jets are characterized by high speeds (large $\lambda$) that can become farther out self-confined thanks to the dominant hoop-stress wrt to both plasma pressure gradient and centrifugal terms. Jets are therefore Poynting-flux dominated flows with $\sigma >1$. On the other hand, winds are low speed MHD flows with small $\lambda$, with almost no collimation besides that introduced by the external pressure. Winds are therefore matter-dominated flows with $\sigma <1$. 

\begin{figure}
    \centering
    \includegraphics[width = 0.35\textwidth]{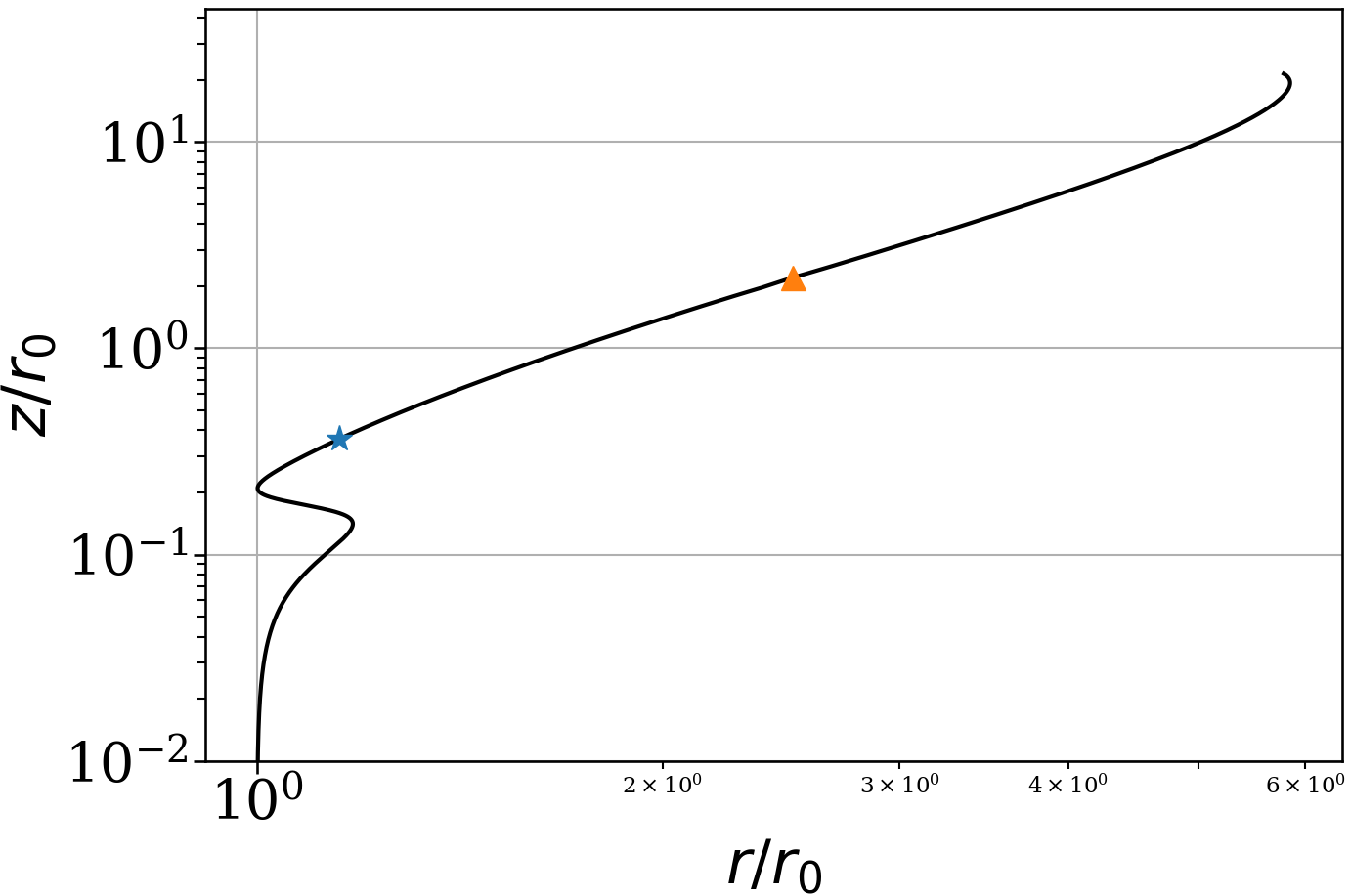}\\
    \includegraphics[width = 0.35\textwidth]{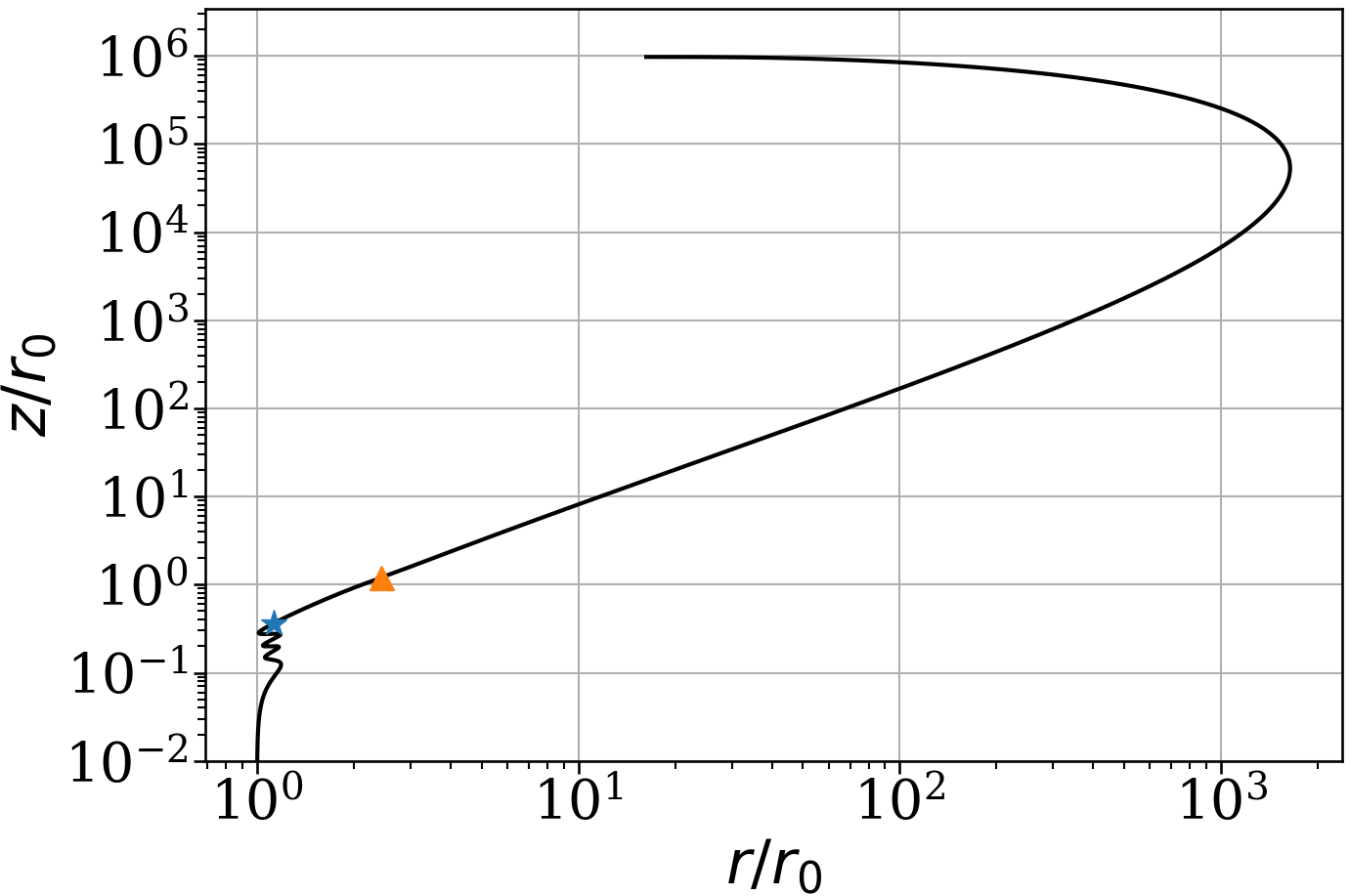}
    \caption{Shape of the poloidal magnetic surfaces for the two weakly magnetized solutions presented in Fig.~\ref{fig:profiles}, both having $\xi=0.1$.  Top: solution with $\mu=6.7\, 10^{-2}$ and $n=1$ spatial oscillation. Bottom: solution with $\mu=5.7\, 10^{-3}$ and $n=3$. The SM point is marked by a blue star and the Alfv\'en point with a red triangle. Both jet solutions open up before recollimating towards the axis.}
    \label{fig:mag_surf}
\end{figure}

We plotted in fig.~(\ref{fig:And_transport}) the contour $\sigma=1$ computed using $\omega=1$. All solutions obtained with $\alpha_m=1$ (left) lie well above this contour and display $\sigma >1$. They are therefore representative of self-confined jets, whatever the disk magnetization $\mu$. Figure~(\ref{fig:mag_surf}) shows the magnetic surfaces of the two low-$\mu$ solutions displayed in Fig.~(\ref{fig:profiles}). It can be seen that their asymptotic behavior is very similar to the high-$\mu$ solutions: the magnetic surfaces first widen before undergoing a recollimation towards the axis (see for instance Fig.~6 and discussion Sect.~5 in \citep{ferr97}). 

But fig.~(\ref{fig:And_transport}) reveals also that some low-$\mu$ solutions obtained with $\alpha_m=8$ do cross the $\sigma=1$ contour. These solutions have $\lambda$ very close to the limiting value $3/2$ and reach $\sigma=1$ because both $\xi$ is quite large and $\bar{\Lambda}$ is small, in agreement with Eq.~(\ref{eq:sigma}). Although these solutions do cross the Alfv\'en point, they meet soon after the modified fast magnetosonic (FM) surface (and stop). However, according to the analysis done in \citet{ferr04}, getting super-FM solutions requires to play with the jet energy equation, which is forbidden with isothermal flows. Moreover, isothermal flows from thin accretion disks are cold and there is no way to provide energy to the outflow when $\lambda$ becomes too small. The only possibility to get winds is then to include relevant thermal effects acting already at the disk surface. Taking into account such a warm corona can be done following the method used in \citet{cass00b}, leading to the build up of a relevant $\Theta$ term in the Bernoulli equation (\ref{eq:Def_en})). Our guess is therefore that warm low $\mu$ solutions obtained with $\alpha_m=8$ will provide proper MHD wind solutions. This is postponed to future work.      

\section{Discussion}

\subsection{Comparison with numerical simulations}
\label{sec:compa_num}

The first paper showing the existence of super-A flows from low magnetized accretion disks was \cite{murp10} and was then extended by a large numerical survey in $\mu$ done by \cite{Stepa2016}. Both works used $\epsilon=0.1$ but while the former included viscosity with $\alpha_v=0.9$ (all stress components), the latter neglected it with ${\cal P}_m=0$, so that the disk angular momentum removal is only done by the jet torque. In terms of magnetic diffusivity, the former used $\chi_m=1$ and $\alpha_m$ starting from 20 and increasing with the radius as $\mu$ decreases, whereas the latter used $\chi_m=2, \alpha_m \simeq 3$ and a constant $\mu$ across the accretion disk. Nevertheless, super-A jets were found in both works with $\mu$ as low as $10^{-4}$, the physics of ejection following the description done here for MRI-like driven outflows. 

The dependences in $\mu$ of the MHD invariants $\kappa, \lambda$ as well as the accretion Mach number $m_s$ shown in \cite{Stepa2016} are also followed by our solutions, although viscosity has been neglected in their work. Indeed, $m_s \propto \mu^{1/2}$ is an analytical result and for $\mu \sim 10^{-3}$, they found jets with $\kappa \simeq 15, \lambda\simeq1.8$, for $\mu \sim 10^{-2}$ $\kappa \simeq 4, \lambda\simeq 2$ and for $\mu \sim 10^{-1}$ $\kappa \simeq 1, \lambda \simeq 3$ (using the proper normalization $\kappa=\tilde k/\mu^{1/2}\epsilon$). These values are consistent with our Fig.~\ref{fig:lam_kap} (derived with $\alpha_m=1, \chi_m=1, {\cal P}_m=1$). Moreover, they find a similar scaling for $|B_\phi/B_r|$ which has lead them to deduce a dichotomy between the magnetic tower and magneto-centrifugal solutions, with a critical magnetization $\mu \simeq 0.01$ separating them. This is also consistent with our work, even though we find that a proper differentiation between solutions should also include the disk ejection efficiency $\xi$ (our Fig.~\ref{fig:Sa_space}). 

The main discrepancy between our semi-analytical work and these numerical "alpha" simulations lies in the disk ejection efficiency. It can be derived knowing $\lambda$ and the jet torque ratio $\bar{\Lambda}$  (Eq.\ref{eq:Def_lam}). Since the simulations of \cite{Stepa2016} have $\bar{\Lambda}/(1+\bar{\Lambda})=1$, we obtain $\xi \sim 0.62, 0.5, 0.25$ for $\mu\sim 10^{-3}, 10^{-2}, 0.1$ respectively (these values are consistent with $\xi$ derived using the independent expression $\kappa \simeq \xi m_s/\mu$). These values of $\xi$ are larger than those obtained in our case. This cannot be an effect of the self-similarity since the simulations themselves do converge to such a situation (see for instance the conical shape of the critical surfaces in their Fig.~1). On the contrary, we believe that it may be due to the fact that these numerical simulations are not cold outflows. Using the same normalization as these authors, the Bernoulli invariant (\ref{eq:Def_en}) writes $e= \lambda- 3/2 + \Theta/2$. Their Fig.~6 clearly shows that the initial thermal content $\Theta$ is non negligible and becomes even of the order unity for several simulations. Since it has been demonstrated that heat deposition at the disk upper layers leads to enhanced mass loss \citep{cass00b}, we argue that this is the main cause of the observed discrepancy in $\xi$ (see also discussion on a possible numerical bias p9-10 in \cite{murp10}). 

For that same reason, we cannot compare our cold solutions with the 3D numerical simulations done by \cite{beth17}, as they assumed the existence of a heating term acting at the disk surface (leading to a huge ejection efficiency $\xi\sim 1$). Furthermore, non-ideal terms (Hall and ambipolar diffusion) have been assumed in the induction equation and it is not clear yet how this affects the generation of the toroidal magnetic field and thereby the whole ejection process.  

The 3D simulations of vertically isothermal, ideal MHD disks with a magnetization $\mu \sim 10^{-3}$ done by \cite{Zhu_Stone} should in principle be comparable to our own work. Indeed, they obtain a MRI active accretion disk giving rise to a super-A flow launched from $r_{i}=0.5$ to $r_{e}=5$ with $\lambda \sim 10-14$ and carrying a mass fraction $2 \dot M_j/\dot M_a \sim 0.4\% \simeq \xi \ln (r_{e}/r_{i})$. These values correspond to a very small ejection efficiency $\xi \sim 1.7-2 \,\times 10^{-3}$ which, since they report $\bar{\Lambda}\sim 5\%$ only of the disk angular momentum transport due to the wind, is indeed consistent with Eq.(\ref{eq:Def_lam}). Within our framework, such a small value of $\bar{\Lambda}$ would require to increase $\alpha_m$ up to $\sim 15$ (according to Eq.~(\ref{eq:av2}), since the authors report $\alpha_v= 0.5$ and ${\cal P}_m \sim 1$). However, Fig.(\ref{fig:Sa_eta}) shows that for $\alpha_m=8$ we obtain even more massive solutions with  $\xi>0.08$. The reason of this discrepancy is probably due to the MHD turbulence itself. Indeed, our semi-analytical solutions are mathematically exact but depend on the vertical profiles that are assumed for all the transport coefficients. 

The work of \cite{Zhu_Stone} but also \cite{Taka2018} show that global 3D simulations that include the wind torque have a turbulent diffusivity and radial angular momentum transport up to almost 10 pressure scale heights. This is quite surprising as the original consensus was that the disk (defined by its scale height) would be responsible for turbulence. It seems on the contrary that MHD turbulence, under the presence of a large scale $B_z$ field at low magnetization levels, is able to persist and self-sustain high above the disk. As a consequence, the resistive disk survives at larger altitudes and ideal MHD ejection occurs only further up, decreasing thereby the disk mass loss $\xi$. This new feature can be easily incorporated within the self-similar framework through the use of different vertical profiles for $\nu_v, \nu_m$ and $\nu'_m$. This deserves definitely further investigation.  
    
\citet{Scepi2018} used local shearing box simulations to infer the angular momentum transfer rate due to a magnetized wind in the context of dwarf novae. The resulting prescription has then been used to predict the secular evolution of such a system \citep{Scepi2019}. \citet{Scepi2018} found that $|B_\phi/B_o| \simeq 30$ (a constant) at the disc surface for all $\mu\gtrsim 10^{-4}$ while we find $| B_\phi/B_o|\propto \mu^{-1/2}$. This difference could have several origins: while \cite{Scepi2018} defines the surface as the altitude where $B_\phi$ is maximum, we define the surface at the SM point. More importantly, the shearing box does not satisfy Eq.~(\ref{eq:Sohm}) since the poloidal field is allowed to be advected radially \citep{lesu13}. This implies that shearing box solutions are not strictly speaking secular stationary solutions, but also that the scaling (\ref{eq:Bphi_SM}) is not valid in a shearing box. This is yet an other illustration of the limitations of shearing box solutions to describe winds. This difference will likely have a significant impact on the dynamics of these systems as our scaling predicts a stronger wind at low magnetization compared to shearing box solutions. 

Finally, we show that it is possible to derive a scaling of the wind stress as a function of the disk pressure from our solutions. We find that $M_{z\phi}\propto \mu^{1/2}$ (Fig.~\ref{fig:vert_torque}), which can be of use in secular disc models to include the effect of a magnetic wind on the disk evolution.

\subsection{Caveats}
\label{sec:Bias}

This work is subject to biases arising from (1) stationarity  (2) our imposed geometry and (3) the prescriptions used for the MHD turbulence. We briefly discuss these important points below.  

The validity of the steady-state assumption depends on the time scales considered. In thin accretion disks, the local dynamical time scale is of the order of the Keplerian period. Another important time scale is the accretion time scale, which is $1/(\epsilon m_s)$ times longer than the dynamical time. Such a huge difference in these two scales allows to make a simple ordering. Our steady-state solutions can thus be considered valid on scales that are longer than the local dynamical time but smaller than the accretion time. On this longer time scale, both mass and magnetic fields can be advected inward/diffused outward, leading to modifications of the radial profiles. This is illustrated for example in \cite{Stepa2016}, where the disk magnetization, $\mu$, is seen to evolve on these long time scales. Note also that one might incorporate magnetic flux advection within a steady-state approach by modifying Eq.~(\ref{eq:Sohm}) (see for example \cite{Conto2017}). It is however unclear why such an effect should also follow a self-similar scaling.

What we call geometry covers actually different assumptions. The first obvious one is self-similarity and has been already discussed extensively in \citet{ferr97, ferr04}. Clearly, jet asymptotics are influenced since neither the inner (jet axis) nor the outer (jet boundary) regions can be described within this mathematical formulation. However, jets should be nevertheless well approximated by these solutions when the critical surfaces are close to cones, a situation which arises whenever the jet emitting disk is established over a large radial extent ($r_e>> r_i$ and $\dot M_a \propto r^\xi$). Note however that most of the mathematical relations between jet and disk parameters are general and can be used to interpret and understand steady-state 3D simulations. For that same reason, we expect our parameter space to be only weakly affected by self-similarity.  
 
The second aspect is the $z$-symmetry imposed on the bipolar magnetic field structure, even for the flux function $a(r,z)$ and odd for $B_\phi$. While such symmetric fields seem to be realized in \cite{Zhu_Stone} simulations, the work of \cite{beth17} shows that accretion disks, at least in the non ideal case, can accommodate a plethora of different symmetries. 

A final geometrical aspect is the boundary conditions imposed at the disk equatorial plane. As discussed  in section \ref{sec:MRI_driven}, the number of allowed MRI-like wavelengths depends critically on them. In our case, the assumed accretion at $z=0$ drives a positive electromotive force ($p$ or $J_\phi >0$) leading to the generation of a positive radial magnetic component $B_r$ within the disk. So, by assumption, solutions with an outward radial motion at the mid plane (hence $B_r <0$) have been discarded. Such solutions could accommodate as well with the required outward bending at the disk surface, but probably with an extra half MRI-like wavelength. This can be seen in \cite{Zhu_Stone} for instance, where $u_r(z=0)>0$ due to a dominant and positive radial magnetic torque at the disk mid plane. We do seem to recover this behavior. Indeed, as $\alpha_m$ increases (hence $\alpha_m \rightarrow 10$ or so, as measured in MRI simulations), both the sonic Mach number $m_s$ and $B_r$ (i.e. the toroidal current $p$) decrease. Our lack of accretion-ejection solutions for large $\alpha_m$ could thus be an indication that the boundary condition for the accretion speed (sign of $p$) must be changed. This slight modification can be easily implemented within a self-similar approach.     

As discussed in Sect.~2.2, our work assumes a magnetic diffusion of turbulent origin. Indeed, the existence of self-confined jets in a wide range of objects advocates for a universal mechanism that would be independent of the physical conditions within disks, and in particular of their ionization degree. The natural source for magnetic diffusion in low ionized plasmas is ambipolar diffusion $\nu_{AD}$. Would it be dominant, the induction equation of the toroidal magnetic field would be deeply affected, modifying significantly the generation of the toroidal field and possibly our results. Ambipolar diffusion can be estimated in the disk mid-plane as $\nu_{AD} \simeq V_A^2/\nu_{ni}$, where $\nu_{ni}= 1/\tau_{ni}$ is the neutral-ion collision frequency. Comparing this expression with our turbulent prescription $\nu_m= \alpha_m V_A h$ shows that $\nu_{AD}$ is negligible whenever $\alpha_m \gg \mu^{1/2} A_m^{-1}$, where $A_m = V_A^2/\nu_{AD} \Omega_K= 1/\Omega_K \tau_{ni}$ is the Elsasser number. Thus, for Elsasser numbers around unity \citep{beth17}, this estimate shows that ambipolar diffusion can be safely neglected with respect to turbulent diffusion (as long as turbulence is going on). This would be in agreement with the universality of accretion-ejection. However, because of the lack of ionisation in certain regions, circumstellar accretion disks are known to harbor dead zones with accretion occurring only at the disk surface \citep{gamm1996,Fle2003,bai13}. We believe that the interdependent accretion-ejection structure will be mostly the same (ie same link between parameters), with an offset from the disk equatorial plane. Note that such a layered accretion structure could be actually described within a self-similar approach, by designing proper vertical turbulent profiles and boundary conditions at the disk equatorial plane. This is left for future work.  

The second major caveat of our work is related to the prescriptions used for the MHD turbulence, namely the viscosity and magnetic diffusivities. As argued before, our choice of $\alpha_v$ and $\alpha_m$ are consistent with our current knowledge of MRI and the way the stress (viscosity) scales with the initial magnetization $\mu$ (\cite{Salvesen_16} and references therein). We would like to stress however that knowledge on the turbulent diffusion of magnetic fields is scarse. Global simulations \citep{Zhu_Stone}) and shearing box studies \citep{lesu09, guan09,from09} report an effective magnetic Prandtl number $\mathcal{P}_m$ of order unity, but this is far from being fully assessed. Besides, the anisotropy $\chi_m$ of MRI turbulence has been measured only in non stratified shearing box setups, i.e. a very idealized configuration.  

Even though our prescriptions and scalings agree qualitatively with MRI turbulence, the vertical profiles of the turbulent coefficients used in our work (a Gaussian $\exp\left(-x^2\right)$), do not seem to be in agreement with recent numerical studies of accretion disk turbulence. Not only the scale height for turbulence is much larger than the disk scale height \citep{Zhu_Stone,Taka2018}, but also the profile of the turbulent "viscous" stress does not seem to simply scale as $\nu_v \rho$ (see for instance Fig.~5 in \cite{fromang_meridional_2011}). In addition, several potentially important processes related to turbulence have been largely ignored in our model and in particular the pressure due to turbulent magnetic fluctuations. This term is known to strongly affect the disc vertical equilibrium \citep{Salvesen_16} for $\mu \gtrsim 10^{-3}$, which could increase dramatically the disc thickness and therefore the quantitative predictions of our model. This advocates therefore for the use of more elaborate closure prescriptions, possibly educated from 3D simulations of MRI turbulence. Note that enhanced diffusion at higher altitudes (due for instance to parasitic instabilities such as Kelvin-Helmholtz \citep{Latter_2010}) could smooth out the spatial oscillations, building up a magnetic configuration closer to that obtained in 3D simulations. We leave these modifications of the self-similar prescriptions for the future.

\section{Conclusion}

Motivated by recent global 3D simulations of accretion disks threaded by a weak vertical magnetic field and showing the launching of jets, we revisited the self-similar accretion-ejection solutions for cold (isothermal) magnetic surfaces. By allowing spatial oscillations of all quantities within the disk, we have been able to extend the previous parameter space by 4 orders in magnitude in the disk magnetization $\mu$, namely from $\mu= 10^{-4}$ to almost unity.   

We recovered the previous solutions and found a new class of MRI-like driven outflows from weakly magnetized disks, in agreements with some simulations. The role of MRI-like spatial oscillations is shown to be essential in order to provide the required bending of the poloidal field lines at the disk surface. Cold outflows from weakly magnetized accretion disks have the tendency to be more massive than their strong field (near equipartition) counterpart, leading to a critical Alfv\'en surface closer to the disk surface.  

There is a continuity in behavior as $\mu$ increases. Low $\mu$ isothermal solutions are quite massive with a typical ejection index $\xi \sim 0.1$ (increasing with $\mu$) and are mostly driven by the pressure of the toroidal field. The previously published high $\mu$ solutions are much less massive, with a typical ejection index $\xi \sim 0.01$ (decreasing with $\mu$), thus faster and mostly centrifugally driven. These are however two manifestations of the same magnetic acceleration process, linking accretion to ejection in an interdependent way.      

It is striking to realize that the confusion between the Shakura-Sunyaev viscosity parameter $\alpha_v <1$ and the turbulence parameter $\alpha_m$ has led to restrict all past self-similar papers to values $\alpha_m \leq 1$. However, modern 3D simulations seem to imply $\alpha_m$ larger than unity instead. We explored, for the first time, the accretion-ejection behavior under such circumstances, allowing to reach the turning point situation where more angular momentum is being transported radially within the disk than vertically into the jets.    

We propose a simple criterion, based on the initial jet magnetization $\sigma$, allowing to discriminate between winds and jets. Solutions with $\sigma>1$ are Poynting flux-dominated jets, representative of fast, tenuous and self-collimated outflows, whereas solutions with $\sigma<1$ are matter-dominated winds, namely massive, slow and weakly collimated outflows. While the first kind (jets) is achievable at all magnetization levels (and are shown here), only weakly magnetized disks ($\mu < 10^{-2}$ or less) could provide winds. However, some energy input must be added in order to provide a positive Bernoulli integral. A further development would thus be to include heating at the disk upper layers, 
as in \citet{cass00b}, mimicking the existence of irradiation from a central source. This is known to dramatically enhance the mass loss $\xi$ as well, further decreasing $\sigma$ and allowing for magneto-thermal winds. 

The disk magnetization $\mu$ appears to be the main control parameter for determining the intrinsic accretion-ejection properties, such as MHD turbulence and the fraction of the disk angular momentum that is transported by the jets. The existence of some external illumination would then be an extra factor allowing to change jet/wind properties through mass loss enhancement ($\xi$). Combining these two properties allows to draw an interesting framework. As argued in \cite{ferr06}, it is reasonable to expect $\mu$ to be a decreasing function of the radius, the magnetic field being dragged in by the  accretion flow. If the innermost regions reach near equipartition ($\mu$ between 0.1 and 0.8), then a proper JED solution can be established, with supersonic accretion and fast self-confined jets (low $\xi$). Such inner regions would have clear astrophysical signatures, in young stellar objects \citep{comb08} and around compact objects \citep{marc18b}. The outer disk regions could have a much lower magnetization $\mu <<1$ and accrete at a subsonic pace, while launching massive winds whenever an efficient irradiation is present. Whether or not such Wind Emitting Disks or WEDs are generic in astrophysics requires further investigation.

\section*{Acknowledgements}
We thank the referee for providing thoughtful comments on the manuscript. This project has received funding from the European Research Council (ERC) under the European Union's Horizon 2020 research and innovation program (Grant agreement No. 815559 (MHDiscs))




\bibliographystyle{mnras}
\bibliography{biblio} 



\appendix

\section{Self-similar equations}
\label{A:self_eq}
For the sake of completeness, the full set of MHD equations solved are reported in this section. We define the self-similar functions $f_i(x)$ with $x= z/h$ and $h=\epsilon r$
	\begin{align}
		\rho&=\rho_o\Tfrac{r}{r_o}^{\zeta_4}f_4\,\nonumber,\qquad\qquad
		P=P_o\Tfrac{r}{r_o}^{\zeta_{10}}f_{10}\,,\nonumber\\
		T&=T_o\Tfrac{r}{r_o}^{\zeta_7}f_7\,,\nonumber\qquad\qquad
		u_z=\epsilon u_o\Tfrac{r}{r_o}^{\zeta_3}f_3\,,\nonumber\\
		u_r&=-u_o\Tfrac{r}{r_o}^{\zeta_2}f_2\,,\nonumber\qquad\qquad
		\Omega=\Omega_o\Tfrac{r}{r_o}^{\zeta_5}f_5\,,\nonumber\\
		B_\phi&=qB_o\Tfrac{r}{r_o}^{\zeta_1-1}f_1\,,\nonumber\quad\quad
		a(r,z)=a_o \Tfrac{r}{r_o}^{\beta}\psi
	\end{align}
where the subscript "o" stand for a quantity evaluated at the disk equatorial plane ($x=0$). Here, $q= \mu_o J_{r0}h/B_0$ is the normalized radial current density, $a_o= B_o r_o^2/\beta$ the magnetic flux with $B_o$ the vertical field component, $\Omega_o= \delta_o \Omega_{Ko}$ the angular velocity, $P_o= \rho_o \Omega^2_{Ko} h^2$ and $u_o= m_s C_s$, with $C_s = \Omega_{Ko} h$ defining thereby the accretion Mach number $m_s$.  
The shape of a magnetic surface anchored at $r_o$ is defined by $a(r,z)=a_o$ and is provided by $r= r_o\psi^{-1/\beta}$.  The three transport coefficients $\nu_v, \nu_m, \nu'_m$ (see Sect.~\ref{sec:Eq_para}) are written $\nu_A= \nu_{Ao}\Tfrac{r}{r_o}^{\zeta_8}f_8$ where the profile is a simple Gaussian\footnote{Note that \citet{murp10} used $f_8(x)=\exp\left({-2x^2}\right)$, while \cite{Stepa2016} used  $f_8(x)=\exp\left({-0.5x^2}\right)$, namely ideal MHD starting sooner. This may explain why the latter found more massive jets than the former.} $f_8(x)=\exp\left({-x^2}\right)$. Inserting these self-similar functions into the set of PDE (\ref{eq:Scon}-\ref{eq:Sohm}) allows to separate them into an algebraic set of equations on the exponents $\zeta_i$ and a set of ODEs on the functions $f_i$. This leads to the unique solution for a near-Keplerian, gas supported, accretion disk
\begin{align}
\beta &= \frac{3}{4}+\frac{\xi}{2},\qquad\qquad \zeta_1=\frac{\xi}{2}-\frac{1}{4}, \qquad\qquad \zeta_2=\zeta_3=-\frac{1}{2} \nonumber\\
\zeta_4 &= \xi-\frac{3}{2}, \qquad\qquad \zeta_5=-\frac{3}{2}, \qquad\qquad\quad \zeta_7=-1\nonumber \\
\zeta_8 &= \frac{1}{2},  \qquad\qquad\qquad  \zeta_{10}= \xi-\frac{5}{2} \nonumber 
\end{align}
where $\xi$ is the exponent of the disk accretion rate $\dot M_a \propto r^\xi$. 	
Defining  $ \tilde{f}_4 = \ln f_4$ and $f_i'= df_i/dx$, allows to express mass conservation and the equation of state as the following ODEs
\begin{align}
\tilde{f}'_4 (f_3 + x f_2)  &=  (\xi - 1)f_2 - f'_3 - x f'_2  \\
f_{10}  &=   f_4f_7 
\end{align}
Similarly, the radial, vertical and toroidal momentum transport equations become respectively
\begin{align}
m_s^2 \epsilon^2 f_4(-\zeta_2 f_2^2 + f'_2 (f_3 + xf_2)) &=  -f_4 \delta_0^2 f_5^2 + \frac{f_4}{(1+x^2\epsilon^2)^{3/2}}\nonumber \\
&\quad + \epsilon^2 ( \zeta_{10} f_{10} - x f'_{10}) \nonumber \\
&\quad + \mu q^2 \epsilon^2f_1 (\zeta_1 f_1 - xf'_1) \nonumber\\
&\quad + \mu \frac{\Delta'\psi}{\beta}\left(\psi - x\frac{\psi'}{\beta}\right) \\
m_s^2 \epsilon^2 f_4(-\zeta_3 f_2 f_3 + f'_3(f_3 + x f_2))  &=  -\frac{x f_4}{(1+ x^2\epsilon^2)^{3/2}} - f'_{10}\nonumber\\
&\quad- \mu q^2 f_1 f'_1 - \mu \psi'\frac{\Delta'\psi}{\beta^2 \epsilon^2} \\
2f_4(f_3 + x f_2)f'_5 - f_2 f_4 f_5  &=   \frac{\Lambda}{1+\Lambda}\left(\psi f'_1 - \frac{\zeta_1}{\beta} f_1 \psi' \right)\nonumber\\
&\quad- \frac{1}{1+\Lambda}f\sub{turb} 
\end{align}
where the modified laplacian (toroidal current density) is
\begin{equation}
\Delta'\psi =  \psi''(1+\epsilon^2 x^2) + \epsilon^2[(2\beta -3)x \psi' - \beta(2-\beta)\psi]
\end{equation}
and $\Lambda=\frac{p}{\mathcal{P}_m\epsilon}-1$ is the ratio of the magnetic to the viscous torque at the disk mid plane. The function 
$f\sub{turb}= f_4f_8$ is the prescription used for the turbulent stress. While the above ODEs are valid both in the disk and in the ideal MHD jet regime, the induction equation requires to deal with each regime in a separate way.  

Within the resistive disk, Ohm's law (\ref{eq:Sohm}) and the induction equation (\ref{eq:Sinduc}) become respectively
\begin{align}
f_8 \Delta'\psi  &=   -\mathcal{R}_m \epsilon^2 ( \beta f_2 \psi - \psi'(f_3 + xf_2)) \\
(f_8 f'_1)'  &= \epsilon^2 x\left(f_8(\zeta_1 f_1 - x f'_1)\right)' -\epsilon^2 f_8(\zeta_1 f_1 - xf'_1)(\beta - \frac{5}{2}) \nonumber \\
& \quad - \chi_m \frac{\mathcal{R}_m\delta_o}{qm_s} \left(\frac{3}{2\beta} \psi' f_5 + \psi f'_5\right)
  + \chi_m \mathcal{R}_m \epsilon^2 \beta f_1 f_2  \nonumber\\
&\quad + \chi_m \mathcal{R}_m \epsilon^2(f_3 + x f_2)(f'_1 - f_1 \tilde{f}'_4)
\end{align}
where $\mathcal{R}_m=\frac{r u_o}{\nu_{mo}}= p/\epsilon$ is the magnetic Reynolds number. When the ideal MHD regime becomes relevant, these equations write respectively
\begin{align}
&(f_3 + xf_2) \psi' = \beta \psi f_2 \\
&(f_3 + xf_2)(f'_1 - f_1 \tilde{f}'_4) = \frac{\delta_o}{qm_s\epsilon^2} \left(\frac{3}{2\beta} \psi' f_5 + \psi f'_5\right) - \beta f_1 f_2 
\end{align}
We need to complement this set of ODEs with an energy equation providing $f_7$. Isothermal magnetic surfaces are represented by $T/T_o= 1$ along each surface anchored at a radius $r_o$, which translates into $f_7= \psi^{-1/\beta}$. The system of ODEs requires the following boundary values 
\begin{align} 
f _ { 1 } ( 0 ) & = f _ { 3 } ( 0 ) = 0 \nonumber\\ 
f _ { 2 } ( 0 ) & = f _ { 4 } ( 0 ) = f _ { 5 } ( 0 ) = \psi ( 0 ) = f _ { 7 } ( 0 ) = f _ { 10 } ( 0 ) = 1\nonumber \\ 
f _ { 2 } ^ { \prime } ( 0 ) & = f _ { 4 } ^ { \prime } ( 0 ) = f _ { 5 } ^ { \prime } ( 0 ) = \psi ^ { \prime } ( 0 ) = f _ { 7 } ^ { \prime } ( 0 ) = f _ { 10 } ^ { \prime } ( 0 ) = 0\nonumber \\ 
f _ { 1 } ^ { \prime } ( 0 ) & = - 1 \nonumber\\ 
f _ { 3 } ^ { \prime } ( 0 ) & = \xi - 1\nonumber
\end{align}

We thus get a complete set of equations that can be formally written as $M.X=P$, where $M$ is a matrix and $P$ a vector depending only on the variable $x$ and the functions $f_i$, while $X$ is a vector of their derivatives $f'_i$. Propagating the equations requires to get $X= M^{-1}P$, where $M^{-1}$ can only be computed as long as the determinant of the matrix $M$ does not vanish. This occurs at the disk equatorial plane $x=0$ (which is a fixed point of nodal type) and at each critical point of the outflow (see \cite{ferr95} for more details). The integration cannot therefore start at $x=0$ and a Taylor expansion must be made.

\section{The Grad-Shafranov constraint}
\label{A:Gra_Shaf}

The Grad-Shafranov equation (\ref{eq:Grad_Shaf}) or GSE requires a regularity condition at the Alfv\'en point, where the Alfv\'enic Mach number $m= u_p/V_{Ap}$ is equal to unity. Its position ($r_A, z_A$) is labelled by the angle $\Psi_A$ such $\cot \Psi_A= z_A/r_A$. This regularity condition provides the value $g_A= g (\Psi_A,\kappa,\lambda, \omega,e)$ of the amount of the poloidal electric current that remains to be used beyond this point, normalized to the current available at the jet base \citep{ferr97}. It is therefore known for a given position $\Psi_A$ of the Alfv\'en point and the MHD invariants $\kappa,\lambda, \omega,e$. The angle $\Psi_A$ is however not known a priori. This is due to the fact that the GSE (\ref{eq:Grad_Shaf}) is a PDE of mixed type, where the shape and position of the Alfv\'en surface need to be imposed. In a time-dependent problem, it would naturally emerge by taking into account all causal connections. Within a self-similar approach of jets, this translates into a conical Alfv\'en surface with a possible choice of the angle $\Psi_A$ (as done for instance in \cite{vlah00}). In our case however, we do have an extra geometrical constraint since the ideal MHD jet is connected to the accretion disk. Our integration starts from the disk equatorial plane and is propagated upwards (increasing variable $x=z/\epsilon r$) so that a trans-A solution can be found only if, at some point  $x_A$, $g(x_A) =g_A$ is verified, which then fixes also $\Psi_A$. So the position $\Psi_A$ of the Alfv\'en surface emerges also as function of the disk parameters. 

\begin{figure}
    \centering
    \includegraphics[width=0.45\textwidth]{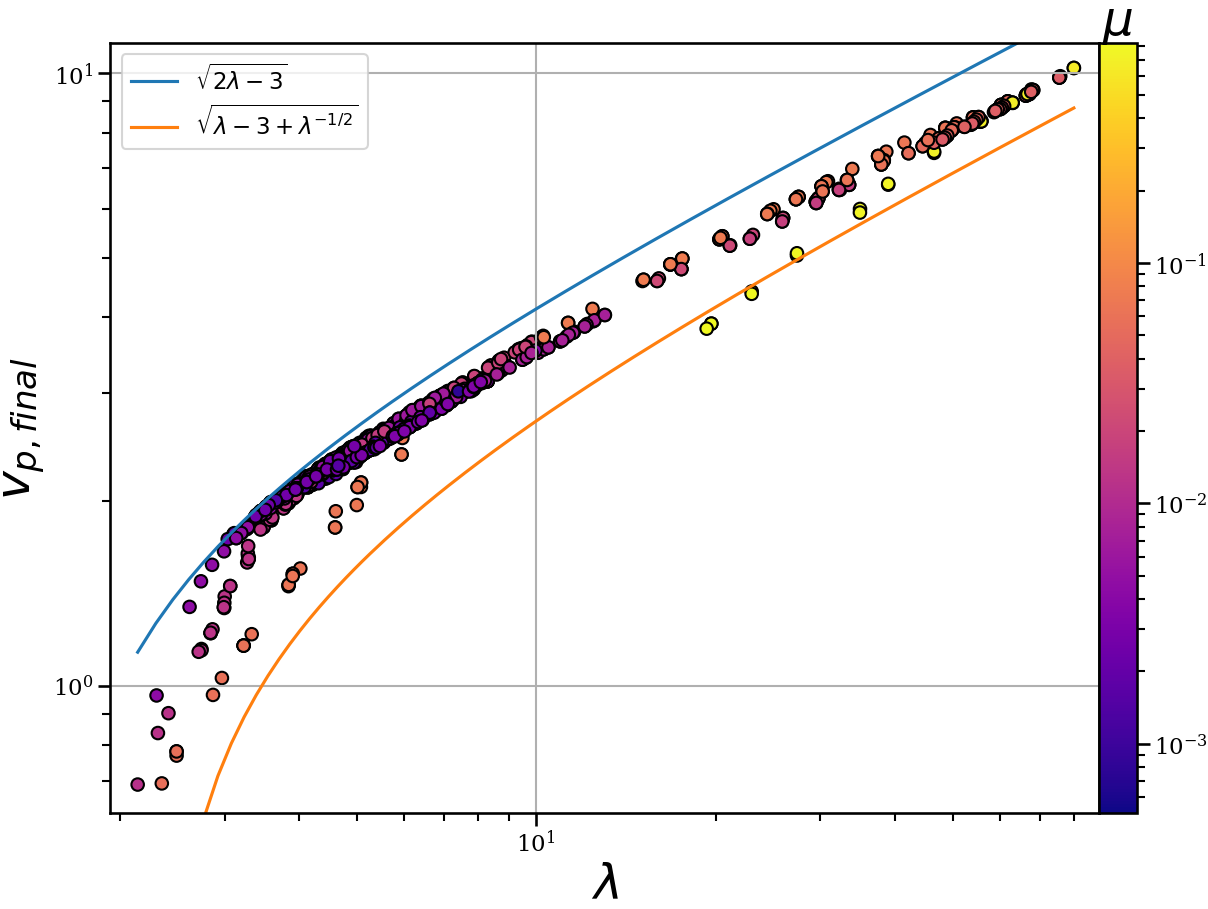}
    \caption{The terminal jet poloidal velocity (in units of the Keplerian speed at the footpoint) as function of the magnetic lever arm $\lambda$ for our super-A solutions found in the fiducial case. The blue and orange solid curves correspond respectively to upper and lower analytical limits (see text). The fact that solutions do not reach the maximum speed indicates that the magnetic field still conserves a fraction of the available energy.}
    \label{fig:up_final}
\end{figure}

Once $g_A$ is known, all jet quantities can be computed at the Alfv\'en surface as function of their mid plane value and jet invariants: $\Omega_A= \omega \Omega_{Ko} (1-g_A)$, $\rho_A= \mu \epsilon^2 \kappa \rho_o$, $u_{pA}=   \varv_A\Omega_{Ko} r_o$, $B_{pA}= \kappa \varv_A B_o$, $|B_\phi/B_p|_A= g_A(\omega \lambda)^{1/2}/\varv_A$.  In these expressions, the value $\varv_A$ gives the poloidal velocity at the Alfv\'en point and is obtained from the Bernoulli invariant (\ref{eq:Def_en}), namely $\varv_A^2= \omega \lambda (g_B^2 - g_A^2)$ where, for a cold flow (negligible enthalpy), one has
 \begin{equation}
 \begin{array}{lcl}
    \label{eq:def_g0}
    g_B^2=1+ \frac{2 e'}{\omega \lambda} &  & e' =- \frac{2+\omega^2}{2} + \sin \Psi_A \sqrt{\frac{\omega}{\lambda}}
\end{array}    
\end{equation}
Here, $g_B$ represents the maximum value for the acceleration efficiency $g_A$ as imposed by energy conservation. Since $g_B^2>0$, this gives a constraint on the minimum energy reaching the Alfv\'en point, namely a minimum magnetic lever arm.  The closer the Alfv\'en surface and the smaller $\lambda$ can be. We found super-A solutions close to the limiting value $\lambda=3/2$. For $\lambda>3$, the position of the Alfv\'en surface plays no role anymore.   
The velocity at the Alfv\'en point can be seen as a lower limit of the jet terminal velocity $u_{p\infty}$, while the upper limit is  $\sqrt{2\lambda - 3}$ for cold flows (in units of the Keplerian speed at the footpoint). These two limits are shown in Fig.~\ref{fig:up_final} for our super-A solutions obtained in the fiducial case. For the lower limit, we used for simplicity $\omega=1$, $\sin \Psi_A=1$ and $g_A=0$. Although rather crude, these two limits do bracket quite well our numerical solutions.  

Writing the GSE at the Alfv\'en point leads to 
\begin{multline}
    \aderiv { \mathcal{E} } - \left . \frac { \vgrad a \cdot \vgrad m^{ 2 } } { \mu _ { o } r_A^2\rho_A }  \right |_{A} + g_A\Omega_{*} \aderiv {\Omega _ { * }r_A^2} + (1-g_A)\Omega_*r_A^2\aderiv{\Omega_*} \\+ \left . \frac { B _ { \phi} ^ { 2 } +B _ { p} ^ { 2 } } { \mu _ { o } \rho_A} \right |_A \aderiv{  \ln \eta }  =0
\end{multline}
where  $\mathcal{E}(a)=E(a)-\Omega_*^2r_A^2$. Computing the derivatives of the MHD invariants is quite simple within the self-similar ansatz, leading to  
\[  \aderiv{A}=\frac{\zeta_A}{\beta}\frac{A}{a_o}=\frac{\zeta_A}{\beta}\frac{A}{B_or_o^2} \]
for an invariant $A$ of radial exponent $\zeta_A$. 
Self-similarity introduces a geometrical constraint by imposing that the Alfv\'en surface is a cone. Along a magnetic surface, one has necessarily $B_z/B_o - \frac{z}{r} B_r/B_o= (r/r_o)^{-2}$. Defining the local jet opening angle as $\tan \theta= B_r/B_z$, allows to write 
\begin{equation}
\cos \theta -  \frac{z}{r} \sin \theta = B_o r_o^2 /B_p r^2
\label{eq:costheta}
\end{equation}
which is verified everywhere along a magnetic surface, and in particular at the Alfv\'en point.  
Making use of this and remembering that $m^2$ is only a function of the self-similar variable $x$, leads after some algebra to 
\begin{equation}
\left . \vgrad a \cdot  \vgrad m^{ 2 } \right |_A = \frac{2}{g_A}\left(\frac{r_A}{r_0}\right)^2\frac{B_{pA}^2}{B_0}\left( \frac{\cos\theta_A \omega}{\kappa\lambda \varv_A}-1\right)
\end{equation}
Inserting this expression into the GS constraint provides
\begin{equation}
    \label{eq:fin_Grad_Shaf}
    g_A(g\sub{GS}-g_A)=(g_B^2-g_A^2)\left (1-\frac{\cos\theta_A \omega}{\kappa\lambda\varv_A} \right )
\end{equation}
where 
\begin{equation}
    \label{eq:def_gmax}
    g\sub{GS}=\frac{3}{4} - \frac{2+\omega^2}{4\omega \lambda} - \frac{\zeta_4}{4} g_B^2
\end{equation}
is another maximal value for $g_A$, imposed by the jet transverse equilibrium. The constant $\zeta_4= \xi -3/2$ is the radial exponent of the density (it comes from the $\aderiv{  \ln \eta }$ term). Noting that the jet opening angle writes  
\[ \cos\theta_A = \frac{\frac{\omega}{\kappa\lambda\varv_A}+\frac{z_A}{r_A}\sqrt{1+\left(\frac{z_A}{r_A}\right)^2-\left(\frac{\omega}{\kappa\lambda\varv_A}\right)^2}}{1+\left(\frac{z_A}{r_A}\right)^2} \]
and inserting it into Eq.~(\ref{eq:fin_Grad_Shaf}) allows finally to express the GS constraint into a quadratic equation on $X= g_A/g_B$
\begin{equation}
\label{eq:Alf_poly}
(k^2c^2 + \cos^2\Psi_A)X^2 - 2cX(k^2 - \sin^2 \Psi_A) + \frac{k^2-1}{k^2}(k^2- \sin^2 \Psi_A) = 0
\end{equation}
where $c= g\sub{GS}/g_B$ and $k^2= \kappa^2/\kappa\sub{min}^2$ with the minimum mass load $\kappa\sub{min}$ defined with
\begin{equation}
    \label{eq:Grad_Cons}
\kappa\sub{min}^2\lambda^3 g_B^2=\omega
\end{equation} 
Which corresponds to the absolute lower limit for getting a super-A flow. Indeed, for $g_A=0$ the GS constraint can only be satisfied for a minimum value $k^2=1$ (see Eq.~\ref{eq:Alf_poly}). Equation (\ref{eq:Grad_Cons}) is none other than a generalization of equation (3.1) in \cite{blan82}. 

\begin{figure}
    \centering
    \includegraphics[width=0.45\textwidth]{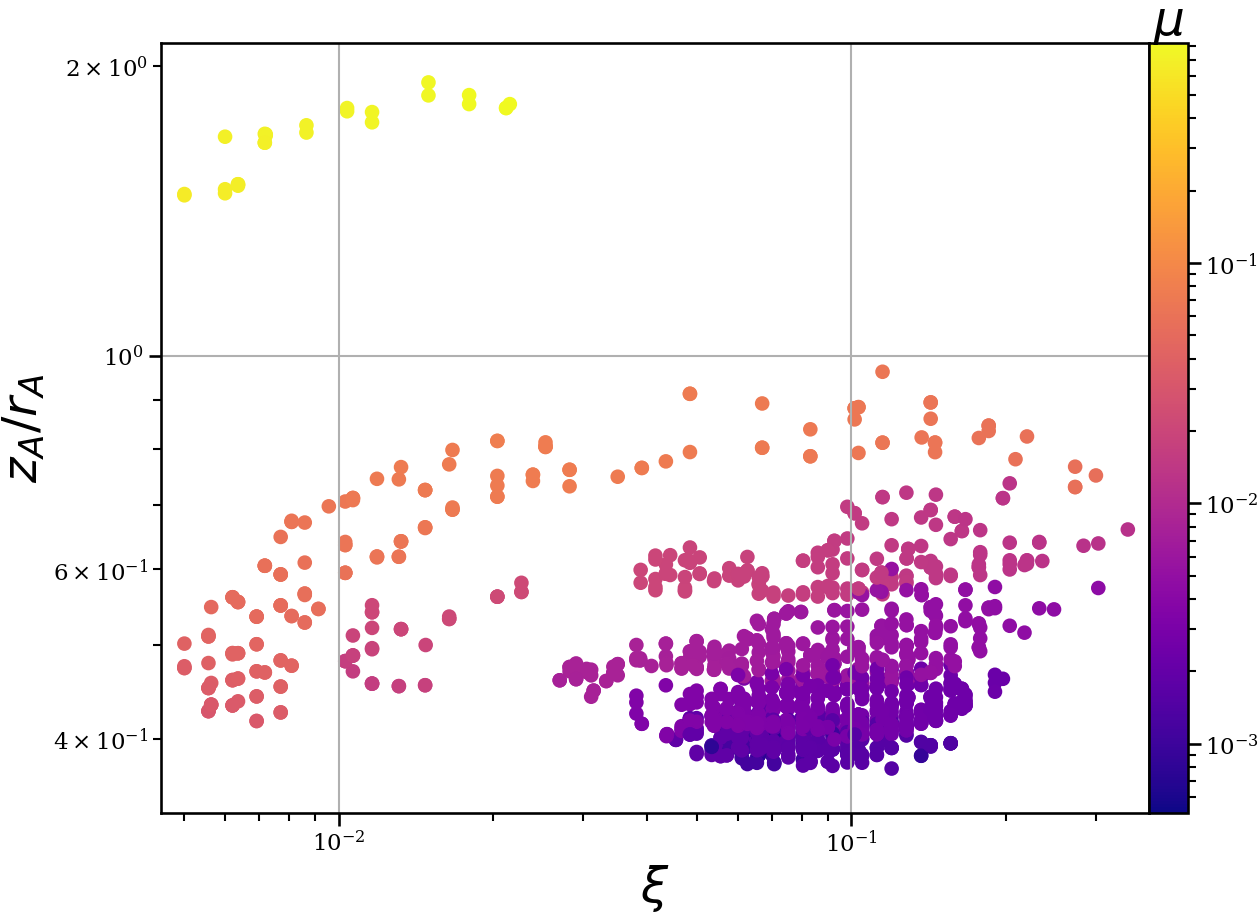}
    \caption{Position of the Alfv\'en point as function of the disk ejection efficiency $\xi$ for our fiducial parameter set. The color scale is the disk magnetization $\mu$. The behavior of the Alfv\'en position is different at high and low disk magnetizations, large $\xi$ requiring both smaller $\mu$ and an Alfv\'en surface closer to the disk surface (see text).}
    \label{fig:A_surface}
\end{figure}

\begin{figure}
    \centering
    \includegraphics[width=0.45\textwidth]{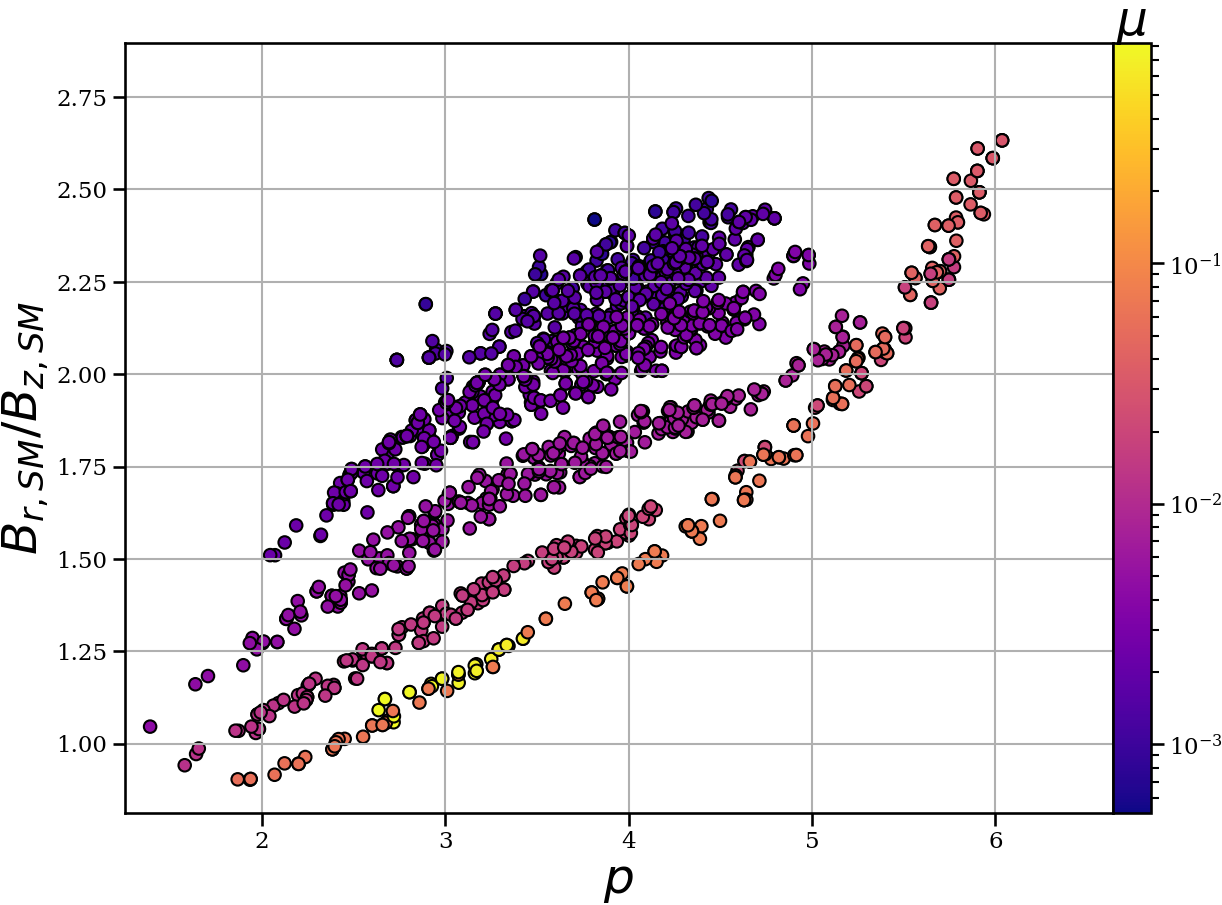}
    \caption{Bending $B_r/B_z$ of the poloidal magnetic field evaluated at the SM point, as function of the parameter $p$
    (toroidal current density at the mid-plane) and the disk magnetization $\mu$ (colors) for our fiducial parameter set. The jet initial opening angle increases monotonously with increasing $p$ and decreasing $\mu$.}
    \label{fig:Br_SM}
\end{figure}

The GS constraint (\ref{eq:Alf_poly}) shows that there are always two positive roots 
\begin{equation}
\left . \frac{g_A}{g_B} \right |_\pm= \frac{  c(k^2 - \sin^2 \Psi_A) \pm \cos \Psi_A \sqrt{(k^2 - \sin^2 \Psi_A)\left(c^2- \frac{k^2-1}{k^2} \right)}}
{k^2c^2 + \cos^2 \Psi_A }
\end{equation}
In the limit $k^2=1$ (super-A solutions found along the solid curve in Fig.~\ref{fig:lam_kap}), the GS constraint provides $g_A=0$ and 
\begin{equation}
 g_A = g_{GS} \frac{2 \cos^2\Psi_A}{c^2 + \cos^2 \Psi_A } 
 \label{eq:ga_gs}
 \end{equation}
When $\lambda$ increases ($\xi$ decreases), both $g_B$ and the contrast factor $c$ tend to unity so that solutions with large $g_A$ become possible. When $\lambda$ decreases ($\xi$ increases) but remains bigger than 3, $g_B$ goes to zero very slowly. Hence,the Alfv\'en surface can still get closer to the pole ($\Psi_A$ decreases), to allow for a more efficient acceleration (larger $g_A$). But when $\lambda\lesssim3$ ($\xi\gtrsim0.25$) and decreases, $g_B$ goes to zero unless the Alfv\'en surface moves closer to the disk ($\Psi_A$ increases, see Eq.~\ref{eq:def_g0}). This non monotonous behavior, seen only for super-A solutions at low magnetization levels, is illustrated in Figure~\ref{fig:A_surface}. It can also be seen that $\Psi_A$ increases as the disk magnetization $\mu$ decreases. For a given $\xi$ (hence $\lambda$), decreasing $\mu$ leads to an increase in $k$ and $g_A$ is then maximized by increasing $\Psi_A$. An intuitive picture would be that the smaller $\mu$, the larger the MRI-like induced magnetic bending at the disk surface (see fig.~\ref{fig:Br_SM}). A larger inclination enhances jet acceleration allowing to meet the Alfv\'en point at a smaller altitude $z_A$.


\section{The toroidal field at the disk surface}
\label{A:ind_eq}

One of the most important quantities in JEDs is the importance of the toroidal magnetic field component at the disk surface (taken here as the SM point). Using the self-similar expression, the magnetic shear writes
\begin{equation}
\left . \frac{B_\phi}{B_z} \right |_{SM} = q f_1 (x_{SM}) = \frac{\mu_o J_{ro} h}{B_o} f_1 (x_{SM})
\end{equation}
showing that it depends on both (1) the amount of the radial current density $J_{ro}$ flowing within the disk and (2) the vertical profile $f_1 (x_{SM})$.  

The conducting disk behaves like an unipolar inductor (Faraday disk), where rotation through a magnetic vertical field gives rise to an electromotive force driving a radial electric current $J_{ro}$. The disk drives therefore two electric circuits, corresponding each to one of the jets. Therefore, the value of $J_{ro}$ (as measured by the shear parameter $q$) is related to the global electric circuit designed by the existence of these two jets. On the other hand, the disk angular momentum equation (\ref{eq:Smonto}) requires that $q= \alpha_m \delta_o (p-{\cal P}_m \epsilon)/2\mu^{1/2}$, which shows already the "natural" tendency to have $\left | \frac{B_\phi}{B_z} \right |_{SM} \propto \mu^{-1/2}$.  

\begin{figure}
    \centering
    \includegraphics[width=0.45\textwidth]{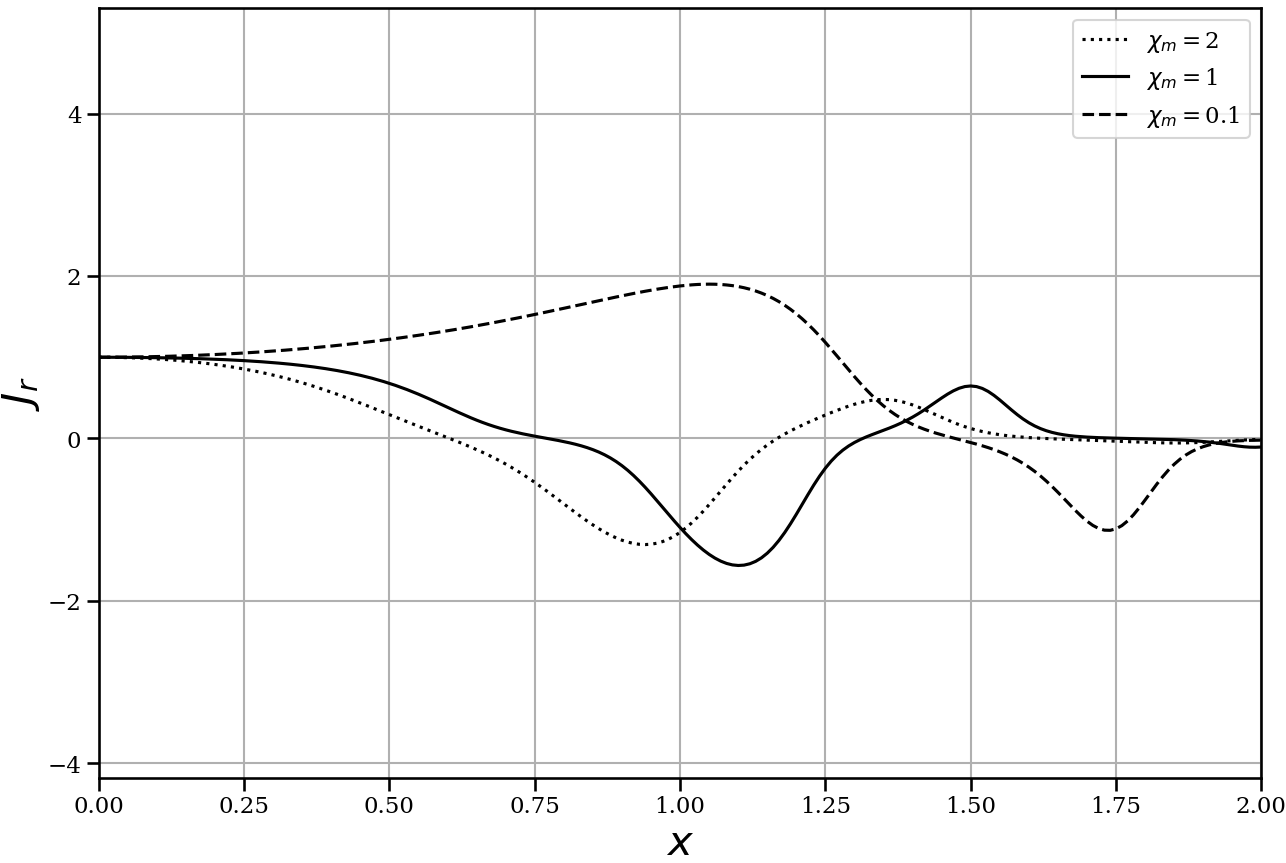}
    \caption{Vertical profiles of the radial electric current density $J_r$ (normalized to $J_{ro}$) for different values of $\chi_m$ and 
    $\Gamma=1.3,0.74,0.62$ . These super-SM solutions were calculated with $\mu\sim2\, 10^{-2}$, $\xi= 0.4$ and $\alpha_m={\cal P}_m=1$.}
    \label{fig:Jr}
\end{figure}

The second important element is the vertical profile of the radial current density $J_r$, which determines the value $f_1 (x_{SM})$.  The induction equation (\ref{eq:Sinduc}) writes
\begin{equation}
\nu'_{m}J_{r}(z) = \nu'_{mo}J_{r0} \, +\, \mu_or\int\limits_{0}^{z}\diff z\,  \Bpo\cdot\vgrad\Omega 
\end{equation}
where $J_z$ and $B_\phi$ advection have been neglected, in agreement with the thin disk approximation. In the absence of any shear (rigid rotation), one would have $\nu'_{m}J_{r}(z) = \nu'_{mo}J_{r0}$ and the vertical profile of $J_r$ would only depend on the profile of the turbulent diffusivity $\nu'_m$. In thin accretion disks, the radial shear is dominant and a Taylor expansion of the rhs of the above equation leads to
\begin{equation}
\nu'_{m}J_{r} =  \nu'_{mo}J_{r0} \left ( 1 - \Gamma x^2 \right ) \quad \mbox{with}\quad  \Gamma=\frac{3}{2}\frac{\chi_{m}}{\alpha_m^{2}}\frac{p}{p-\mathcal{P}_m\epsilon }
\label{eq:Jr_Ga}
\end{equation}
The differential rotation is therefore counteracting the $J_{ro}$ electric current, which is crucial to deviate the current towards the disk surface and allow current closure in the jets. Now, the amplitude of this effect, as measured by $\Gamma$, is challenged  by the vertical decrease of the turbulent diffusivity $\nu'_m$. Indeed, if $\Gamma>1$, $J_r$ is going to quickly tend to zero, despite the decrease of $\nu'_{m}$. On the contrary, if $\Gamma<1$,  $J_r$ is going first to reach a plateau (or may even increase) before decreasing to 0 (at a higher altitude)\footnote{Note that we used a profile $\exp\left(-x^2\right)$ for $\nu'_m$. Using $\exp\left(-\tau x^2\right)$ would have lead instead to a comparison between $\Gamma$ and $\tau$, $\Gamma> \tau$ giving rise to a small value of $f_1 (x_{SM})$ (see Appendix B in \citep{ferr97}).}. In the former case, $f_1 (x_{SM})$ is small (remember that $J_r = - \partial B_\phi/\partial z$), whereas in the latter case $f_1 (x_{SM})$ can be quite large. This is illustrated in Fig.~\ref{fig:Jr}, where different values of $\Gamma$ have been obtained by playing with different $\chi_m$. Note that in our approach, $\Gamma$ is not free but depends on the turbulence parameters $\alpha_m, {\cal P}_m$ and $\chi_m$. 

As argued in \cite{ferr95}, the magnetic (jet) torque $F_\phi= J_z B_r - J_r B_z \simeq - J_r B_z$ must change sign around the disk surface so that magnetic acceleration can take place. This requires therefore $\Gamma$ to be of order unity, allowing $J_r$ to conveniently decrease to zero neither too close to the equatorial plane, nor too far away.  Assuming $\Gamma \sim 1$ then leads to $\chi_m \sim \alpha_m^2$ providing $\nu'_m \sim \alpha_m^{-1} V_A h$. But this is some optimal estimate as solutions can be found with $\Gamma \neq 1$.  

For a given value of the ratio $\chi_m/\alpha_m^2$, the toroidal current parameter $p$ can be adapted in order to get super-SM solutions. When $\chi_m$ decreases (and/or $\alpha_m$ increases), solutions with $\Gamma$ of order unity require $p \rightarrow p\sub{min}$, with $p\sub{min}={\cal P}_m \epsilon$. Rather small values of $\chi_m$ are thus allowed since the ratio $p/(p-p\sub{min})$ can be very large. On the contrary, when $\chi_m$ increases (and/or $\alpha_m$ decreases), $p$ needs to increase but the ratio  $p/(p-p\sub{min})$ is bounded by 1. This explains why the parameter space is disappearing so abruptly in this case.  

\section{Effect of the turbulence anisotropy $\chi_m$}
\label{A:chim}

\begin{figure}
    \centering
    \includegraphics[width=0.45\textwidth]{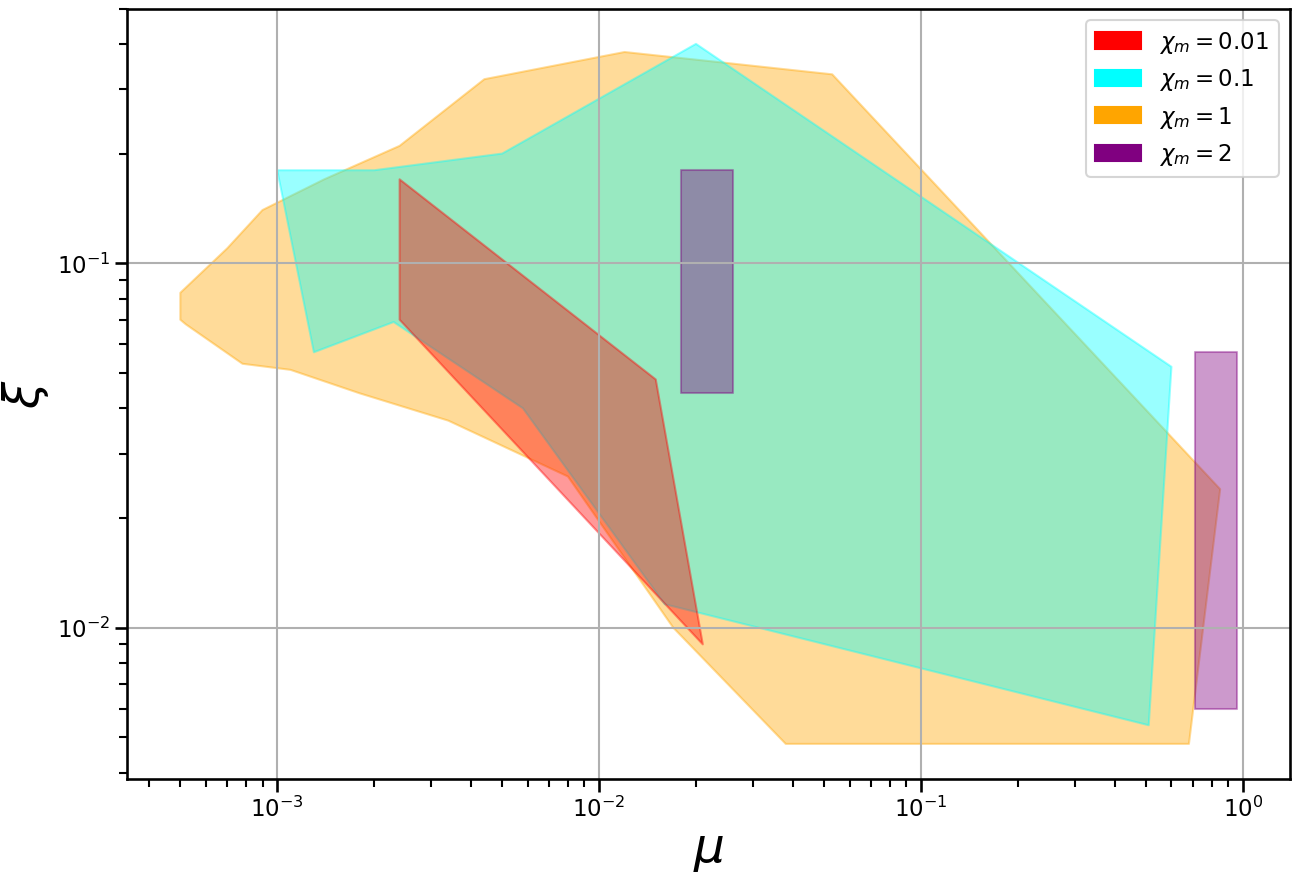}
    \caption{Effect of the anisotropy $\chi_m$ of the turbulent magnetic diffusivity on the parameter space of super-A flows for $\epsilon=0.1, \alpha_m=1, {\cal P}_m=1$. A value $\chi_m < 1$ corresponds to a stronger diffusion of the toroidal magnetic field wrt to the poloidal field.}
    \label{fig:Sa_chim}
\end{figure}

Figure~\ref{fig:Sa_chim} shows the effect of $\chi_m$ on the existence of super-A flows obtained with $\epsilon=0.1, \alpha_m=1, {\cal P}_m=1$. The parameter space shrinks when $\chi_m$ decreases below unity, with again a displacement to smaller values of $\mu$, but with a more complex behavior on $\xi\sub{min}$ and $\xi\sub{max}$. For $\chi_m=2$, we find the same behavior as for $\alpha_m=0.8$, namely only two sets of separated solutions, one at $n=0$ and the other at $n=3$. Above this value, we found no super-A solution.

Increasing $\chi_m$ leads to the same kind of behavior as reducing $\alpha_m$. Hence, the parameter space of $\chi_m=2$ and $\alpha_m=0.8$ are similar (see fig.~\ref{fig:Sa_eta}). Using the same argument, increasing $\chi_m$ forbids the outflow acceleration and  the majority of solutions are wiped out. 

Decreasing $\chi_m$ has however a much less pronounced effect as increasing $\alpha_m$. This is due to the fact that the latter is controlling all magnetic field components while the former only affects the toroidal field. The parameter space for $\chi_m=0.1$ appears quite similar to the fiducial case. However, it can be seen that decreasing $\chi_m$ to 0.01 leads to a much reduced parameter space, shifted to lower $\mu$ and smaller $\xi\sub{max}$. A magnetic shear $|B_\phi/B_z|$ too large produces a strong vertical pinch on the disk so that solutions tend to have both smaller $\xi$ and $\mu$ (large $\xi$ become forbidden for a given $\mu$). However, solutions with a magnetization $\mu > 10^{-2}$ become now impossible. Indeed, the only possibility to lower the magnetic compression at high $\mu$ would be to reduce also the magnetic bending, namely $B_r/B_z$. 

In practice, our super-A solutions require a toroidal current parameter $p \rightarrow p\sub{min}={\cal P}_m \epsilon$ (see Appendix~\ref{A:ind_eq}). This points to a situation where the field lines would have a different curvature at the disk equatorial plane, namely $J_\phi <0$ (thus $u_r >0)$ at $z=0$. This situation is forbidden by our choice of boundary conditions and hints to a clear bias of our solutions for $\chi_m <<1$. Whether or not such anisotropy is physically relevant remains however to be assessed.      

\bsp	
\label{lastpage}
\end{document}